\documentclass{aa}  

\usepackage{graphicx}
\usepackage{txfonts}
\usepackage{enumitem}
\usepackage{amsmath}
\usepackage{subcaption}
\usepackage{alphalph}
\usepackage{morefloats}
\usepackage{caption}
\usepackage{placeins}
\usepackage{float}
\usepackage{xtab,afterpage}
\usepackage{longtable,booktabs,chngcntr}

\let\OLDthebibliography\thebibliography
\renewcommand\thebibliography[1]{
  \OLDthebibliography{#1}
  \setlength{\parskip}{0pt}
  \setlength{\itemsep}{0pt plus 0.3ex}
}

\renewcommand\thesubfigure{\alphalph{\value{subfigure}}}

\makeatletter
\def\blfootnote{\xdef\@thefnmark{}\@footnotetext}
\makeatother

\begin{document}

   \title{COBRaS: The e-MERLIN 21\,cm Legacy survey of Cygnus OB2}
   \titlerunning{The COBRaS 21\,cm Survey of Cyg OB2}

   \subtitle{}

   \author{J. C. Morford\inst{1}, D. M. Fenech\inst{1, 2}, R. K. Prinja\inst{1}, R. Blomme\inst{3}, J. A. Yates\inst{1}, J. J. Drake\inst{4}, S. P. S. Eyres\inst{5,}\inst{6}, A. M. S. Richards\inst{7}, I. R. Stevens\inst{8},  N. J. Wright\inst{9}, J. S. Clark\inst{10}, S. Dougherty\inst{11, 12}, J. M. Pittard\inst{13}, H. A. Smith\inst{4} \and J. S. Vink\inst{14}
          }
    
   \authorrunning{J. C. Morford et al.}

   \institute{Department of Physics {\&} Astronomy, University College London, Gower Street, London WC1E 6BT, UK\\
         \and
                 Astrophysics Group, Cavendish Laboratory, University of Cambridge, UK
                 \email{dmfd2@cam.ac.uk}
         \and
             Royal Observatory of Belgium, Ringlaan 3, 1180 Brussels, Belgium
         \and
             Harvard-Smithsonian Centre for Astrophysics, 60 Garden Street, Cambridge, MA 02138, USA
         \and
             Jeremiah Horrocks Institute, University of Central Lancashire, Preston, PR1 2HE, UK
         \and
             Faculty of Computing, Engineering and Science, University of South Wales, Pontypridd, CF37 1DL, UK
         \and
             UK ARC Node, JBCA, Alan Turing Building, University of Manchester, M13 9PL, UK
         \and
             School of Physics and Astronomy, University of Birmingham, Edgbaston, Birmingham, B15 2TT, UK
         \and
             Astrophysics Group, Leonard-Jones Building, Keele University, Staffordshire, ST5 5BG, UK
         \and
             Department of Physical Sciences, The Open University, Walton Hall, Milton Keynes, MK7 6AA, UK
         \and
             Dominion Radio Astrophysical Observatory, National Research Council Canada, PO Box 248, Penticton, BC V2A 6J9, Canada
         \and
             ALMA, Alonso Cordoba 3107, 7630355 Vitacura, Chile
         \and
             School of Physics and Astronomy, E. C. Stoner Building, The University of Leeds, Leeds, LS2 9JT, UK
         \and
             Armagh Observatory, College Hill, Armagh, BT61 9DG, Northern Ireland, UK}

   \date{Received June 15, 2017; accepted December 26, 2019}

 
  \abstract
   {The role of massive stars is central to an understanding of galactic ecology. It is important to establish the details of how massive stars provide radiative, chemical, and mechanical feedback in galaxies. Central to these issues is an understanding of the evolution of massive stars, and the critical role of mass loss via strongly structured winds and stellar binarity. Ultimately, and acting collectively, massive stellar clusters shape the structure and energetics of galaxies.}
   {We aim to conduct high-resolution, deep field mapping at 21\,cm of the core of the massive Cygnus OB2 association and to characterise the properties of the massive stars and colliding winds at this waveband.}
   {We used seven stations of the e-MERLIN radio facility, with its upgraded bandwidth and enhanced sensitivity to conduct a 21\,cm census of Cygnus OB2. Based on 42 hours of observations, seven overlapping pointings were employed over multiple epochs during 2014 resulting in 1$\sigma$ sensitivities down to $\sim$ 21\,$\mu$Jy and a resolution of $\sim$ 180\,mas.}
   {A total of 61 sources are detected at 21\,cm over a $\sim$ 0.48$^{\circ}$ $\times$ 0.48$^{\circ}$ region centred on the heart of the Cyg OB2 association. Of these 61 sources, 33 are detected for the first time. We detect a number of previously identified sources including four massive stellar binary systems, two YSOs, and several known X-ray and radio sources. We also detect the LBV candidate (possible binary system) and blue hypergiant (BHG) star of Cyg OB2 \#12.}
   {The 21\,cm observations secured in the COBRaS Legacy project provide data to constrain conditions in the outer wind regions of massive stars; determine the non-thermal properties of massive interacting binaries; examine evidence for transient sources, including those associated with young stellar objects; and provide unidentified sources that merit follow-up observations. The 21\,cm data are of lasting value and will serve in combination with other key surveys of Cyg OB2, including Chandra and Spitzer.}


   \maketitle
%

\section{Introduction}
\label{introduction}

The Cygnus OB2 Radio Survey (COBRaS) is an e-MERLIN Legacy project\footnote{http://www.e-merlin.ac.uk/legacy/projects/cobras.html}  conducting an extensive radio survey of the central region of the Cygnus OB2 association at L- (1.5 GHz) and C-band (5 GHz), with a total time allocation of 294 hours. This project exploits the substantially upgraded e-MERLIN facility, with up to a factor of 30 increase in sensitivity and increased bandwidths. In this paper we report on the L-band dataset of Cygnus OB2.

In the northern hemisphere constellation of Cygnus lies the massive star forming region of Cygnus X. At relatively close proximity \citep[0.7 - 2.5 kpc;][]{uyaniker_etal_2001}, the entire Cygnus region contains nine OB associations and at least a dozen young open clusters \citep{mahy_etal_2013}. Numerous surveys of the entire region (from the X-rays through to the radio) have already been conducted, leading to the discovery of hundreds of H\textsc{ii} regions and several supernova remnants \citep{uyaniker_etal_2001}. Moreover, X-ray observations have revealed the presence of a `super-bubble' spanning 450 pc and attributed to the past events of 30-100 supernovae \citep{cash_etal_1980}. At the heart of this bubble, and located behind the `Great Cygnus Rift', lies the tremendously OB-rich, massive stellar association of Cyg OB2. First identified by \citet{munch_morgan_1953}, this stellar system is amongst the most massive observed in the Galaxy \citep{knodlseder_2000, hanson_2003, wright_drake_2009} with an estimated mass of $10^4 - 10^5$ M$_{\odot}$ and containing $\sim$ 2600 $\pm$ 400 OB stars \citep{knodlseder_2000, drew_etal_2008, wright_etal_2010a, wright_etal_2015}. This, in conjunction with its close proximity\footnote{Whilst distance estimates span between 0.9 and 2.1 kpc, here we adopt a distance of 1.4 kpc in line with some of the more recent estimates, see e.g. \citet{rygl_etal_2012} and \citet{kiminki_etal_2015}.}, makes Cyg OB2 a unique case for studies of massive stellar clusters, star formation, and stellar evolution within the Galaxy.

The association is known to suffer from a substantial amount of (variable) visual extinction due to a combination of the large amounts of dust found within its parental cloud and also within the Great Cygnus Rift lying within our line of sight. In a recent study by \citet{wright_etal_2015}, the visual extinction towards 164 of its OB stars ranged from $A_V$ = 2.2$^m$ to 10.2$^m$ (where the superscript $m$ denotes magnitude). This makes Cyg OB2 ideal for studies within the long wavelength (radio) regime, as extinction at these wavelengths is negligible.

In addition to a high stellar density, Cyg OB2 also boasts a diverse range of stellar objects. Its central 0.5 deg$^2$ has been found to contain approximately 8000 X-ray point sources, with many of them considered to be young stellar objects (YSOs) and T Tauri stars (at least 444 have been diagnosed to be stars with discs; \citealt{guarcello_etal_2015}). Additionally, in and around the immediate vicinity of the association there are a number of Be stars, two Wolf-Rayet stars (WR145, WR146), two candidate luminous blue variable (LBV) stars (G79.29+0.46, Cyg OB2 \#12), a red supergiant (IRC+40427), a B[e] star (MWC 349), and a high-energy $\gamma$-ray source (TeV J2032+4130).

\citet{wright_etal_2016} recently carried out a high-precision proper motion study to  derive the 3D velocity dispersion ($\sigma_{3D}$ = 17.8 $\pm$ 0.6\,kms$^{-1}$) using 873 Cyg OB2 member stars, concluding that the association is gravitationally unbound but shows no signs of previous gas expulsion. This argues against the proposition that OB associations are the expanded remnants of disrupted star clusters. Instead, Cyg OB2 was likely born with a considerable amount of physical and kinematic substructure and underwent gas exhaustion, managing to form massive stars in a low density environment by the direct collapse of gas and dust onto a stellar protostar. Table \ref{table:cygob2_characteristics} gives a summary of the main physical characteristics of the Cyg OB2 association.

Since the mid-eighties, numerous surveys have imaged the association at radio wavelengths. The following is a list of some of the main surveys and their findings.
\begin{itemize}
\item{W84: \citet{wendker_1984}, at 4800 MHz; limiting flux density 50 mJy; angular resolution $2'.6$.}
\item{Z90: \citet{zoonematkermani_etal_1990} at 1400 MHz; limiting flux density 25 mJy; angular resolution 20".}
\item{W91: \citet{wendker_etal_1991}, at 408 and 1430 MHz; limiting flux densities 150 mJy and 45 mJy; angular resolutions $3'.5$ $\times$ $5'.2$ and $1'.0$ $\times$ $1'.5$ respectively.}
\item{T96: \citet{taylor_etal_1996} at 327 MHz; limiting flux density 10 mJy; angular resolution $1'.0$}
\item{T03: \citet{taylor_etal_2003} at both 408 and 1420 MHz; angular resolutions $5'.3$ and $1'.6$ respectively.}
\item{SG03: \citet{setiagunawan_etal_2003} at 350 and 1400 MHz; limiting flux densities of 10-15 mJy and 2 mJy; angular resolutions 13" and 55" respectively.}
\end{itemize}

\begin{figure}[htbp]
\begin{center}
\includegraphics[width=0.5\textwidth]{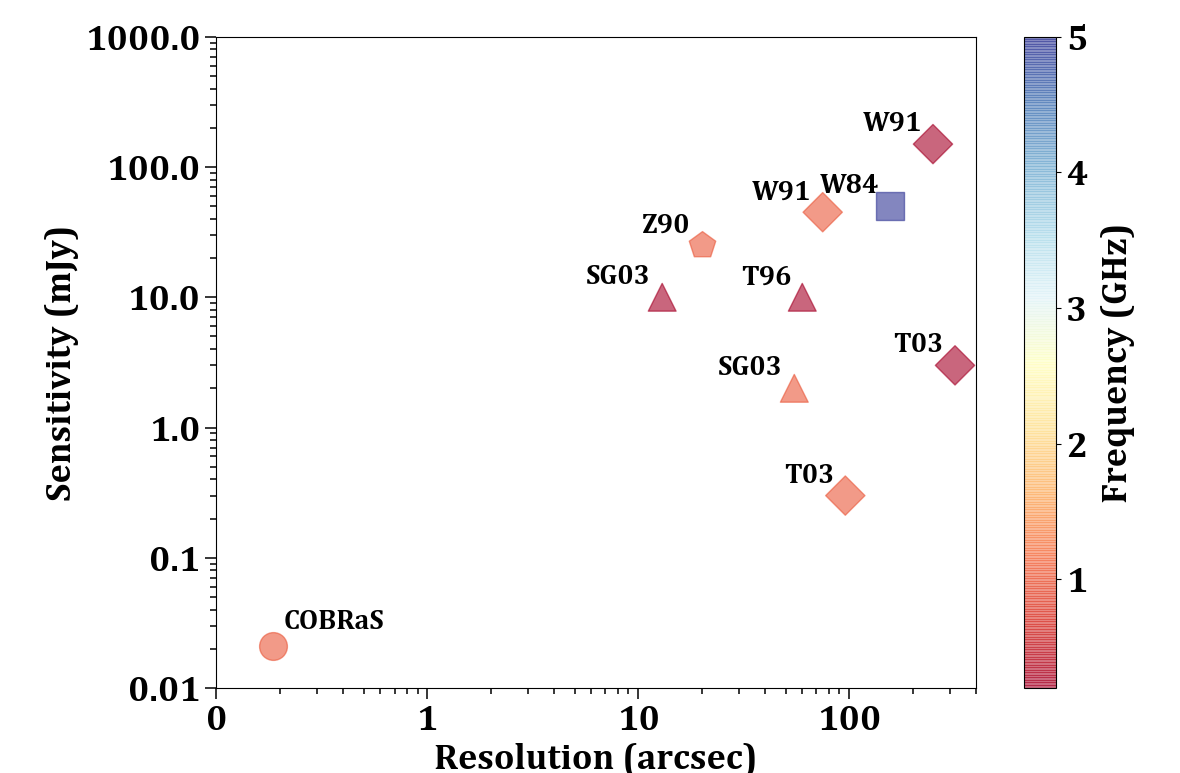}
\caption[Survey comparison figure]{Resolution and sensitivity of the COBRaS e-MERLIN observations presented in this paper, in comparison to previous radio surveys covering the Cyg OB2 association. The colour scale gives the frequency of a given survey whilst the different shapes represent the different instruments used: MPIfR 100\,m (squares); VLA (pentagons); DRAO (diamonds); WSRT (triangles) and e-MERLIN (circles). References relating the shorthand labels can be found in Section \ref{introduction}.}
\label{fig:survey_comp}
\end{center}
\end{figure}

To highlight the difference in sensitivity and resolution of these previous radio surveys in comparison to the COBRaS observations presented in this paper, Figure \ref{fig:survey_comp} shows a log--log plot of resolution versus sensitivity.

By taking advantage of the enhanced sensitivity and high resolution of the e-MERLIN array, COBRaS will deliver the most detailed and sensitive radio census of the region to date. The Legacy project aims to reach 1-$\sigma$ flux density limits of $\sim$ 3 and $\sim$ 10\,$\mu$Jy, and angular resolutions of 40\,mas and 150\,mas for C- and L-band respectively. COBRaS was awarded a total allocation of 252 hours and 42 hours at C- (6\,cm) and L-band (21\,cm), respectively, and the project aims to investigate several of the current astrophysical problems related to massive stars and clusters, including the mass-loss and evolution in massive stars; the formation, dynamics, and content of massive OB associations; the frequency of massive binaries; and the incidence of non-thermal radiation.

Massive stars have a significant influence in many areas of astrophysics. They are major contributors to galactic evolution as they return prodigious amounts of mass, momentum, and energy to the interstellar medium (ISM). These massive stars lose mass via stellar winds over the course of their lifetime. Previous results have shown that the current estimates of mass-loss rates in such stars derived from different observational diagnostics are discrepant by up to an order of magnitude \citep[e.g.][]{puls_etal_2006, fullerton_etal_2006, sundqvist_etal_2010}. This may have profound implications for broad astrophysical domains, including stellar evolution and the mass-loss process across the H-R diagram, and the injection of enriched gas into the ISM. High sensitivity is required to detect as much of the O and B stellar population as possible, and to study the role of clumping in stellar mass loss.


The high sensitivity of e-MERLIN enables the direct detection of massive binary systems within Cyg OB2. In a binary consisting of two early-type stars (e.g. O+O), the stellar winds collide. Around the shocks in the colliding-wind region, electrons are accelerated to relativistic velocities. These electrons then emit synchrotron radiation which can be detected at radio wavelengths \citep[e.g.][]{dougherty_williams_2000}. The e-MERLIN data from this project will allow us to: (1) obtain a better determined binary frequency, an important constraint for evolutionary population synthesis models, which will significantly improve our understanding of galactic chemical evolution; (2) study statistically the colliding-wind phenomenon and better understand its dependence on stellar and binary parameters; and (3) improve our understanding of the first-order Fermi mechanism responsible for the particle acceleration.


In combination with current multi-waveband surveys of the Cyg OB2 association (the INT Photometric H$\alpha$ Survey (IPHAS): H$\alpha$; Spitzer: near-IR; Chandra: X-ray) and indeed future surveys (e.g. JWST, Gaia), COBRaS will also deliver data on YSOs, transient sources, and background galaxies.
This paper presents the results gained from the complete reduction and analysis of the COBRaS L-band (21 cm) observations and is organised as follows. Section \ref{datareduction} describes the 21\,cm e-MERLIN observations and their subsequent reduction. Section \ref{imagefitting} begins with the introduction of the novel Source Extraction Algorithm for COBRaS (SEAC) and goes on to describe the steps taken to extract the total number of sources and their flux densities found within the seven L-band pointings. Section \ref{results} presents the detected sources within the 21\,cm observations including information regarding those identified with previous observations. Section \ref{analysis} gives a discussion into the types of sources detected within the data, focussing on those objects with previous identifications. Finally, Section \ref{conclusions} is used to summarise the main findings of the COBRaS 21\,cm Legacy survey.

\begin{table}
\begin{center}
\begin{footnotesize}
\caption[Fundamental physical parameters describing the Cygnus OB2 association]{Fundamental physical parameters describing the Cygnus OB2 association. The RA and DEC describe the centre of the association as chosen by numerous authors based on the position of the main concentration of OB stars.}
\begin{tabular}{llc}
 & & Ref \\
\hline
\hline
RA (J2000) & $20^h 33^m 12^s$ & 1 \\
DEC (J2000) & $+41^{\circ} 19\arcmin 00\arcsec$ & 1 \\
Stellar mass & $16500^{+3800}_{-2800}$ $\rm{M_\odot}\,$ & 2 \\
Virial mass & $(9.3 \pm 0.8) \times 10^5$ $\rm{M_\odot}\,$ & 3 \\
Volume density & $\sim$ 100 stars pc$^{-3}$ & 4 \\
Age & 1-7\,Myr (Peak 4-5\,Myr) & 2 \\
Distance & 1.4 $\pm$ 0.08\,kpc & 5 \\
Visual extinction, A$_V$ & $4^m - 7^m$ (IQR$^{\star}$) & 2 \\
OB members & 2600 $\pm$ 400 & 6 \\
Binary fraction & 55\% & 7 \\
3D velocity dispersion & 17.8 $\pm$ 0.6\,${\rm km\,s}^{-1}$ & 3\\ 
Half-light radius & 10.1 $\pm$ 0.9\arcmin (4.1\,pc) & 3 \\
Observational diameter & $\sim$2$^{\circ}$ ($\sim$49 pc) & 6 \\
\hline

\end{tabular}
\label{table:cygob2_characteristics}
\end{footnotesize}
\end{center}

\vspace{0.1cm}
\footnotesize{References: {\bf 1} \citealp{wright_etal_2014a}, {\bf 2} \citealp{wright_etal_2015}, {\bf 3} \citealp{wright_etal_2016}, {\bf 4} \citealp{wright_etal_2014b}, {\bf 5} \citealp{rygl_etal_2012}, {\bf 6} \citealp{knodlseder_2000}, {\bf 7} \citealp{kobulnicky_etal_2014}.\\
$^{\star}$Interquartile range (IQR).}
\end{table}%


\section{Observations and data reduction}
\label{datareduction}

\subsection{Observations}

\begin{figure}[htbp]
\begin{center}
\includegraphics[width=0.85\linewidth]{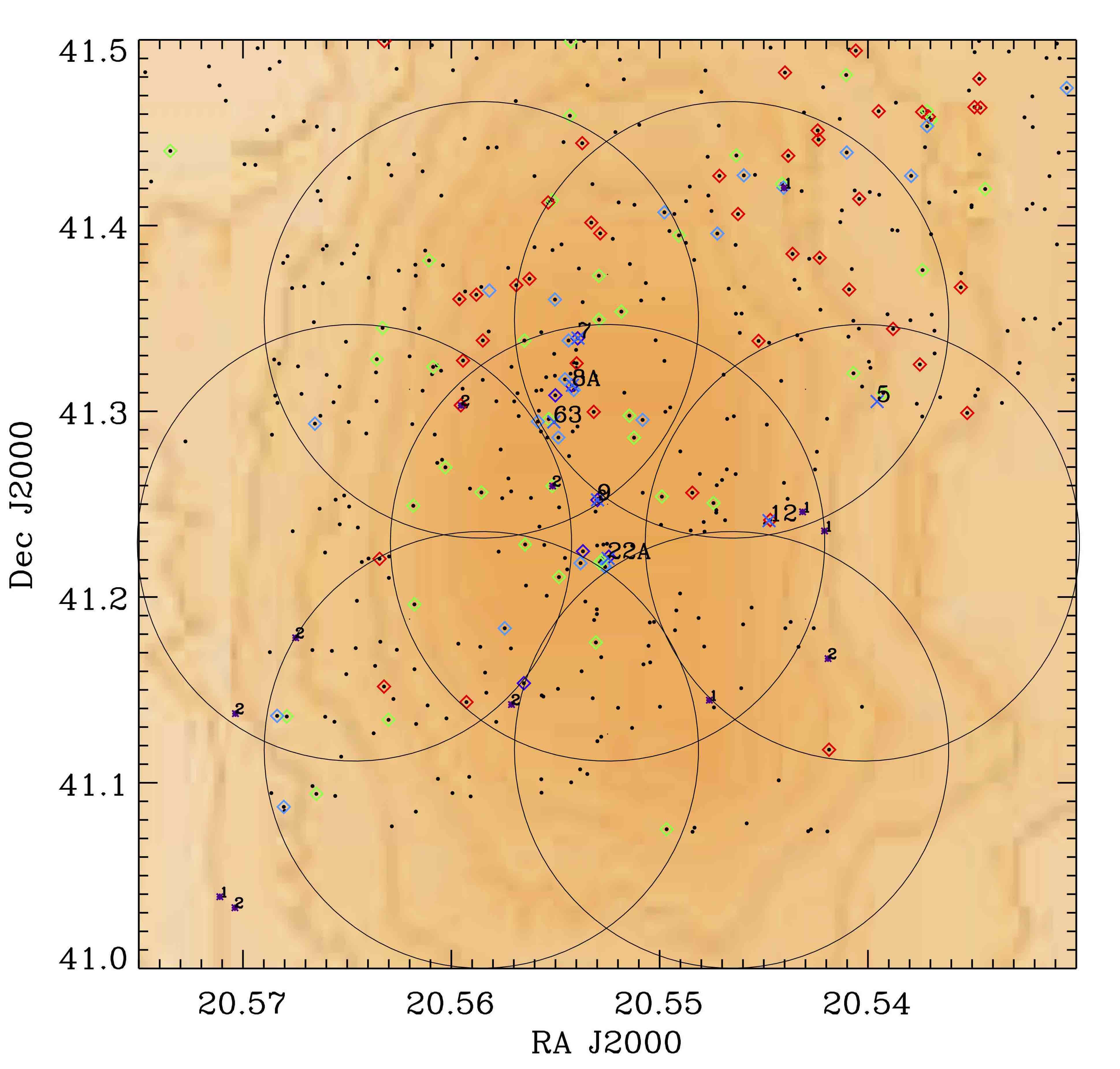}

\caption[COBRaS positions for the seven L-band (21\,cm) COBRaS pointings]{COBRaS positions for the seven L-band (21\,cm) pointings imaged with e-MERLIN interferometer, based on the primary beam of the Lovell antenna shown as large circles. The background colour figure is the stellar density distribution from the 2MASS survey as given in \citep{knodlseder_2000}. The black dots represent OB stars from \citet{massey_thompson_1991}, the coloured diamonds highlight OB binary systems from \citet{kiminki_etal_2007}, and the star symbols indicate massive stars taken from \citet{comeron_etal_2002}. The numbers identify some of the well-studied Cyg OB2 stars.}
\label{fig:cobras_pointings}
\end{center}
\end{figure}

The L-band (21\,cm) observations presented here were obtained as part of the COBRaS e-MERLIN Legacy project. The bulk of these observations ($\sim$ $75\%$ of the total allocation) were completed over a three day period from April 25 to 27,  2014 (from here after denoted as April 26), whilst additional observations, contributing to the final quarter of the total allocation at L-band were taken on April 11, 2014 (from here after April 11). A total of seven (overlapping) pointings were required in order to cover the central area of the Cyg OB2 association ($\sim 0.48$ deg$^2$), due to the primary beam size of the e-MERLIN array (based upon the Lovell antenna) at 21\,cm. This observation strategy can be seen in Figure \ref{fig:cobras_pointings}, which shows the seven overlapping pointings chosen to cover the highest concentration of stellar sources as determined from \citet{knodlseder_2000}. The observations were cycled into two seven-minute scans per pointing in order to provide a good hour-angle coverage and to maintain as similar a $u,v$-coverage for all pointings as possible. 

The point source J2007+404 was used to perform cycled phase calibration scans during the observations. In total, each pointing was observed for approximately five hours on source, with the data from April 11 and 26 combined. The observations were made using full stokes parameters at a central frequency of {1.51}\,GHz using 512\,MHz bandwidth split over eight intermediate frequencies (IFs) and 512 channels per IF. A complete summary of the 21\,cm COBRaS observations is given in Table \ref{table:obs_info}.

\begin{table}[ht]
\begin{footnotesize}
\caption[Observational summary of the COBRaS L-band Legacy data set]{Observational summary of the COBRaS L-band Legacy data set. The total integration time and data size includes all of the seven target fields and the observed calibration sources over both epochs (April 11 and 26). We note that the given integration time and data size describes that of the raw data set, i.e. before any loss of data due to RFI, noisy data, or `off-source' visibilities.}
\begin{center}
\begin{tabular}{ll}
\hline \hline
Total integration time & $\sim$ 48 hrs \\
Data size & $\sim$ 940 GB \\
Number of pointings & 7 \\
FoV (diameter) per pointing$^\dagger$ & $\sim 10.95'$ \\
Central frequency & 1.5103374 GHz \\
Frequency range & 1.2543374 - 1.7663374 GHz \\
Number of baselines & 20 \\
Number of IFs & 8 \\
Number of channels (per IF) & 512 \\
Channel increment & 125 kHz \\
Polarisations (stokes) & RR, LL, RL, LR \\
Predicted resolution & 183 mas \\
Predicted sensitivity & $\sim$ 13\,$\mu$Jy for 3.5\,hrs on source \\
\hline
\end{tabular}
\label{table:obs_info}
\end{center}
\footnotesize{$^\dagger$As measured from the central frequency (listed) of the band}
\end{footnotesize}

\end{table}

\subsection{Data reduction}

The complete reduction of the COBRaS L-band data set was carried out using the Astronomical Image Processing Systems (AIPS). Where possible, scripting procedures within the \textsc{parseltongue}/\textsc{python} environment were used to increase the overall efficiency of data handling processes.

\subsubsection{Radio frequency interference}

A significant portion of the observable bandwidth suffers from contamination by Radio frequency interference (RFI). The observed RFI was unpredictable and varied significantly in intensity (up to three to four magnitudes greater than that observed from the astronomical source) across both time and frequency space. Due to the extremity of the RFI, a manual flagging procedure was necessary as a first pass throughout the RFI mitigation procedure. A data visualisation programme SPPlot\footnote{available at https://github.com/jackmorford/SPPlot} (SPectral Plot) was used to inspect the data, from which large chunks of RFI-affected data could be identified and flagged accordingly due to their location in time and frequency space. The data were then edited using the RFI-mitigation software SERPent (\citealp{peck_fenech_2013}), a programme developed for e-MERLIN that utilises the {\textsc{parseltongue}} scripting environment, as well as editing tasks within AIPS (Astronomical Image Processing System). In total, approximately 25-30\% of the data for each pointing were removed because of RFI alone. A further $\sim$ $25\%$ of the data were lost due to corrupt radio visibilities as a result of instrumental errors both internally (electronics, correlator) and externally (antennae pointing errors). Across both observation epochs, $\sim 50 \%$ of the data were lost and were unable to undergo a successful calibration.

\subsubsection{Calibration}

\begin{figure*}[htbp]
\begin{center}
\includegraphics[width=\textwidth]{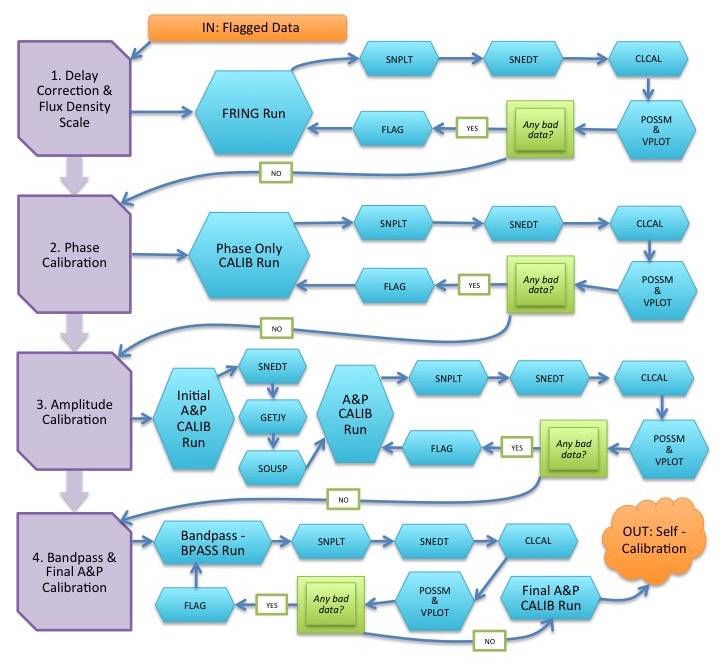}
\caption[Steps taken in the calibration of the COBRaS Legacy L-band data]{Steps taken in the treatment of the COBRaS Legacy L-band data, through to the point at which the data is ready for self-calibration. Steps carried out with the use of AIPS tasks are highlighted by cyan-coloured hexagons.}
\label{fig:calibration_flow}
\end{center}
\end{figure*}

The calibration procedure was based upon the official e-MERLIN cookbook (version 3.0 - February 2015)\footnote{http://www.e-merlin.ac.uk/}. Whilst several calibration pipelines exist for the treatment of e-MERLIN data (see \citealt{argo_2014} and Chapter 3 of \citealt{peck_thesis_2014}), strong inconsistencies (and variations in data quality) between the COBRaS L-band Legacy data sets necessitated a tailor-made calibration for each epoch. At each step of the calibration routine, the data were visually inspected with the possibility of applying further flags. The flow diagram in Figure \ref{fig:calibration_flow} highlights the main steps taken to complete the calibration, after removing any RFI and corrupt data, through to the point of self-calibration. 

The flux calibrator 1331+305 (3C286), a bright quasar of known flux, is used to set the fluxes for the remaining target and calibration sources. At L-band, 1331+305 is resolved by the longer baselines of e-MERLIN and has a steep spectral index. The flux densities must be set for each IF to account for this variation across the wide bandwidth. This was accomplished using a \textsc{parseltongue} script called \texttt{dflux.py} that calculates the flux of 1331+305 at the central frequency of each IF. The calculation uses a polynomial expression derived from the most recent spectral flux densities of 1331+305 from \citet{perley_butler_2013}, which is then corrected for the higher resolving power of e-MERLIN over the JVLA. The output values from the script were then entered into the source (SU) table using the AIPS task SETJY.

Fringe fitting was performed by the AIPS task FRING, which performs a least squares fit to the phase, delay, and fringe rate to remove the variable delay offsets and fringe rates. Delay offsets within e-MERLIN data can be as large as several micro-seconds and are generally greater on longer baselines; they can vary in time either as a gradual drift due to changes in the temperature affecting the length of the fibres connecting the antennae, or as sharp changes originating within the correlator.

After every stage of the calibration process, the data were inspected in both frequency and time using the AIPS tasks POSSM and VPLOT, respectively. This inspection is crucial to ensure the derived solutions have been correctly implemented and the visibilities are in the desired state to move onto the next stage in the data reduction process.

J2007+404 was chosen as a phase calibrator source as it is bright, lies within close proximity (within $\sim 2^{\circ}$) of our target fields, and even on the longest e-MERLIN baselines at L-band is not resolved enough to introduce significant phase errors. The AIPS task CALIB was used to compute antenna-based gain solutions by initially solving for the phases only. The phase solutions are calculated across each of the calibration sources, however only the solutions gained from the phase calibrator, J2007+404 are used to calibrate the phases of the target fields.

In applying the amplitude calibration, two initial runs of CALIB are used to solve for both the amplitudes and phases of firstly the flux-calibrator 1331+305, and secondly of the point source and phase calibrators 1407+284 and J2007+404 respectively. The resulting solutions are used to set the absolute flux scale of 1407+284, J2007+404, and the target fields by using the AIPS task GETJY. Furthermore, the AIPS task SOUSP is used to fit to the derived flux values over each IF to determine the spectral indices of the point source and phase calibrator. The final flux densities for 1407+284 and J2007+404 from the April 26 observations can be seen in Table \ref{tab:getjy_25th}. In the case of J2007+404, a large amount of RFI (and therefore significantly less data in comparison to the other spectral windows) within IF1 led to overestimation of its derived flux density. This value was not included within the subsequent fit with SOUSP to avoid `skewing' the fit.

\setlength\tabcolsep{2pt}
\begin{table}[htp]
\begin{footnotesize}
\caption[Final derived fluxes of the point and phase calibration sources of the COBRaS L-band Legacy Data set]{Final derived flux densities of the point and phase calibration sources of the COBRaS L-band Legacy data set. These values were derived using the April 26 observations and were applied to those from April 11. The final values used are those under the SOUSP column in either case.}
\begin{center}
\begin{tabular}{cccccc}
\hline \hline
\textbf{IF} & \textbf{Frequency} & \multicolumn{2}{c}{\textbf{1407+284}} & \multicolumn{2}{c}{\textbf{J2007+404}}  \\
 & (GHz) & GETJY (Jy) & SOUSP (Jy) & GETJY (Jy) & SOUSP (Jy) \\
\hline
1 & 1.286 & 0.77$\pm$0.04 & 0.77 & 4.58$\pm$2.38 & 2.21 \\
2 & 1.350 & 0.83$\pm$0.02 & 0.82 & 2.25$\pm$0.06 & 2.24 \\
3 & 1.414 & 0.88$\pm$0.01 & 0.88 & 2.27$\pm$0.03 & 2.28 \\
4 & 1.478 & 0.93$\pm$0.01 & 0.93 & 2.33$\pm$0.03 & 2.32 \\
5 & 1.542 & 0.96$\pm$0.02 & 0.99 & 2.34$\pm$0.02 & 2.36 \\
6 & 1.606 & 1.05$\pm$0.01 & 1.04 & 2.37$\pm$0.02 & 2.39 \\
7 & 1.670 & 1.11$\pm$0.01 & 1.10 & 2.44$\pm$0.02 & 2.42 \\
8 & 1.734 & 1.16$\pm$0.01 & 1.16 & 2.47$\pm$0.03 & 2.46 \\
\hline
\multicolumn{2}{c}{Sp. Indx, $\alpha$} & \multicolumn{2}{c}{1.34} & \multicolumn{2}{c}{0.35}   \\
\multicolumn{2}{c}{RMS error} & \multicolumn{2}{c}{0.01} & \multicolumn{2}{c}{0.01}
\end{tabular}
\end{center}
\label{tab:getjy_25th}
\end{footnotesize}
\end{table}%


The April 26 observations constituted a larger portion of the entire Legacy data set than those of April 11 and were of better quality. Whilst the flux density values derived from the April 11 dataset were within 10\% of those shown in Table \ref{tab:getjy_25th}, flux density values from the April 26 data were deemed more reliable. We therefore chose to use the values listed in Table \ref{tab:getjy_25th} to set the flux densities of both the April 11 and April 26 data.

The AIPS task BPASS was used to correct for the response of the interferometers' receivers, which cause complex amplitude and phase gain variations as a function of frequency. In order to solve for the variations across the frequency band, BPASS uses a point)source calibrator that is both bright and has a flat spectrum across the band. Here, 1407+284 was used as the initial bandpass calibrator, however for both observation epochs the solutions from BPASS did not correctly remove the amplitude and phase variations across the entire frequency band for a number of baselines. The bandpass response showed changes as a function of time, and therefore the phase calibrator was used to correct for the bandpass response, solving for each ($\sim$ 2-min) scan.

Self-calibration was applied to the phase calibrator J2007+404 for the observations taken on April 11 and made significant improvements to the phase solutions. However, for the April 26 epoch,  self-calibration on the phase calibrator made no significant improvement and was left out of the data-reduction routine. Self-calibration was also applied to each of the seven target field pointings from both the April 26 and 11 datasets. Using approximately six sources per field (with flux $\gtrsim 1$ mJy), CALIB calculated solutions over a ten-minute time interval and by averaging the RR and LL polarisations.

\subsubsection{Imaging}

\begin{figure}[htbp]
\begin{center}
\includegraphics[scale=0.35]{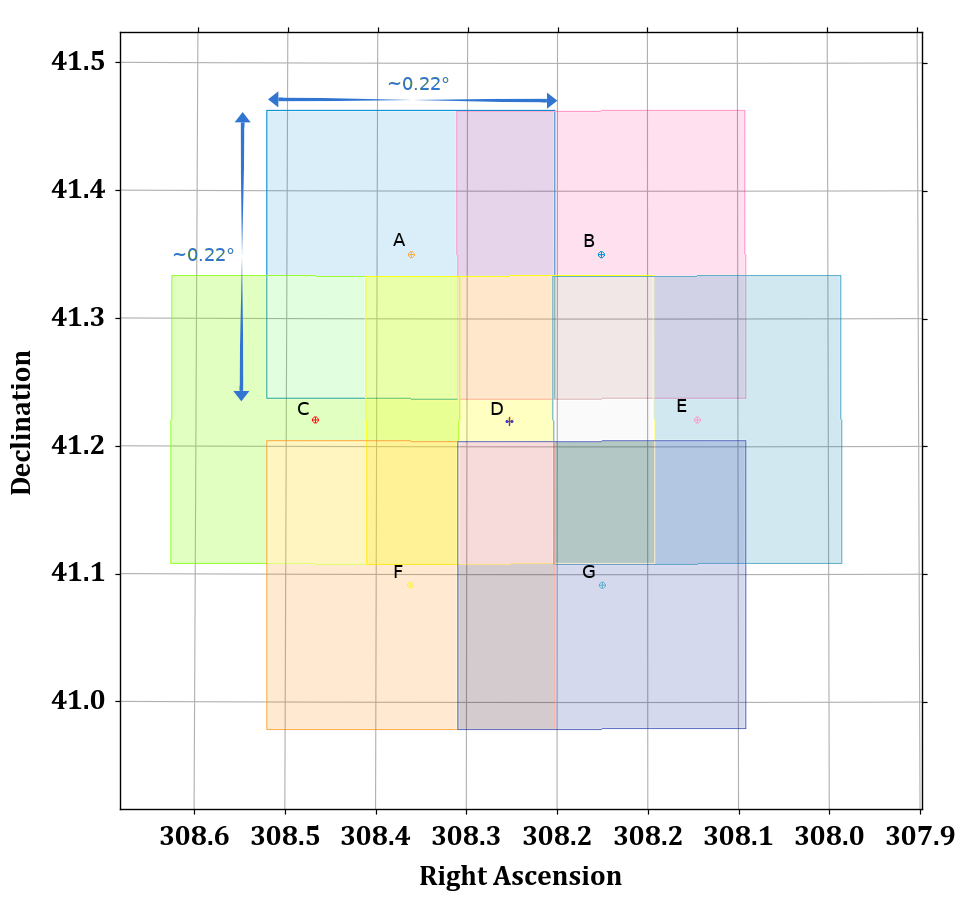}
\caption[Regions covered by the seven wide-field COBRaS L-band images on the sky]{Regions covered by the seven wide-field COBRaS L-band images on the sky. Each pointing centre is highlighted by a cross and labelled alphabetically from A to G.}
\label{fig:pointingcentres}
\end{center}
\end{figure}

In order to perform wide-field imaging and mosaicing on the data, they were first translated to measurement set format for use with the wide-field imaging software WSClean \citep{offringa_2014}. Each epoch was imaged and analysed independently in order to search for any variable emission. During this process the data were re-weighted to account for the different antenna sizes across the e-MERLIN array. The AIPS task {\textsc WTMOD} was first used to set the weights to unity. Following this, the data were loaded from {\textsc FITS} format into the radio data-processing software {\textsc CASA} \citep{mcmullin_2007} to create a measurement set. The CASA task {\textsc STATWT} was used to re-weight the data according to the relative sensitivity of the antennas in the array prior to imaging. The WSClean software package performs w-stacking to correct for the w-term when imaging wide fields in order to account for any effects from sky curvature and non-coplanar baselines. In order to maximise the $u,v$-coverage, we used pseudo-I imaging to include any parallel-hand polarisation data present where the other had been flagged. As this option is currently unavailable with WSClean, a version of the data utilising the unflagged data for both polarisations was used to recreate the process, as is done in other imaging tasks such as AIPS IMAGR. 
Each pointing from both the April 11 and 26 data was imaged independently using WSClean with natural weighting. The expected field of view of each pointing (based on the size of the Lovell antenna) was $\sim 657$" in diameter at the central frequency of the observations. In order to maximise the field of view and sensitivity but limit the required computation, we therefore chose to image 20000 $\times$ 20000 pixel images, large enough to cover the primary field of view of the Lovell antenna at the very top of the observing band (i.e. those depicted in Figure \ref{fig:pointingcentres}) using a pixel size of 0.04\,arcsecs. An auto-threshold of 3.5\,$\sigma$ was used to limit the cleaning along with a maximum of 50000 iterations using an mgain of 0.4. In order to mosaic the images together in the image plane, an averaged restoring beam was used for all pointings in each epoch. These were 305$\times$194 (PA -12$^\circ$) and 198$\times$158\,mas (PA 36$^\circ$) for the April 11 and 26 data respectively. When imaged, each of the resulting wide-field maps covered an area on the sky of $\sim 0.22^{\circ}$ in RA and $\sim 0.22^{\circ}$ in declination. The positions of the seven pointing centres and the area on the sky covered by the wide-field images can be seen in Figure \ref{fig:pointingcentres}.
Once complete, the output FITS images were transferred back to AIPS to perform the mosaicing. The {\textsc FLATN} task was used to combine the images into a single field at each epoch and apply a primary beam correction to account for the changing sensitivity across the field of view as a function of the primary beam response. The resulting full-field images for each epoch are approximately 55000$\times$55000 pixels.

The single-pointing wide-field images achieved noise levels in the central regions of approximately $\sim$30-35 and 20-22\,$\mu$Jy/beam for the April 11 and 26 respectively. For approximately 3.5\,hr on-source integration time the expected noise level for the April 26 portion of the data was 13\,$\mu$Jy/beam, with 20\,$\mu$Jy/beam expected for the corresponding 1.5 hr on-source in the April 11 data. The noise levels achieved fall somewhat short of those expected compared to the theoretical noise, which is likely due to a combination of factors. These observations were taken during the later commissioning stages of e-MERLIN when the data quality was still undergoing improvements. In addition, a number of issues with telescopes during the observations meant they were not all observing for the full time period. These data were also significantly affected by RFI resulting in further losses where this was excised. As a result of these problems, on average approximately 50\% of the data was removed during processing (and in some cases slightly more). Applying this to the theoretical noise calculation we would expect around 28 and 19\,$\mu$Jy/beam for the April 11 and 26 respectively. We achieve noise values approaching these levels in the wide-field images suggesting that the known issues likely account for the difference between those expected and realised. 

\section{Image fitting and flux extraction}
\label{imagefitting}

\subsection{Source Extraction Algorithm for COBRaS (SEAC)}
\label{seac}

SEAC was originally based on a flux extractor script written by Luke Peck (see \citealt{peck_thesis_2014}). Having extended and improved the code it is now the main tool used to compile the source population catalogue for the COBRaS L-band observations. The code is maintained by J. C. Morford and D. M. Fenech and is completely open source, made available via GitHub \footnote{\url{https://github.com/daniellefenech/SEAC} 
}.

The program has a number of user-defined parameters to control different aspects of the code which are edited at the beginning of the script, and performs the following steps:

\begin{enumerate}
\item It reads an astronomical image into a two-dimensional \textsc{numpy} array and calculates the beam size from the image header and the initial RMS of the entire image.
\item If required, it calculates a noise map across the image at a user-defined resolution to account for the variation in the noise across the image.
\item It then implements the floodfill algorithm (see Section \ref{floodfill}) in order to find the positions of the various pixels associated with the astronomical sources or `islands' within the image. 
\item Having obtained a given number of islands, the next part of the code calculates the integrated flux, the maximum pixel, the weighted position in sky coordinates, and the local noise level, as well as their associated uncertainties using one of two possible methods:
        \begin{enumerate}
        \item Using the AIPS task JMFIT to fit up to four Gaussian components to each island, from which the various statistics are calculated;
        \item or from within the \textsc{python} environment using a pixel-by-pixel method, summing the pixel flux densities for the island and then subtracting the local background to obtain the integrated flux density, finding the RMS from around the immediate vicinity of the source and calculating their related uncertainties.
        \end{enumerate}
\item It then outputs the integrated flux density, positional information, and their related uncertainties to a text file, whilst various output plots show the position of each source within the input image, and (in the case of the pixel-by-pixel method) the area of the source detected by the floodfill algorithm.

\end{enumerate}

\subsubsection{The floodfill algorithm}
\label{floodfill}

The floodfill algorithm was deemed as the best option in extracting sources from the COBRaS observations due to its low rate of false detections and its flexibility in terms of finding both resolved and unresolved sources. The algorithm takes two user-defined inputs; the seed threshold ($\sigma_S$) and the flood threshold ($\sigma_F$). The seed threshold defines the level at which any pixel across the entire image can be first considered as an island. The algorithm then takes each seed pixel on an individual basis and tests the level of the adjacent pixels surrounding it to see if they are above the flood threshold. This is implemented in SEAC using a scan-line stack-based approach. Every connected pixel above $\sigma_F$ is considered a part of the island. This iterative process is continued until all the  pixels adjacent to the island are lower than the flood threshold level. Since we are working with radio maps that have been convolved with a synthesised beam, an island is only generally considered an island if the number of pixels representing it is larger than the number of pixels that make up the beam area\footnote{This was chosen to be 99\% of the beam in order to account for low level point like sources within the data.}. The seed threshold must be larger than or equal to that of the flood threshold with typical values of $\sigma_S = 5\sigma$ and $\sigma_F = 4\sigma$. For a complete discussion on the range of possible values, we direct the reader to the work of \citet{hancock_etal_2012} and \citet{hales_etal_2012}, who implement the floodfill algorithm within the source-extraction programs \textsc{aegean} and \textsc{blobcat} respectively. Within SEAC, both $\sigma_S$ and $\sigma_F$ are manually set by the user to give an added flexibility to the program and the specific settings used for analysing the data are discussed in section \ref{sourceext}.

\subsubsection{Background noise map}

The flood and seed threshold parameters affect the results from SEAC to the greatest extent. However, these parameters represent multiples of the background RMS within the image. When dealing with small images (e.g. 512 $\times$ 512 pixels or $\sim$ 20 arcseconds in diameter), the variation in the noise level across the image is insignificant. However, the wide-field images produced in the reduction of the COBRaS 21 cm data span $\sim 15$ arcminutes in diameter, covering the entire field of view of the e-MERLIN array at L-band. In correction for the response of the primary beam across each the full-field image, the noise level varies significantly from the centre to the outer edges. The variation can be as much as a factor of  approximately three meaning a single measurement of the noise level across the entire image will not accurately represent the localised RMS of each source. 



To compensate for this variation in noise, SEAC can create a separate noise map that mimics the size and shape of the image array, with pixel values corresponding to the local noise. The user can input the resolution of the noise map by choosing the number of boxes in the $x$ and $y$ directions from which to calculate the noise level. This resolution is required to be sufficiently small such that the noise is unlikely to vary significantly between each grid cell, yet large enough in order for a reliable estimate of the noise level to be made, even if bright sources are included within a given grid cell. As the algorithm cycles through each pixel within the image array, their amplitude is compared to the noise level of the same pixel position in the noise array, thus accounting for the variation in noise level across large images. In this way each source detected is more fairly represented by its local background.

\subsubsection{Source position determination}
\label{sourceposition}

Considering the high resolution obtained from these e-MERLIN 21\,cm observations, a precise source position determination of each detected source is required. Within SEAC, there exists two methods to measure the position of each detected source or island. Firstly, when using the AIPS task JMFIT to fit Gaussian models to each island by a least squares method, the task will return the integrated flux of the island and the peak position of the fitted model. Alternatively, the fluxes and positions of each island can be obtained using a pixel-by-pixel method (i.e. from an analysis of the island pixels, conducted within \textsc{python}). With this method, the user can choose to either return the coordinates of the peak pixel position of each island, or the coordinates of the weighted mean pixel position. The flux-weighted mean pixel position will obtain the most accurate position for resolved, non-Gaussian sources. Using the peak pixel position is sufficient to accurately represent the position of point-like, unresolved sources. Each method will return the $x$ and $y$ pixel position of each source, which SEAC then converts to the true position on the sky in astronomical coordinates.

%
%
%


\subsubsection{Source flux determination} 

The source integrated flux density and its associated error can either be derived using the AIPS task JMFIT or via an analysis of the detected island pixels. The flux calculated with JMFIT can vary significantly depending on the input parameters. The AIPS task benefits hugely from some prior information regarding the general source shape and is therefore tailored towards point-like, unresolved sources. In regards to resolved (non-Gaussian like) sources, JMFIT struggles to accurately represent the source and generally results in an unreliable integrated flux density. Great care must be taken if using JMFIT to determine the source fluxes of resolved or extended sources. Thus, in the case of resolved sources, a better option is to use an analysis of the determined island pixels. The integrated flux of a given island is calculated as follows:
\begin{equation}
F = \frac{1}{N_{beam}} \left( \sum\limits_{i,j=1}^{N_{src}} F(x_i,y_j) \right) -  \frac{\overline{x}_{LB}N_{src}}{N_{beam}}
\label{eqn:integratedflux}
,\end{equation}
where $\overline{x}_{LB}$ is the mean of the local background, $N_{beam}$ is the number of pixels representing the size of the beam, and $N$ is the number of pixels belonging to the island. The local background should represent the general noise level in the immediate vicinity of the source. Within SEAC, this can be calculated individually for each source, or over the entire image. Calculating the mean and RMS of the local background in the immediate vicinity of each source will give a much more accurate representation of the true background for each source. SEAC includes an option to change the size of the area used to calculate the mean and RMS of the local background but as a default uses a box of $4 \times 4$ arcseconds centred around each island. We note that the mean and RMS noise calculated from the local background do not include the pixels that make up the island. The associated error $\delta F$ in the integrated flux density calculation is given by:
\begin{equation}
{ 
{\delta F}^2 = \frac{\sigma_{LB}^2 N_{src}}{N_{beam}} + \frac{N_{src}^2\sigma_{LB}^2}{N_{LB}N_{beam}^2}
}
\label{eqn:integratedfluxerror}
,\end{equation}
{ where the $\sigma_{LB}$ is the root mean square noise calculated from the local background and $N_{LB}$ is the number of pixels used to calculate the mean of the local background}. { We note that this does not account for any amplitude calibration error. This is combined with the SEAC error after source-extraction.} The reliability of the integrated flux density of a given source, calculated in Equation \ref{eqn:integratedflux}, depends purely on the reliability of the floodfill algorithm. Each detected pixel contributing to the integrated flux density must give a fair representation of the true source on the sky. As a result, the integrated flux density measurement via this method has a dependence on the seed and flood threshold levels which dictate the detected pixel area of a given source.

\subsection{Catalogues for cross-correlation}

For the purpose of source identification, we required the use of previously published catalogues containing sources within the area of the sky covered by the Cyg OB2 association, from which a cross-correlation procedure could be performed. In the literature, a total of 36 published catalogues (including large-scale surveys) were found to harbour sources covering the same part of the sky as the COBRaS L-band observations (references regarding these 36 catalogues can be seen in Table \ref{tab:catalogues}). For ease of use, these were amalgamated into a single catalogue titled the Cyg OB2 super catalogue. Due to the vast number of sources found within catalogues 32 to 36 (as listed in Table \ref{tab:catalogues}), these five catalogues were not incorporated into the Cyg OB2 super catalogue. The total number of sources within the concatenated super catalogue from the remaining 31 catalogues is 14355 and covers an area 3$^{\circ}$ in radius, centred on the coordinates of the Cyg OB2 association (as given in Table \ref{table:cygob2_characteristics}). This provided six individual catalogues (2MASS, WISE, HERSCHEL, SPITZER, G12, and the Cyg OB2 super catalogue) and over 10$^5$ sources with which to cross-correlate the sources found in our observations.

\begin{table}[htp]
\begin{footnotesize}
\caption[A list of the previous catalogues containing information on the members of Cyg OB2]{A list of the previous catalogues found using the VizieR Service, all of which include sources from the region covered by the Cyg OB2 association.}
\begin{center}
        
        \begin{tabular}{clcc}

        \textbf{\#} & \textbf{Reference} & \textbf{Waveband} & \textbf{\# of Sources} \\
        \hline \hline
        1 &\citealt[][(MT91)]{massey_thompson_1991} & Visible & 801 \\
        2 &\citealt[][(C98)]{condon_etal_1998} & Radio & 72 \\
        3 &\citealt[][(PK98)]{pigulski_kolaczkowski_1998} & VIR & 288\\
        4 &\citealt[][(C01)]{comeron_toora_2001} & JHK & 320  \\
        5 &\citealt[][(C02)]{comeron_etal_2002} & KH & 85 \\
        6 &\citealt[][(S03)]{setiagunawan_etal_2003} & Radio & 239 \\
        7 &\citealt[][(W07)]{wolff_etal_2007} & Visible & 13 \\
        8 &\citealt[][(C07a)]{colombo_etal_2007b} & X-Ray & 1003  \\
        9 &\citealt[][(C07b)]{colombo_etal_2007a} & X-Ray & 147  \\
        10 &\citealt[][(M07)]{marti_etal_2007} & Radio & 153 \\
        11 &\citealt[][(K07)]{kiminki_etal_2007} & Visible & 303 \\
        12 &\citealt[][(K08)]{kiminki_etal_2008} & Visible & 17 \\
        13 &\citealt[][(V08)]{vink_etal_2008} & Visible & 54  \\
        14 &\citealt[][(K09)]{kiminki_etal_2009} & Visible & 22 \\
        15 &\citealt[][(W09)]{wright_drake_2009} & X-Ray & 1696 \\
        16 &\citealt[][(S09)]{skiff_2010} & Various & 150 \\
        17 &\citealt[][(K10)]{kobulnicky_etal_2010} & Visible & 17 \\
        18 &\citealt[][(R11)]{rauw_2011} & X-Ray & 453 \\
        19 &\citealt[][(K12a)]{kiminki_etal_2012a} & Visible & 21 \\
        20 &\citealt[][(K12b)]{kiminki_kobulnicky_2012} & Visible & 46 \\
        21 &\citealt[][(K12c)]{kobulnicky_etal_2012} & Visible & 28 \\
        22 &\citealt[][(C12)]{comeron_pasquali_2012} & Visible & 240 \\
        23 &\citealt[][(G13)]{guarcello_etal_2013} & Various & 1843 \\
        24 &\citealt[][(K14)]{kobulnicky_etal_2014} & Visible & 50 \\
        25 &\citealt[][(W14)]{wright_etal_2014a} & X-Ray & 7924 \\
        26 &\citealt[][(W15)]{wright_etal_2015} & Various & 167 \\
        27 &\citealt[][(W16)]{wright_etal_2016} & Various & 873 \\
  28 &\citealt[][(B18)]{berlanas_etal_2018} & Visible & 223 \\
        29 &Radio Master Catalogue$^*$ & Radio & 2850 \\
        30 &Galactic O Star Catalogue$^{**}$ & Various & 15 \\
        31 &Simbad Database$^{***}$ & Various & 2077 \\
        \hline
        32 & \citealt[][(2MASS)$^{\dagger}$]{cutri_etal_2003} & JKH & 27090 \\
        33 & \citealt[][(HERSCHEL)$^{\dagger}$]{poglitsch_etal_2010} & IR & 1346 \\
        34 &\citealt[][(SPITZER)$^{\dagger}$]{beerer_etal_2010} & IR & 69043 \\
        35 &\citealt[][(WISE)$^{\dagger}$]{wright_etal_2010} & Radio & 10579 \\
        36 & \citealt[][(G12)]{guarcello_etal_2012} & RIZ & 58415 \\
        \hline
        
        \end{tabular}
        
\label{tab:catalogues}
\end{center}
\footnotesize{
        $^{*}$Radio Master catalogue is a combination of two previous catalogues from \citet{dixon_1970} and \citet{kuehr_etal_1979}.\\
        $^{**}$Galactic O star Catalogue was taken from \citet{maiz-appelaniz_etal_2004}.\\
        $^{***}$Simbad Database catalogue was constructed using a VO search with a radius of 24 arcminutes, centred on the Cyg OB2 association.\\
        $^{\dagger}$Each of these catalogues was derived from its respective large-scale survey using the NASA/IPAC Infrared Science Archive. The number of sources represent those within a 24 arcminute radius of the centre of the Cyg OB2 association.
}
\end{footnotesize}
\end{table}

\subsection{Source-extraction methodology}\label{sourceext}

We ran SEAC independently on the April 11 and 26 full-field images. As the final full-field images are large, for efficiency of processing they were divided into subimages to run SEAC. Each subimage was 2600$\times$2600 pixels with an overlap of 100 pixels on each edge to ensure no sources were missed close to the edge of the subimage. The following describes the process applied in both cases to derive the final source list.

\begin{enumerate}
\item{SEAC was run on each subimage with a seed threshold, $\sigma_s$ = 5.0 and a floodfill threshold, $\sigma_f$ = 3.0, with a minimum source size of 99\% of the beam via an iterative script. The local noise map was produced at a resolution of 200$\times$200 pixels. For the majority of the field this ensured the best trade-off between testing potential islands against the RMS calculated from their local pixels, thereby accounting for the change in noise level across the full field, and ensuring the area considered had a sufficient source-free contribution to provide a suitable RMS.}
\item{The SEAC output from each subimage was combined into one master table, resulting in a total of 52 and 92 islands in the April 11 and 26 images respectively.}
\item{These results were then inspected individually to assess the detection. In each case a number of islands were discarded because they contained potential artefacts in the image or were associated with areas of steep changes in the noise. The presence of the strong radio emission from Cyg X-3 beyond but close to the edge of the Lovell primary beam resulted in aliasing artefacts in the contribution from pointing G. Seven and four islands  were discarded from April 11 and 26 data, respectively, as they appeared to be associated with these artefacts. There are regions within the full-field images where the noise level changes significantly over a relatively small region. Whilst this is mitigated in most cases by the use of the local noise background, there are areas where this resulted in potential false detections. In particular, regions where the local noise map covered the transition between the contributions from multiple overlapping images to a singular pointing in some cases resulted in pixels from regions of higher noise being identified against a relatively low local noise background calculation. This was also the case towards the edge of the full field where the primary beam response changes more rapidly. As a result, 9 and 17 islands were discarded from the April 11 and 26 data, respectively, (including one duplicate at the overlapping edge of two subimages). A further two islands were also discarded from the April 26 data as they were identified on the extreme outer edge of the full-field map as well as one from the April 11 data for the same reason.}
\item{Three and six islands were identified as duplicates from the April 11 and 26 data, respectively, where the same source had been identified in the overlapping region of two sub-images.}
\item{Another three and eight were identified from the April
11 and 26 data, respectively, as being associated with other islands and part of the same source.}
\item{Whilst many of the sources are identified in both epochs, as SEAC was run independently on each epoch, several were detected in only one. For the majority, the single detections were made in the April 26 image as this is the most sensitive; however three were identified only in the April 11 image.}
\item{Further SEAC runs were performed with lower seed thresholds of 4.5$\sigma$ and 4$\sigma$ (with flood thresholds of 3 and 2.6$\sigma$ respectively) to search for the counterparts of these sources, successfully detecting four in the April 11 image. This also identified a further three islands in the April 26 images with comparative properties to those identified in the primary SEAC run which have therefore been included in the final list.} 
\item{The final catalogue contains 61 sources and is shown in Fig. \ref{tab:clasc_main}.}

\end{enumerate}

For an image containing 20000 $\times$ 20000 pixels (i.e. 4$\times10^8$), with noise properties which can be represented by a Gaussian profile, approximately $ 225$ of these pixels will have a value larger than 5$\sigma$. Furthermore, $\sim1.056\times10^6$ pixels will have a value large than 3$\sigma$. Given the large image sizes, a certain number of false detections are expected particularly when considering the full-field mosaiced image. The floodfill algorithm within SEAC is expected to limit the number of false detections through the combination of the thresholding, source size, and local background calculation and this is further aided by the mosaiced combination of the images. The parameters used were chosen to minimise the number of false detections whilst detecting all sources present as determined by previous Monte-Carlo studies of the algorithm performance \citep{peck_thesis_2014}. As a result the majority of the 
observed false detections were found towards the outer edges of the full-field images where the noise level increases due to the response of the primary beam. 

\begin{figure*}[!htbp]
\begin{center}
\includegraphics[scale=0.5]{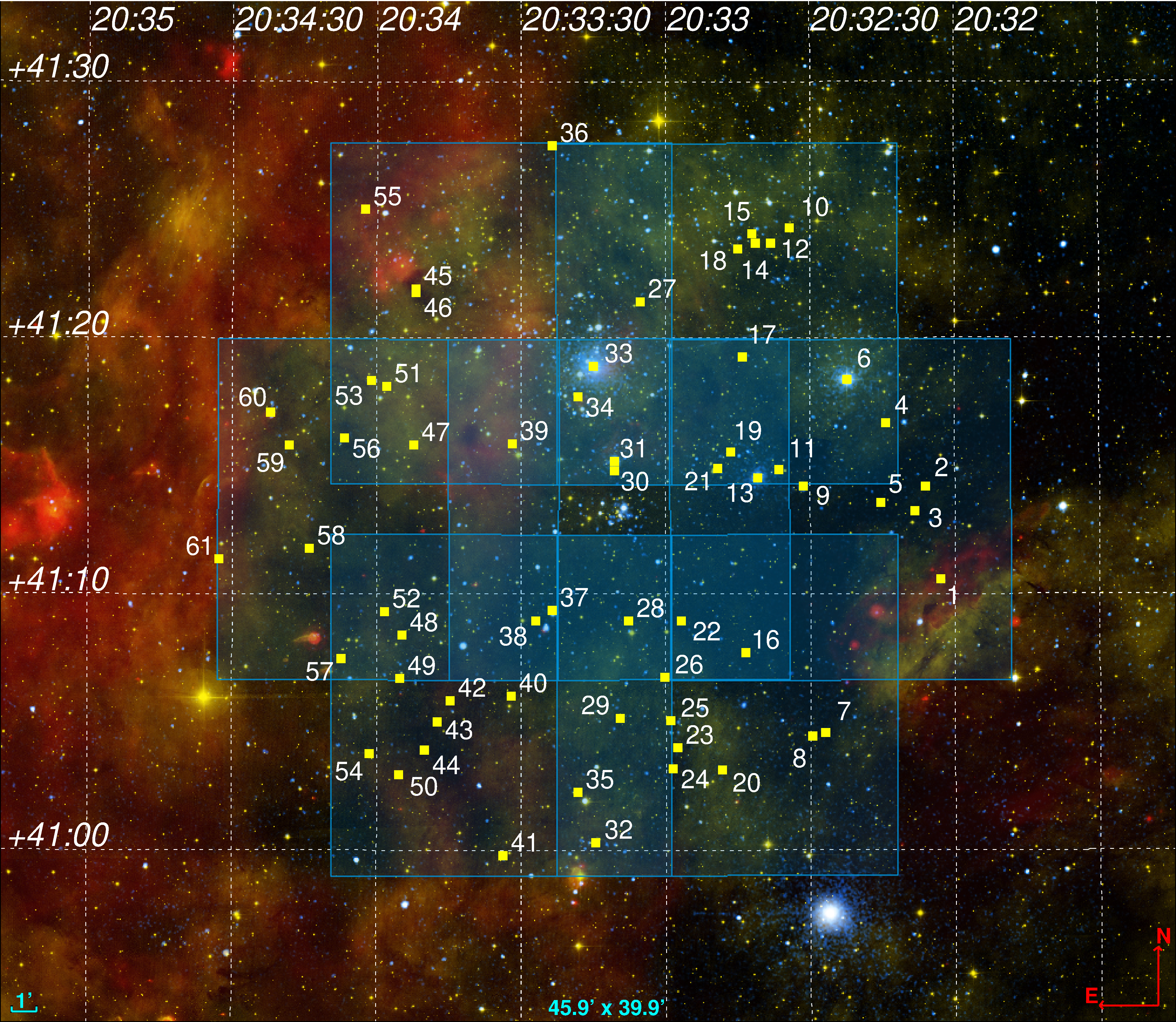}
\caption[Positions of the 61 sources (shown as yellow squares) derived from the CLASC]{Positions of the 61 sources (shown as yellow squares) in the CLASC, overlaid onto the areas as marked by the seven COBRaS L-band pointings. The background shows a composite of images from Chandra X-ray (NASA/CXC/SAO/J.Drake et al), Isaac Newton Telescope (Optical: Univ. of Hertfordshire/INT/IPHAS), and Spitzer IR (NASA/JPL-Caltech) observations.}
\label{fig:aladin_clasc}
\end{center}
\end{figure*}


{\renewcommand{\arraystretch}{1.2}
\setlength\tabcolsep{4pt}

\normalsize


\section{Results}
\label{results}

\subsection{COBRaS L-band All Source Catalogue}

The COBRaS L-band All Source Catalogue (CLASC) consists of 61 objects from the maps produced from the COBRaS L-band observations. Of these 61 objects, 58 were detected in the April 26 images and 31 were detected in the April 11 data;
9 were found to have more than one component as detected with SEAC; 29 were found in both epochs; 3 were detected in only the April 11 data; and 30 were detected only in the April 26 data.

The positions of the sources are plotted in Figure \ref{fig:aladin_clasc} which shows the region covered by the COBRaS L-band observations as well as the areas imaged within each pointing. Images of each of the identified sources is presented in Figures \ref{fig:both_1} and \ref{fig:both_app} where the April 11 and April 26 observations are shown on the left and right respectively. Contours are plotted at -1, 1, 1.4, 2, 2.8, 4, 5.6, 8, 11.3, and 16 $\times\, {\rm{3\sigma}}$  and the colour scale ranges from ${\rm{3\sigma}}$ to the source maximum for each source and epoch. For sources detected in only one epoch, images are shown in Figure \ref{fig:11th_1} for April 11 and Figures \ref{fig:26_1} and \ref{fig:26_app} for April 26. The positions, integrated flux densities and their associated uncertainties, signal-to-noise ratio (S/N), and local RMS of each object detected within the COBRaS L-band observations can be found in Table \ref{tab:clasc_main}. The flux density errors are those taken from the SEAC output combined in quadrature with the amplitude calibration uncertainty, taken to be 10\%. The positional uncertainties have been calculated from a combination of the error in measuring the flux density-weighted mean pixel position of each source and the positional error inherent to the radio interferometer. Where the object was detected in both epochs the position is taken from the April 26 detection.




\begin{figure*}[!htbp]
\begin{center}
\begin{subfigure}[b]{\textwidth}
\includegraphics[scale=0.82]{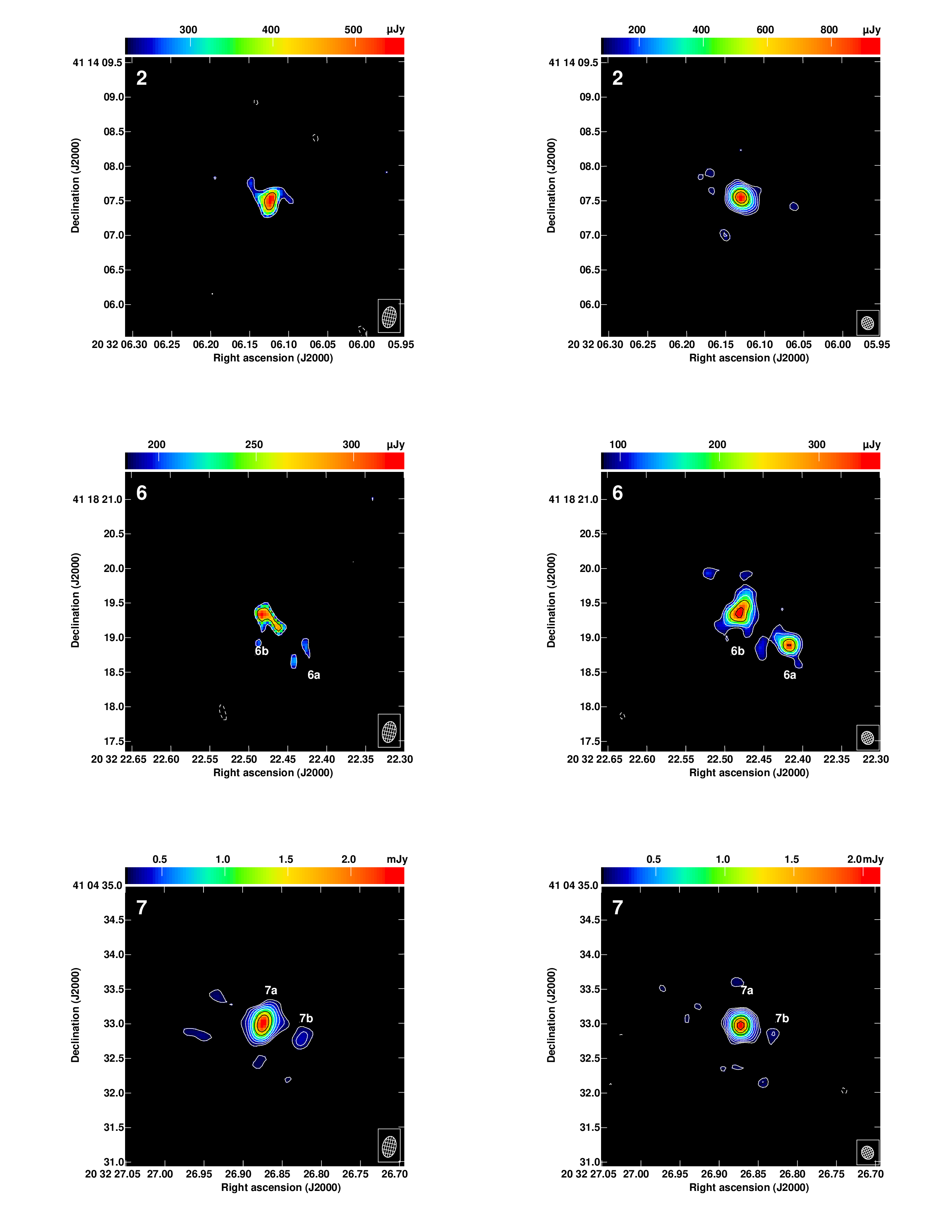}
\caption{}
\end{subfigure}
\caption[CLASC sources detected in both epochs]{A selection of CLASC sources detected in both April 11 (left) and 26 (right) epochs. The colour scale ranges from the 3$\times \sigma_{RMS}$ to the peak flux density. The contours are plotted at -1, 1, 1.4, 2, 2.8, 4, 5.6, 8, 11.3, 16) $\times$ 3$\sigma_{RMS}$. Images of the remaining sources detected in both epochs can be found in Appendix \ref{fig:both_app}.}
\label{fig:both_1}
\end{center}
\end{figure*}

\begin{figure*}[htp]
\begin{center}
\includegraphics[scale=0.85]{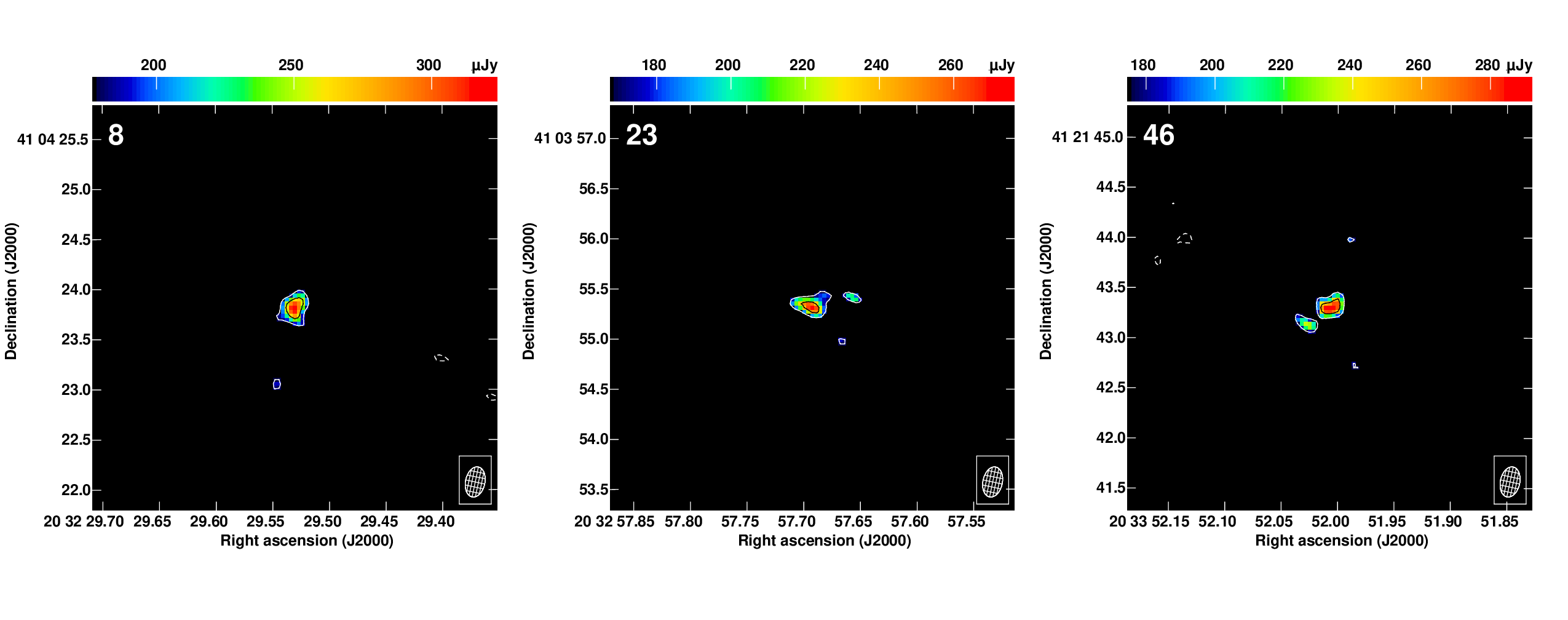}
\caption[CLASC sources detected in only the April 11 epoch]{CLASC sources detected in only the April 11 epoch. The colour-scale ranges from the 3$\times \sigma_{RMS}$ to the peak flux density. The contours are plotted at -1, 1, 1.4, 2, 2.8, 4, 5.6, 8, 11.3, and 16 $\times$ 3$\times\sigma_{RMS}$.}
\label{fig:11th_1}
\end{center}
\end{figure*}


\begin{figure*}[htp]
\begin{center}
\begin{subfigure}[b]{\textwidth}
\includegraphics[scale=0.82]{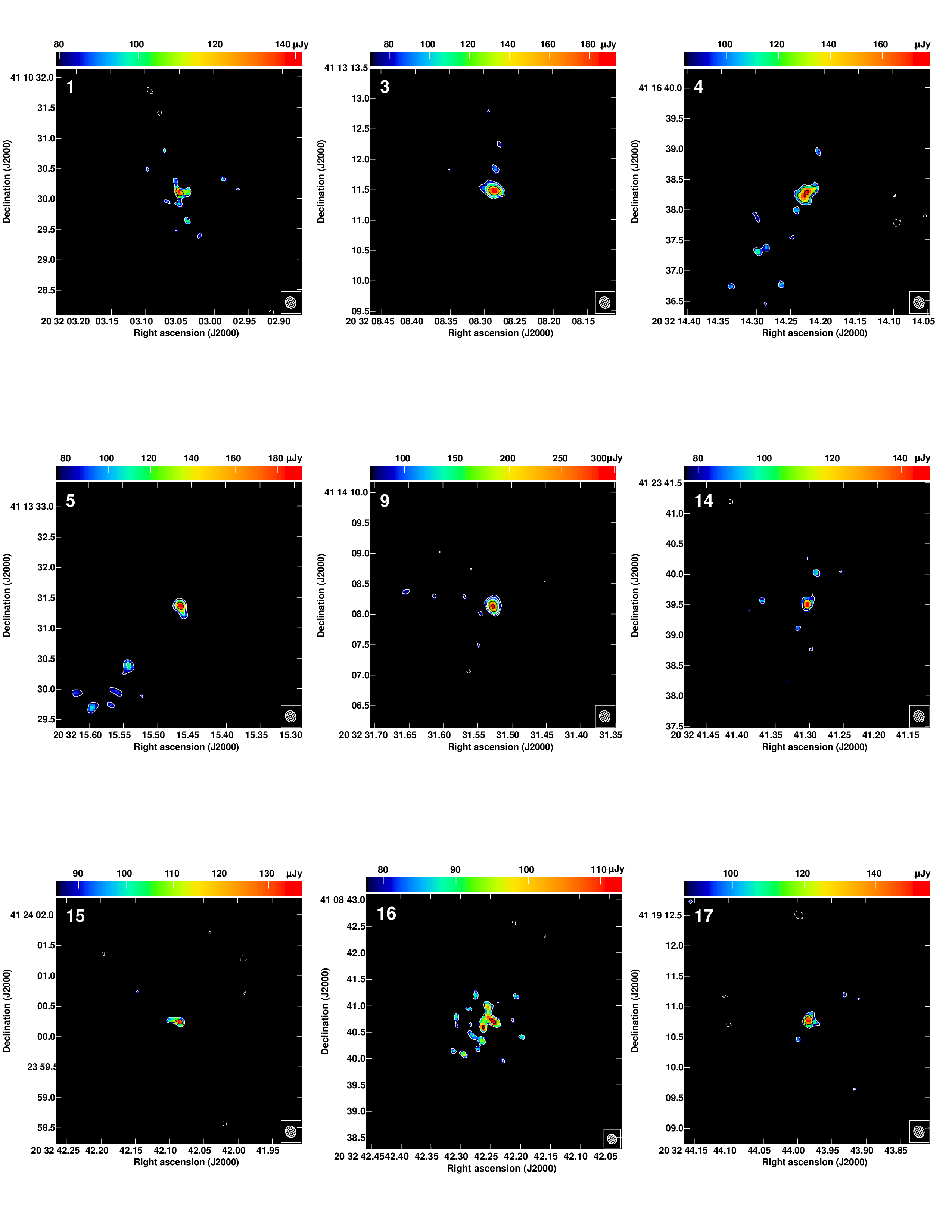}
\caption{}
\end{subfigure}
\caption[CLASC sources detected in only the April 26 epoch]{CLASC sources detected in only the April 26 epoch. The colour-scale ranges from the 3$\times \sigma_{RMS}$ to the peak flux density. The contours are plotted at -1, 1, 1.4, 2, 2.8, 4, 5.6, 8, 11.3, 16) $\times$ 3$\times\sigma_{RMS}$. Images of the remaining sources detected in only the April 26 epoch can be found in Appendix \ref{fig:26_app}.}
\label{fig:26_1}
\end{center}
\end{figure*}

\subsection{Source identification}
\label{source_identification}

Each of the 61 detected objects was cross-correlated in turn with the Cyg OB2 super catalogue, the 2MASS, WISE, SPITZER, and G12 \citep[][]{guarcello_etal_2012} catalogue. Each source was cross-matched to within an initial accuracy of 3.0 arcseconds in order to account for the large variety of observations that contributed to the Cyg OB2 catalogue. 
From the initial match, a total of 32 objects from CLASC were found to have positions within 3.0 arcseconds of at least one previously found object in the three cross-correlation catalogues. To validate these matches a method originally developed by \citet{deruiter_willis_arp_1977}, and also used by \citet{setiagunawan_etal_2003}, was employed to find the likelihood ratio ($LR$) of a given match. The $LR$ is defined as the ratio between the probability distribution of position differences between a source and its identification, $dP_{id}$, and the probability distribution of the background objects, $dP_{bg}$. Whilst we refer the reader to the full text within \citet{deruiter_willis_arp_1977}, the least likelihood ratio, $LR$, is given as:

\begin{equation}
LR(r) = \frac{1}{2\lambda} exp\{ [r^2(2\lambda-1)/2] \}
\label{eqn:lr_xc}
,\end{equation}
where $r$ is the normalised separation between a source and its identification given by 
\begin{equation}
r = \left[ \frac{\Delta\alpha^{2}}{\sigma_{\alpha}^{2}}+\frac{\Delta\delta^{2}}{\sigma_{\delta}^{2}}\right] ^{1/2}
\label{eqn:norm_sep}
,\end{equation}

where $\Delta\alpha$ and $\Delta\delta$ are the separation of the source and their COBRaS position in RA and Dec. Here, $\lambda = \pi \sigma_\alpha \sigma_\delta \rho(b)$, which is the number of sources within an area corresponding to the size of the combined positional error in RA, $\sigma_{\alpha}^{2} = \sigma_{\alpha CLASC}^{2} + \sigma_{\alpha ID}^{2}$ and in DEC $\sigma_{\delta}^{2} = \sigma_{\delta CLASC}^{2} + \sigma_{\delta ID}^{2}$. In order to derive the source number densities, $\rho(b)$, and the positional uncertainties, the survey or observation associated with each identification was considered. Here, $\rho(b)$ is the source density and is calculated in each case by taking the number of sources within the restricted catalogue and dividing by the area covered. In most cases, the catalogues for cross-correlation were limited to a radius of 24\,arcmins centred on the central COBRaS pointing centre. Where this was greater than the area covered by the catalogue, $\rho(b)$ was calculated using the full catalogue list and the area covered. Large values of $LR$ imply a more probable detection.

The initial cross-correlation gave one or more identifications to 32 sources within CLASC. The well-studied sources within the sample (Cyg OB2 \#5, A11, Cyg OB2 \#12, Cyg OB2 \#9 and Cyg OB2 \#8A) had numerous identifications from within the Cyg OB2 super catalogue and only the most recent identifications from a given wavelength range (i.e. X-ray, IR, radio) were chosen.

Having calculated the likelihood ratio for each initial identification, those with a value of $LR \lesssim 1$ were discarded. 
This limiting value of $LR \approx 1$ was also used in the Westerbork radio survey by \citet{setiagunawan_etal_2003} and represents a good compromise between claiming a false identification and missing a true identification. 
We note that all sources originally identified with those from \citet{guarcello_etal_2012} have been discarded since for these identifications, $LR << 1$, which is due to a combination of the large number density of the survey and the small positional errors. Having discarded all identifications with $LR \lesssim 1$, a total of 68 remained, corresponding to 27 of the 61 sources from CLASC. These are shown in Table \ref{tab:identifications}.

\subsection{Source counts}


A source count was performed on the CLASC (see Table \ref{tab:clasc_main}) by first binning each of the 61 sources by their flux densities (according to that derived from the April 26 observations except for those only detected in the April 11 images). The flux density bins were defined by first starting at the lowest flux density source (CLASC \#39 with F$_{21cm}$ = 68 $\pm$ 19\,$\mu$Jy) and incrementing in bins of 0.5 in $log$ space (as was also used by \citealt{hopkins_etal_1998}), which provided a good resolution whilst maintaining a suitable number of sources within each bin. The final bin however was made much larger to incorporate source CLASC \#22, the highest flux density source in the sample at F$_{21cm} = 15.55\pm1.57$ m\,Jy. All of the bins are shown in Table \ref{tab:sourcecounts} alongside the number of sources within them, N$_{\nu}$. This represents the number of sources within a given flux density range observed in a $\sim$ 0.48 $\times$ 0.48 deg area of the sky (i.e. that covered by the COBRaS L-band observations). This number is then converted into the number of sources per steradian, and divided by the bin width to obtain the differential source count d$N$/d$S$, which is normalised to those expected in a Euclidean geometry (i.e. an isotropic, static universe with a non-evolving population) by dividing by $S^{-5/2}$ where S is the central flux density. This final value is shown in the last column of Table \ref{tab:sourcecounts}.

For comparison, these differential source counts are plotted alongside those obtained from previous radio surveys at 21\,cm in Figure \ref{fig:sourcecounts}. These include those from the Westerbork continuum survey of Cyg OB2 by \citet[][blue triangles]{setiagunawan_etal_2003} and from a number of VLA observations presented in \citet[][yellow squares]{condon_mitchell_1984}. For sources below 1 mJy, the e-MERLIN 21\,cm COBRaS observations led to significantly lower counts than those from \citet{condon_mitchell_1984}. This is likely due to the high resolution obtained with the e-MERLIN instrument meaning that these COBRaS observations are less sensitive to low-surface-brightness sources which are `resolved out'. For sources with a flux density larger than 1\,mJy, the derived source counts from the CLASC are in much better agreement with both of those from \citet{condon_mitchell_1984} and \citet{setiagunawan_etal_2003}. 

Additionally, it is worth considering whether or not some of the `missing' sources could potentially include the large population of YSOs likely to inhabit the Cyg OB2 association, which in contrast tend to have an angular size of up to a few astronomical units.
The large population of stars in Cyg~OB2 with masses down to below $1M_\odot$ revealed by the {\it Chandra Cygnus OB2 Legacy Survey} \citep{wright_etal_2014} has been comprehensively characterised in the optical and near infrared by \citet{guarcello_etal_2012} and \citet{guarcello_etal_2013}. It is of interest to understand whether or not any of the T~Tauri stars in the region should have been detected by our survey.

The $3\sigma$ limit of {$\sim$ 58\,$\mu$Jy/beam} in the central regions of the field of view for our observations is equivalent to $\sim 2\times10^{17}$ ergs s$^{-1}$ Hz$^{-1}$. This limit for the Cyg~OB2 distance of 1.4~kpc converted to the distance of Orion \citep[414~pc][]{menten_2007} is $2.3\times 10^{18}$ ergs s$^{-1}$ Hz$^{-1}$. Comparing this with the Orion population of non-flaring low-mass objects studied at 4.7 and 7.3 GHz by \citep[][their Fig.~15,]{forbrich_etal_2016} we see that only three stars of their sample of 556 radio detections have peak radio luminosities that exceed this threshold. Moreover, the brightest objects all have spectral indices indicative of thermal emission. Extrapolating to the lower frequency L-band of the survey reported here would further reduce the expected observable flux densities. Our L-band survey thus appears to fall just short of the detection threshold of the large population of T~Tauri stars in Cyg~OB2 suggesting that they are unlikely to form part of the missing population of sources $<$ 1\,mJy. Extrapolating from Orion, the brighter, thermally emitting objects would be expected to be picked up in future higher frequency and higher sensitivity C-band observations.

Given the area covered by this survey, we expect to detect a number of background galaxies based on previous L-band extragalactic surveys. It is possible to estimate the expected number within our field of view; however precise predictions are difficult given the variation in noise level across the field and the relatively high resolution of the COBRaS observations in comparison to most extragalactic surveys at this frequency (e.g. \citealt{miller_etal_2013}). The MERLIN+JVLA survey of the Hubble Deep Field (HDF) by \citet{muxlow_etal_2005} detected 92 galaxies within a 10 $\times$ 10\,arcmin field with resolutions of $\sim$ 0.2 - 1.0". Assuming an average noise level of 29\,$\rm{\mu Jybeam^{-1}}$ from the COBRaS observations results in 17 HDF sources with flux densities $\geq 5\,\sigma$, implying that we would detect approximately 100 background sources within the total COBRaS field of view. However, the majority of the sources detected by \citet{muxlow_etal_2005} are resolved and taking this into account reduces this number to approximately 10 - 30 sources. \citet{chi_etal_2013} also performed VLBI observations of the HDF and flanking fields (covering a slightly larger region than \citealt{muxlow_etal_2005}) detecting a total of ten sources $\geq 5\,\sigma$. As the resolution of these observations is much higher, and assuming an equivalent distribution of sources in the COBRaS field, we would expect to detect all of these sources. Extrapolating this to the full COBRaS field would suggest we would detect approximately 32 extragalactic sources.

\begin{table}[hb]
\begin{footnotesize}
\caption[Source counts from the CLASC]{Source counts from the CLASC. The uncertainty in the normalised differential source count dN/dS was derived assuming a Poisson process, i.e. $\sqrt{N}$.}
\label{tab:sourcecounts}
\begin{center}
\begin{tabular}{cccc}
\hline\hline
Flux density range & N$_{\nu}$ & N$_{\nu}$  & dN/dS $\times$ S$^{5/2}$  \\ 
($\mu$Jy) & & (sr$^{-1}$) & (sr$^{-1}$ Jy$^{1.5}$)    \\ 
\hline
68-112  & 9 & 128235  & 0.22$\pm$0.07 \\
112-184 & 8 & 113986  & 0.42$\pm$0.15 \\
184-303 & 13  & 185228  & 1.44$\pm$0.40 \\
303-500 & 10  & 142483  & 2.34$\pm$0.74 \\
500-825 & 5 & 71241 & 2.48$\pm$1.11 \\
825-1360  & 9 & 128235  & 9.45$\pm$3.15 \\
1360-2241 & 6 & 85490 & 13.33$\pm$5.44  \\
2241-3696 & 7 & 99738 & 32.93$\pm$12.45 \\
3696-15550  & 4 & 56993 & 43.67$\pm$21.83 \\
\hline
\end{tabular}
\end{center}
\end{footnotesize}
\end{table}

\subsection{Resolution analysis}

With the high resolution offered by e-MERLIN, it is possible to determine the spatial extent of the identified sources. A complete analysis regarding the resolved nature of each of the 61 objects found within the CLASC was undertaken in order to further understand their spatial properties. A number of commonly applied methods have been used as resolution indicators and compared with the measured sizes in Table \ref{tab:resolution_analysis}, enabling a test of their suitability in determining resolution (something that is only feasible with a relatively low source count survey where individual source sizes can be obtained). 

A comparison of the number of pixels in the source as measured by SEAC (to 3$\sigma$) and the number of pixels in the convolving beam size (measured at full width at half maximum; FWHM) is presented. Following \citet{setiagunawan_etal_2003}, we took sources with ${\rm{Area}}/\theta_{beam}$ > 1.9 as being resolved as indicated in Table \ref{tab:resolution_analysis}.

\begin{figure}[h]
\begin{center}
\includegraphics[scale=0.31]{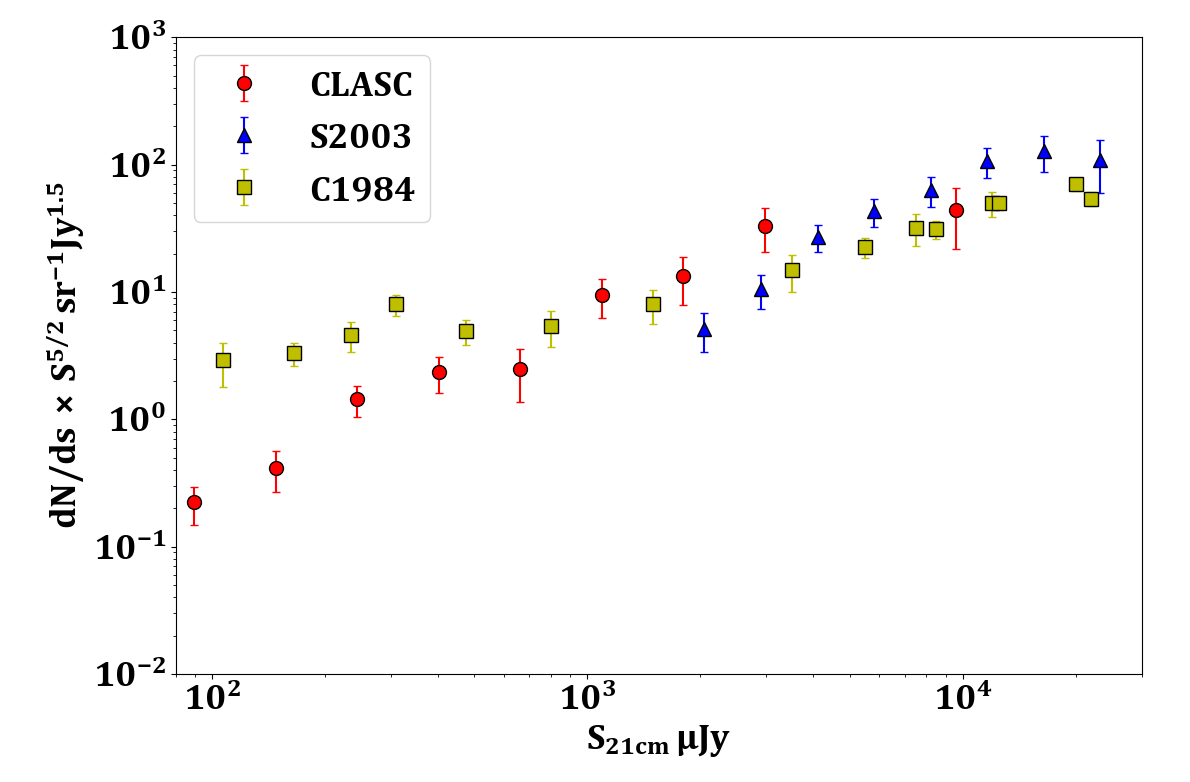}
\captionof{figure}{Normalised differential source counts. CLASC sources are represented by red circles and are compared against the source counts from other surveys at 21\,cm by \citet[][blue triangles]{setiagunawan_etal_2003} and \citet[][yellow squares]{condon_mitchell_1984}.}
\label{fig:sourcecounts}
\end{center}
\end{figure}

The ratio of the integrated flux density to the peak flux density is often used in large radio surveys as a measure of the resolution of the observed sources, following the relation \begin{equation} \frac{S_{int}}{S_{peak}} = \frac{\theta_{maj}\theta_{min}}{b_{maj}b_{min}} \label{eqn:resln} ,\end{equation} \citep[see e.g.][]{huynh_etal_2005, moss_etal_2007, miller_etal_2013} where $\theta_{maj}$ and $\theta_{min}$ represent the FWHM of the source and $b_{maj}$ and $b_{min}$ represent the FWHM of the image restoring beam. The measured ratios for each of the sources for both epochs are listed in Table \ref{tab:resolution_analysis} and are plotted in Fig. \ref{fig:resolution_total} versus S/N ($S_{peak}/\sigma$). The upper and lower panels show the April 26 and April 11 data respectively. To determine which of our sources are resolved, we follow the methodology presented in \citet{huynh_etal_2005} and define a lower envelope that contains $\sim $90\% of the sources with $S_{int}/S_{peak} < 1$. 
The resulting line is mirrored around $S_{peak} = S_{int}$ and is shown with a red dash line in Fig. \ref{fig:resolution_total}. The upper envelope for both epochs is given by \begin{equation} \frac{S_{int}}{S_{peak}} = 1 + \frac{100}{S/N^{m}} \label{eqn:resln_envelope} ,\end{equation} where $m$ is 3.6 and 3.1 for the April 11 and 26 epochs respectively. Sources lying outside of this envelope are considered resolved.

The sizes of the sources in both epochs have also been determined using the {\textsc{AIPS}} task {\textsc{JMFIT}} to perform a 2D Gaussian fit to each source. The deconvolved FWHM values are listed in Table \ref{tab:resolution_analysis}. Where the source is heavily resolved or deviates significantly from a Gaussian morphology a largest angular size (LAS) has been measured (these values have not been de-convolved), measured to and from the 3$\sigma$ level across the largest extent of the source.
In order to provide a comparative metric to the other resolution estimates, the ratio of the measured area divided by the primary beam (i.e. the right-hand side of Equation \ref{eqn:resln}) of each source is also listed. For sources with sizes measured from Gaussian fitting, the convolved major and minor axis have been used. For those sources with LAS measurements, these values were calculated using an average source size. Along with LAS, this was determined using a secondary measurement perpendicular to the LAS covering the smallest angular extent of the source. The full resolution information upon each of the CLASC sources can be found in Table \ref{tab:resolution_analysis}.

\begin{figure}[h]
\begin{center}
\includegraphics[height=7cm,width=0.47\textwidth]{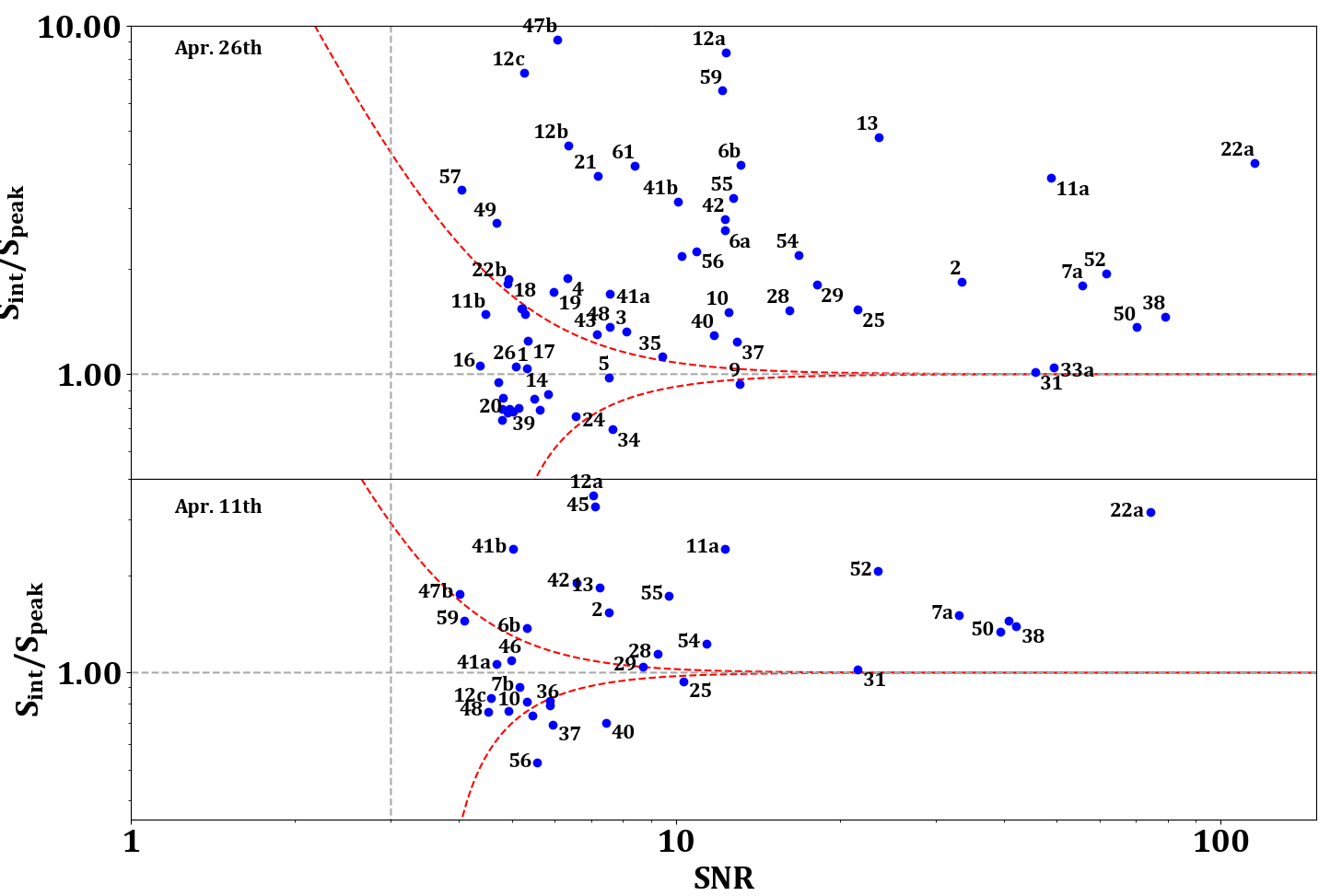}
\captionof{figure}{Ratio of the integrated flux density to the peak pixel flux ($S_{int}/S_{peak}$; see Table \ref{tab:resolution_analysis}) plotted against the S/N of each of the CLASC sources for the April 26 (top) and April 11 (bottom) epochs. The red dotted lines represent the boundary between  resolved (outside the red dotted lines) and unresolved (inside the red dotted lines) according to the criterion described above. Several CLASC IDs have been omitted for clarity.}
\label{fig:resolution_total}
\end{center}
\end{figure}

As expected, both methods for determining the resolution by directly examining the spatial extent as a function of beam size (i.e. via pixel ratios and size measurement ratios to the primary beam) provide close agreement of resolved and unresolved sources in general. If the same criteria of sources > 1.9\,$\theta_{beam}$ was applied (as is indicated in Table \ref{tab:resolution_analysis}) then a number of sources would be considered unresolved via the size measurements but resolved via the pixel count. For the April 26 observations, these two methods provide a total of 43 and 50 source components as resolved respectively. This tends to be the case for the compact sources with sizes measured via {\textsc{JMFIT}} and is possibly a reflection of the difference between measuring to the FWHM (for {\textsc{JMFIT}}) and to the 3$\sigma$ level for the SEAC pixel count. If instead the sources with ${\rm{Size}}/\theta_{beam}$ > 1 are considered as resolved, as is more commonly the case, then the numbers of resolved sources would agree more closely. The use of the ratio of $S_{int}/S_{peak}$ as a measure for resolution is commonly used for large surveys containing hundreds to thousands of sources and the portion of resolved sources within any given survey is likely to better reflect that of the whole population. We consider it a suitable approach here, though with a total of 61 sources in the COBRaS April 26 data (31 for April 11) we will potentially suffer from low number statistics when applying this analysis. However, we find that the percentage of resolved sources using this approach compared to that from pixel counting is in very good agreement.

\section{The COBRaS 21\,cm detections}
\label{analysis}

\subsection{Identified objects}
\label{objectsinclasc}

A total of 61 objects were detected in these observations, 27 of which were matched to previously identified sources at another waveband. These sources are discussed in turn below and grouped where possible according to any pre-existing classification. Previous radio observations from a Westerbork survey \citep{setiagunawan_etal_2003} enable a comparison for a number of the identified sources and the observed flux densities have been summarised in Table \ref{tab:SBHW_sources}.


\subsubsection{Single massive stars: CLASC \#13}

Radio observations can be used to determine the mass-loss rates of single massive stars whilst offering constraints on the degree of structure (or wind clumping) within their powerful stellar winds. This in turn can help to resolve current discrepancies in OB star mass-loss rates as inferred from different diagnostics \citep{fullerton_etal_2006, puls_etal_2006, sundqvist_etal_2010}. 

Of the 27 previously identified objects detected within the COBRaS L-band observations, the only potential\footnote{Secondary and tertiary components have since been found to be associated with Cyg OB2 \#12 by \citet{caballeronieves_etal_2014} and \citet{maryeva_etal_2016} respectively, however they have been deemed as too faint to significantly alter the derived luminosity of Cyg OB \#12.} single massive star detected is the hypergiant (and LBV candidate) Cyg OB2 \#12, which is discussed in detail in \citet{morford_etal_2016}. The mosaiced combination of the April 26 images presented here appears to reveal extended low-surface-brightness emission surrounding Cyg OB2 \#12 that was not previously seen (see Fig. \ref{fig:both_3}).

Based on the L-band dataset, \citet{morford_etal_2016} presented constraints on the 21\,cm flux densities of O3 to O6 supergiant and giant stars within the Cyg OB2 association to less than $\sim$ 70 $\mu$Jy. Using the relation derived in \citet{wright_barlow_1975} and \citet{panagia_feli_1975}, these fluxes are translated into `smooth' wind mass-loss upper limits of $\sim$ 4.4 - 4.8 $\times$ 10$^{-6}$ M$_{\odot}$ yr$^{-1}$ for the O3 supergiants and < 2.9 $\times$ 10$^{-6}$ M$_{\odot}$ yr$^{-1}$ for B0 to B1 supergiants. The constraints support the model in which the (outer) wind regions sampled by these radio observations are less clumped than the inner wind regions (close to the stellar surface) that see the production of H$\alpha$ emission. See \citet{morford_etal_2016} for further details.

\setlength\tabcolsep{6pt}

\begin{table*}[ht]
\begin{footnotesize}
\caption[Positions and flux densities of four PACWB systems detected in the COBRaS 21\,cm observations]{Positions and flux densities of the four massive star binaries as derived from the COBRaS L-band observations. The positions listed are a weighted position calculated by SEAC and taken from the April 26 observations. Similarly, the fluxes listed across both observation epochs were calculated using SEAC with seed and flood threshold levels, $\sigma_s$ and $\sigma_f$ equal to 5.0 and 3.0, respectively.}
\begin{center}
\begin{tabular}{cccccc}
\hline \hline
\label{tab:cwb_objects}

CLASC & Source & RA & DEC & Spectral & Period  \\
ID & Name & (J2000) & (J2000) & Type & (days)  \\
\hline
6a & Cyg 5 (SW) & 20 32 22.42 & 41 18 18.89 & (O6.5-7I + O5.5-6I)$^{1}$ + late O/early B$^2$ & 2445.5 $\pm$ 109.5$^2$ \\
6b & Cyg 5 (NE) & 20 32 22.48 & 41 18 19.38 & + early B$^{3}$ & $\sim$ 9200$^{4, \star}$ yrs \\
8 & A11 & 20 32 31.53 & 41 14 08.14 &  O7.5III-I + O/B$^{5}$ & 15.511 $\pm$ 0.056$^{5}$ \\
21 & Cyg 9 & 20 33 10.73 & 41 15 08.14 & O5-5.5I + O3-4III$^6$ & 860 $\pm$ 3.7$^6$ \\
22 & Cyg 8A & 20 33 15.07 & 41 18 50.43 & O6If + O5.5III$^{7}$ & 21.908$^8$ \\
\hline

\end{tabular}

\end{center}
\end{footnotesize}
\footnotesize{References: \textbf{1} \citet{rauw_etal_1999}, \textbf{2} \citet{kennedy_etal_2010}, \textbf{3} \citet{contreras_etal_1997}, \textbf{4} \citet{linder_etal_2009}, \textbf{5} \citet{kobulnicky_etal_2012}, \textbf{6} \citet{blomme_etal_2013},  \textbf{7} \citet{debecker_etal_2006}, \textbf{8} \citet{blomme_etal_2010}.}
\footnotesize{\\$^{\star}$This has been estimated assuming a distance of 1.7 kpc to the object.}
\end{table*}

\subsubsection{Massive star binaries: CLASC \#6, 9, 31, and 33}

From the identifications made in Table \ref{tab:identifications}, four of the objects detected within the COBRaS 21\,cm observations are well-studied, multiple massive star systems. These include the O7.5\textsc{iii}-\textsc{i} + O/B system A11, the quadruple stellar system of Cyg OB2 \#5, and the massive O star binaries of Cyg OB2 \#8A and Cyg OB2 \#9. All of these objects have previously been classified as particle accelerating colliding wind binary (PACWB) systems. Their spectral types and periodicities are shown in Table \ref{tab:cwb_objects} (and are given by the references listed). The observed 21\,cm emission from these objects is likely due to the production of non-thermal (synchrotron) emission as a result of a colliding wind region between the stellar winds of the two (or more) components within the system. Their detection within these COBRaS L-band observations highlights one of the great advantages of observing at such long wavelengths (21\,cm), showing how sensitive this method can be to the non-thermal emission produced within a colliding wind binary. A complete discussion on the 21\,cm non-thermal emission detected from the massive multiple stars within these observations will feature in a separate paper.

\subsubsection{Infrared counterparts}
\label{sec:IR_match}

A number of the CLASC sources appear to have IR counterparts predominantly found in the Spitzer catalogue. Whilst several analyses based on colour--colour and colour--magnitude information have been performed previously \citep[see e.g.][]{wright_etal_2010a,guarcello_etal_2013}, it is useful to consider this specifically for the group of CLASC sources to provide further insight into the source type in each case. A total of 15 sources were identified as having Spitzer counterparts CLASC \#6, 9, 13, 25, 28, 31, 33, 34, 36, 48, 50, 53, 56, 59, and 60, though Spitzer magnitudes are not available in all bands for all sources \citep{beerer_etal_2010}. Of these sources, 6, 9, 13, 31, and 33 have been identified as binary systems.
\citet{guarcello_etal_2013} present several colour--colour and colour--magnitude diagrams as part of their multi-wavelength analysis to distinguish disc stars and classify YSOs. They present a [3.6]–[5.8] versus [4.5]–[8.0] diagram (see their Fig. 2) distinguishing discs, photosphere stars, and galaxies. Reproducing this diagram for the CLASC sources was possible for CLASC sources 6, 9, 31, 33, 34, and 56. This shows \#56 occupies a region of the diagram identified as containing disc stars and through the analysis in \citep{guarcello_etal_2013} was in fact identified as a class {\textsc I} YSO. Interestingly, this analysis would also suggest source \#34 \citep[identified as W14 4790 from][]{wright_etal_2014} lies in the overlapping region of photosphere and disc stars, with the remaining sources lying within the region considered to represent normal photosphere stars. Additionally, as part of their full analysis, \citet{guarcello_etal_2013} also identify CLASC \#60 as a class {\textsc II} YSO.
Assessing the stellar sources further and following the K-diagram presented in \citep[][see their Fig. 1]{comeron_etal_2002}, CLASC \#34 appears to occupy a region well below that identified for the spectral type B0,  suggesting that this is potentially a cooler-type object.

\citet{guarcello_etal_2013} also present a 4.5 versus [4.5-8.0] colour--magnitude diagram (see their Fig. 4). Reproducing this analysis for the available CLASC sources appears to suggest that \#25, 28, 50, and 53 would be situated inside the locus identifying AGNs, suggesting that these could in fact be extragalactic sources.

\subsubsection{CLASC: \#3}

Previously detected at a wavelength of 86\,cm with a flux of 960\,$\mu$Jy \citep{marti_etal_2007}, this source is detected in the April 26 observations with a flux density of $255\pm43$\,$\mu$Jy, providing a steep spectral index of $\alpha=-1.66$. The object is marginally resolved within the COBRaS April 26 observations with a slight elongation in the east--west direction. An extragalactic or Galactic origin for this source cannot be determined from these COBRaS L-band observations alone.


\subsubsection{CLASC: \#7}

This relatively bright radio object has been detected with a 21\,cm flux density of {$3.64\pm0.39$ mJy and $3.81\pm0.40$ mJy for the main component (7a) in the COBRaS April 26 and 11 observations, respectively. This suggests no variability in the 21\,cm emission of this object over the 15 day period. This source has previously been detected at both 6 and 21\,cm with the VLA at flux densities of $1.9\pm0.4$ mJy and $3.3\pm0.5$ mJy, respectively \citep{marti_etal_2005}, in good agreement with the 21\,cm flux densities found here.} \citet{marti_etal_2005} were searching for possible `hot spots' associated with the relativistic jets produced from the `nearby' micro-quasar Cygnus X-3. These latter authors propose that the emission detected as source \#7 in this work is indeed a hot spot candidate (HSC) and is associated with the X-ray binary despite lying at an angular distance of $7.07'$ to the north. \citet{marti_etal_2005} also detect a second HSC lying $4.36'$ to the south of Cyg X-3 (and therefore not within the field of view covered by these COBRaS observations), which in combination with source \#7 forms an almost perfect alignment with the micro-quasar at a position angle of $1.8^{\circ} \pm 0.4^{\circ}$. This is in agreement with the position angle of the inner arcsecond radio jet ($2.0^{\circ} \pm 0.4^{\circ}$ as measured by \citealt{marti_etal_2001}). Moreover, the asymmetry observed between the north and south HSC is also similar to that observed in the inner jet components \citep{marti_etal_2005}.

The CLASC detection of this source has two potential components. The secondary component (7b) is detected at $>5\sigma$ only in the April 11 image, though appears to be associated with low-level emission surrounding the main component (7a), which is present in both epochs. 7a is offset by only $0.144$" and within the $\sim 0.2$" uncertainty for the VLA observations taken in 2000 \citep{marti_etal_2005}.
Cyg X-3 is outside the primary beam of the Lovell antenna covered by COBRaS 21\,cm G pointing and does not feature in the full-field image. As a result, its positional information cannot be determined, with which the proper motion of the HSC in relation to Cyg X-3 could have been investigated. From the images in Fig. \ref{fig:both_1}, it is clear that the main component of \#7 appears to be fairly resolved within both epochs of the COBRaS L-band observations. However, the morphology represents a spherically symmetric source. If this source is indeed associated with the micro-quasar Cyg X-3, then the detected 21\,cm emission (or at least part of the emission) must be the result of synchrotron radiation produced from the particles within the relativistic jets of a micro-quasar that lies $\sim 14.8$ parsecs (at a distance to Cyg X-3 of 7.2 kpc; \citealt{ling_nan-zhang_tang_2009}) from its position. Whilst the symmetry with a second HSC to the south and their relation to the properties of the inner jets of Cyg X-3 corroborate this hypothesis, these 21\,cm COBRaS observations do not provide a conclusion as to the true nature of this object. However, given the resolution of these observations and the potential for the full jet region to contain large areas of low-surface-brightness emission, it is possible in this case that the emission in \#7 represents only the strongest region of emission.

\subsubsection{CLASC: \#10}
This object is detected in both epochs, though with a significant change in flux density over the 15-day separation changing from $\sim$222 to 597\,$\mu$Jy.
Whilst falling short of the required likelihood ratio in the cross-correlation ($LR$=0.4, required $>$ 1), this source lies close to a previously detected radio source from \citet[][offset by 1.97$''$]{marti_etal_2007}. If these are indeed the same source, the 86\,cm flux density of $\sim$1.45\,mJy \citep{marti_etal_2007}, in combination with the 21\,cm COBRaS flux density, implies a negative spectral index of $\alpha = -0.6$. However, this assumes that the object is not radio variable. Additionally, source \#10 has been matched to a source from the Chandra X-ray observations of \citet{wright_etal_2014a} with a value of $LR = 2.3$. The object therefore emits at both X-ray and radio wavelengths. Its radio spectrum is likely non-thermal suggesting the presence of synchrotron radiation whilst its 21\,cm emission has been observed to vary over a period of $\sim 15$ days. The strong non-thermal radio spectral index and the lack of a 2MASS counterpart would seem to argue that this is an extragalactic source rather than a massive star binary system.

\subsubsection{CLASC: \#11}

This object has been identified with source SBHW~81 from the previous radio observations by \citet{setiagunawan_etal_2003} with a very high $LR$. Source \#11 (SBHW 81 hereafter) is resolved in both of the COBRaS epochs with a secondary component detected in the April 26 data at $>5\sigma$. The main component appears to be extended with a slightly elongated structure to the northwest and southeast (see Fig. \ref{fig:both_2}). With no detection at 86\,cm from the observations of \citet{setiagunawan_etal_2003}, those authors classified SBHW~81 as type S/O (stellar/other) suggesting a Galactic origin. Furthermore, they detected this object at 21\,cm with a flux density of $3.6\pm0.5$ mJy. The two 21\,cm fluxes measured with e-MERLIN vary over the two epochs with the lower value in the first and higher value in the second epoch \citep[please see Table \ref{tab:SBHW_sources}, which shows the comparison of all sources coincident with SBHW sources from][]{setiagunawan_etal_2003}. Despite the differences in sensitivity and likely in $u,v$ coverage between these three observations, the variation between the flux densities is significant.

\def\arraystretch{1.2}
\vspace{0.3cm}
\begin{table}[htp]
\begin{footnotesize}
\caption[CLASC and SBHW comparison.]{CLASC source comparison with cross-matched SBHW catalogue sources from \citet{setiagunawan_etal_2003}. Where sources have been detected with multiple components (11, 12, and 47), a combined flux density is included. The spectral index is that calculated between the $S_{86\,cm}$ flux density and the COBRaS April 26 $S_{21\,cm}$ flux density, except for CLASC \#52 which lists the upper limit from \citet{setiagunawan_etal_2003}.}
\label{tab:SBHW_sources}
\begin{center}
\begin{tabular}{ccccccc}
\hline \hline
\multicolumn{3}{c}{CLASC}& \multicolumn{4}{c}{SBHW}\\ 
\vspace{0.1cm}
ID & $F_{21cm}^{11^{th}}$  & $F_{21cm}^{26^{th}}$  & ID & $S_{21cm}$ & $S_{86cm}$ & $\alpha$ \\ 
 & mJy & mJy & & mJy & mJy &  \\
 \hline 
 11 & 1.90$\pm$0.24 & 4.35$\pm$0.44 & 81 & 3.6$\pm$0.5 & -- & -- \\
 12 & 1.78$\pm$0.72 & 6.31$\pm$0.48 & 83 & 8.4$\pm$0.5 & 62$\pm$3 & -1.60 \\
 22 & 13.5$\pm$1.37 & 15.9$\pm$1.6 & 90 & 14.7$\pm$0.8 & 54$\pm$4 & -0.89 \\
 47 & 0.32$\pm$0.08 & 3.69$\pm$0.17 & 109 & 3.0$\pm$0.6 & -- & -- \\
 50 & 3.86$\pm$0.42 & 3.83$\pm$0.39 & 110 & 2.2$\pm$0.2 & -- & -- \\
 52 & 2.43$\pm$0.28 & 3.18$\pm$0.33 & 112 & 2.2$\pm$0.3 & -- & >\,0.37 \\ 
\hline
\end{tabular}
\vspace{0.1cm}
\footnotesize{}
\end{center}
\end{footnotesize}
\end{table}%

\subsubsection{CLASC: \#12}

CLASC \#12 has also been identified with a source detected from \citet{setiagunawan_etal_2003} known as SBHW~83. Detected in both observation epochs of the COBRaS L-band data as shown in Fig. \ref{fig:both_2}, this source shows a huge amount of extended structure. The observations taken on April 26 resolve three different components to the system. 
Summing the measured flux densities of each of the detected components results in a total 21\,cm flux density of $6.31\pm0.48$ mJy. Whilst readily detected in the April 11 observations they only resolve the main component (CLASC \# 12a) and partially detect component 12c with a lower total flux density (see Table \ref{tab:SBHW_sources}); though if taken to be the same components in both epochs the peak positions have shifted in both cases. As the source is clearly resolved, the difference between the two epochs is potentially due to the different sensitivities resulting in missed lower-surface-brightness emission in the April 11 epoch. It is also possible that the difference in spatial scales sampled by their $u,v$ coverages, particularly for the inner baselines (largely as a result of observing time differences and missing telescopes), contributes to this effect. 


The observations of \citet{setiagunawan_etal_2003} report flux densities at 21\,cm and 86\,cm of $8.4\pm0.5$ mJy and $62\pm3$\,mJy, respectively, for SBHW~83. Whilst this 21\,cm flux density is comparable to that found from the April 26 COBRaS observations (we highlight the fact that the object is unresolved in the observations from \citealt{setiagunawan_etal_2003}), the high 86\,cm flux density in combination with the 21\,cm flux density derived here results in a negative spectral index, $\alpha = -1.6$; we note that this is likely a lower limit for the spectral index since the COBRaS observations are still likely missing flux on large spatial scales in comparison to the observations from \citet{setiagunawan_etal_2003}. This source was also detected in the GMRT 150\,MHz survey with a flux density of 96.5$\pm$8.5\,mJy \citep{intema_etal_2017}. Taken alongside these observations, this also implies a steep spectral index of $\sim -1.2$. The non-thermal radio spectrum and extended morphological features suggest that SBHW~83 is most likely extra-galactic in origin as it appears to be lacking any infrared or optical counterpart. More explicitly and with reference to Fig. \ref{fig:both_2}, from a morphological perspective, the source appears to show a bright core (i.e. ID: \#12a) and further emission representing bipolar jets to both the north and the south.
 

 \subsubsection{CLASC: \#17}

This is a relatively faint point-like source as seen in the COBRaS April 26 observations and has been identified with a previous radio detection from \citet{marti_etal_2007}. Combined with the 0.67\,mJy flux density at 49\,cm, this suggests a steep spectral index of $\sim$-1.4. Given the low flux density observed in the April 26 observations, this source was unlikely to be detected in the April 11 data as a result of the higher noise. Therefore, whilst it cannot be ruled out completely, the observations do not suggest any significant variability on these timescales.

\subsubsection{CLASC: \#22}

CLASC \#22 has also been detected previously by \citet{setiagunawan_etal_2003} and is identified as SBHW~90 with measured flux densities at 21\,cm and 86\,cm (see Table \ref{tab:SBHW_sources}). This is the brightest source detected in these COBRaS observations with a total flux density of $\sim$15.9\,mJy in the April 26 data. Whilst the April 11 flux density  is slightly lower, it is well within the errors of the two observations and is therefore considered consistent. A secondary small component to the east of \#22a is however only detected in the April 26 data and is likely a result of the higher sensitivity (see Fig. \ref{fig:both_3} ).  The main component of this source is highly resolved in these observations, showing significant elongation from the northwest to southeast direction, highly indicative of a potential jet structure.

The observed flux density is in good agreement with that of the 21\,cm detection from the lower resolution observations in \citet{setiagunawan_etal_2003}, suggesting that there is no strong long-term variability and/or that we are recovering the majority of the emission. 


\citet{setiagunawan_etal_2003} determine a 21\,cm flux density of 19\,mJy when convolved with the same beam size as their 86\,cm observations and hence derive a negative spectral index, $\alpha = -0.77$ for SBHW~90. Using our 21\,cm flux density instead only confirms the likely steep spectrum nature of this source, providing an estimate of $\alpha = -0.89$. This source was also detected in the GMRT 150\,MHz survey though below the 7$\sigma$ threshold \citep[24.5\,mJy$bm^{-1}$;][]{intema_etal_2017}. Similarly to ID\#12 (SBHW~83), the highly resolved nature of CLASC \#22 (SBHW~90) alongside its negative spectral index and lack of IR or optical counterpart suggest an extragalactic origin for this source.

\subsubsection{CLASC: \#25}

CLASC \#25 is detected in both of the COBRaS 21\,cm epochs as a compact source varying slightly from $\sim$ 0.57 to 0.90\,mJy. The object appears marginally resolved in both epochs with $S$ to beam ratios of 1.4 and 4.4 from April 11 and 26 data, respectively (Fig. 6c). In combination with potential differences in $u,v$ coverage sampled by either observation and the small difference between the flux densities, this implies that this source is not significantly variable over the 15-day interval. CLASC \#25 has only previously been detected within the X-ray observations of \citet{wright_etal_2014a} as a point source within their Chandra Legacy survey of the Cyg OB2 association. With no optical counterpart and a hard X-ray spectrum (as inferred from \citealt{wright_etal_2014a}), this object is likely to be extragalactic in origin, which is supported by its relative positioning on the IR colour--magnitude diagram (see section \ref{sec:IR_match}).

\subsubsection{CLASC: \#28}

This object has previously been detected with the Spitzer Space Telescope at wavelengths of 4.5, 5.8, 8.0, and 24 $\mu$m. It is detected in the COBRaS observations as a marginally resolved compact source showing no variability with an average flux density of $\sim$0.45\,mJy. Since it appears both faint and compact in these COBRaS observations (see Fig. \ref{fig:both_4}) and has been detected in the IR, it is possible that this object is situated within the Galaxy and is a potential member of the Cyg OB2 association. However, when plotted on an IR colour--magnitude diagram, it is situated in a region associated with AGN, and therefore could instead be extragalactic (see section \ref{sec:IR_match}).

\subsubsection{CLASC: \#34}

CLASC \#34 is a faint detection in only the April 26 data with a flux density of $\sim$100\,$\mu$Jy. It is likely too faint to be detected in the April 11 data. However, this source has been matched to a number of different catalogues with small positional offsets and high likelihood ratios (see Table \ref{tab:resolution_analysis} for details), suggesting that it has optical IR and X-ray counterparts. Combined with the low radio detection, this source is likely of Galactic origin.

\subsubsection{CLASC: \#36}

CLASC \#36 appears to be partially resolved in both epochs and changes significantly in flux density over the 15-day period from $\sim$0.4 to 1.1\,mJy. Furthermore, whilst not detected at $>5\sigma,$ there appears to be potential lower surface brightness emission associated with this source in both epochs at the level of $\sim4\sigma$, suggesting that higher sensitivity observations may reveal a larger source extent (see Fig. \ref{fig:both_6}).
Additionally, this object has been detected in the IR by the Spitzer Space Telescope at both 3.6 and 4.5 $\mu$m, suggesting a potential Galactic origin for this object. 

\subsubsection{CLASC: \#40}

This object has also previously been detected at X-ray wavelengths during the observations of \citet{wright_drake_2009} and \citet{wright_etal_2014a} and is observed here with 21\,cm flux densities of $\sim$0.23 and 0.35\,mJy in these COBRaS observations. Whilst differing slightly, with a relatively low flux density, this object does not appear to show any significant variation over the timescale of 15 days. Moreover, the object appears only partially resolved in the April 11 observations and slightly more so in the those of April 26. Similarly to CLASC \#25, this object has no optical or NIR counterpart; this, in combination with its hard X-ray spectrum suggests that this source is extragalactic in origin.

\subsubsection{CLASC: \#47}

CLASC \#47 is another object with a previous radio identification; SBHW~109 from \citet[][see Table \ref{tab:SBHW_sources} for a comparison]{setiagunawan_etal_2003}. It is only fully detected in the April 26 epoch of the COBRaS observations and is found to contain two components, one to the south (47a) at a 21\,cm flux density of $\sim$2.3\,mJy and another to the north (47b) at a flux density of $\sim$1.3\,mJy. 
The position given by \citet{setiagunawan_etal_2003} lies approximately equidistant between these two components. With a derived 21\,cm flux density of $3.0\pm0.5$ mJy from \citet{setiagunawan_etal_2003}, it is clear that the COBRaS observations are resolving out the majority of the radio emission associated with the source. The images shown in Fig. \ref{fig:both_8} clearly supports this, showing the emission to be highly distributed in both components; this is likely the reason for the only marginal detection in the lower sensitivity April 11 epoch. With no known detection of the source at other radio frequencies no spectral index information can be obtained. From a morphological perspective, this source resembles the radio emission from bi-polar jets associated with an  AGN and is therefore likely of extragalactic origin.

\subsubsection{CLASC: \#50}

This object is another that was identified within the radio observations of \citet{setiagunawan_etal_2003}. Otherwise referred to as SBHW~110, this source is detected in both epochs of the COBRaS observations (see Fig. \ref{fig:both_8}). The 21\,cm flux density of SBHW~110 is stable over the 15-day period with an average flux density of $\sim$3.8\,mJy. \citet{setiagunawan_etal_2003} measure a lower 21\,cm flux density (see Table \ref{tab:SBHW_sources}), suggesting potential long-term variability which could also explain the lack of detection at 86\,cm. The authors suggest the emission is likely thermal in nature and postulate that the source is of a Galactic origin. Within the COBRaS observations, SBHW~110 appears to be almost spherically symmetric and is relatively well resolved with $S$ to beam ratios of 2 and 5 in April 11 and 26 data, respectively. There also appears to be some low-level emission within the immediate vicinity of the source; however this is very close to the noise. There currently exists no other detections of this source at any other frequencies within the literature. With no spectral index information and no significant variability over a short timescale (i.e. 15 days), it is difficult to determine whether or not this object is associated with our Galaxy. However, according to the Spitzer magnitude data, it potentially sits within a region occupied by extragalactic sources (see section \ref{sec:IR_match}).

\subsubsection{CLASC: \#51}

CLASC \#51 is detected with a relatively modest flux density of $\sim$90\,$\mu$Jy in the April 26 observations, though has been identified with a hard X-ray source from \citet{wright_drake_2009}. With no optical or IR counterpart this would potentially suggest an extragalactic origin; though with such little information, this is not conclusive, especially considering the low radio flux density observed.

\subsubsection{CLASC: \#52}

CLASC \#52 appears to be resolved in both of the COBRaS observation epochs and shows a slight deviation from spherical symmetry due to elongation to the northwest and northeast within the April 11 and April 26 observations, respectively. This source has been identified with SBHW~112, another object first detected within the radio observations of \citet{setiagunawan_etal_2003}. The 21\,cm flux density of SBHW~112 is found to vary over the 15-day interval from around 2.4 to 3.2\,mJy. \citet{setiagunawan_etal_2003} find a slightly lower 21\,cm flux density (see Table \ref{tab:SBHW_sources}) suggesting potential longer term variability as well; however they are unable to detect the source at 86\,cm. The authors conclude that the object is likely to be found within the Galaxy and of a stellar nature.

\subsubsection{CLASC: \#53}

CLASC \#53 is detected as two compact components in the April 26 observations both with flux densities around $\sim$200\,mJy. Whilst relatively faint, the lack of detection in the April 11 data does hint at potential variability. Furthermore, this object has been identified with an IR counterpart detected at 4.5 and 8.0\,$\mu$m suggesting a potential Galactic origin for this source. 

Similar two-component systems have been observed within these COBRaS observations, such as the PACWB system of Cyg OB2 \#5 for example. With an angular distance between the components of $\sim$1.0$''$, this object could indeed resemble a multiple star system (such as Cyg OB2 \#5), where the two components are in fact non-thermal radiation from a CWR. However, additional radio observations (i.e. COBRaS 6\,cm) are clearly required to determine the spectral index of the system in order to further investigate the nature of this object.

It is worth noting however that an extragalactic origin cannot be ruled out for this source as when plotted on an IR colour--magnitude diagram, this source is potentially situated alongside AGN sources (see section \ref{sec:IR_match}).

\subsubsection{CLASC: \#56}

This object has been identified as a source from the catalogue from \citet{guarcello_etal_2013} with a $LR\approx 533$. This catalogue of proto-planetary disc stars within the Cyg OB2 association was compiled from a combination of observations from various IR surveys including in particular those from the Spitzer space telescope at 3.6$\mu$m, 4.5$\mu$m, 5.8$\mu$m, 8.0$\mu$m, and 24$\mu$m \citep{beerer_etal_2010}. \citet{guarcello_etal_2013} determine an IR spectral index of $\alpha_{IR} = 0.547$ for the source and classify it as a class \textsc{i} YSO through colour--colour analysis (see section \ref{sec:IR_match} for further discussion).


CLASC {\#56} is detected in the COBRaS observations in both epochs, though was detected with a low SEAC seed threshold of 4$\sigma$ in the April 11 data and appears to increase in flux density from $\sim$0.17 to 0.65\,mJy between the two observations.  In the April 26 image it appears relatively well resolved with a source size to beam ratio (${\rm{Size}}/\theta_{beam}$) of 4.39 and shows a slightly elongated morphology to the east (see Fig. \ref{fig:both_10}). The faint detection in the April 11 observations appears to be accompanied by potential further emission to the north of the source. 
There is a slight difference between the $u,v$ coverage of pointing C (providing the main contribution) between the two epochs and as a result the variation in the flux density between the two epochs may not be significant. Further observations are required to derive its radio spectral index, from which a comparison can be made to that expected of a class \textsc{i} YSO.

\subsubsection{CLASC: \#57}

CLASC \#57 has been detected in the COBRaS April 26 observations with a 21\,cm flux density of 419$\mu$Jy, using a SEAC seed threshold of 4.5$\sigma$ courtesy of its low peak flux density. It is not detected in the April 11 epoch courtesy of the decreased sensitivity of these observations and the inherently low 21\,cm flux density associated with the source (and thus no variability can be found between the two epochs). Furthermore, as can be seen in Fig. \ref{fig:26_1} the object appears to be a heavily resolved low-surface-brightness source and its true integrated flux density may well be much greater than measured here. 
This source has been identified with an X-ray counterpart within the observations of \citet{wright_etal_2014a} and also as a 2MASS object \citep{cutri_etal_2003}, both with a high likelihood ratio. From the 2MASS observations, its position on the J-H versus H-K plot does not lie within the region covered by hot (OB) stars and therefore suggests a cooler type of object. However, the resolved nature of the radio emission may not be consistent with a pure stellar identification. 
It is likely however, that this source is of Galactic origin and possibly a member of the Cyg OB2 association.

\subsubsection{CLASC: \#59}

This object has been detected in the April 26 epoch of these observations with a flux density of $\sim$2.5\,mJy and appears highly resolved (see Fig. \ref{fig:both_10}). Surprisingly, the object is only detected within the April 11 epoch with a flux density of $\sim$0.35\,mJy (detected using a SEAC seed threshold of 4.5$\sigma$). This short-term, large-scale variability implies this object is a transient source (see Section \ref{transients} for further discussion). Additionally, CLASC \#59 has also been detected with the Spitzer Space telescope at 3.6 and 4.5 $\mu$m. The lack of any X-ray counterpart and a previous IR detection argue in favour of a Galactic origin for this object, which is perhaps also a member of the Cyg OB2 association.

\subsubsection{CLASC: \#60}

This source is only detected in the April 26 data as a result of its relatively low flux density of $\sim$150\,$\mu$Jy. It is however identified with a source from the proto-planetary disc star catalogue from \citet{guarcello_etal_2013} with a high likelihood ratio $LR\approx 533$. This catalogue combines a number of IR surveys that detect this source at 5.8, 8.0, and 24\,$\mu$m giving an IR spectral index of $\sim$1.9. The authors classify this object as class \textsc{ii} YSO, one of around 1300 identified in their survey.

\subsubsection{CLASC: \#61}

CLASC \#61 has been identified with a source from the X-ray point source catalogue of \citet{wright_etal_2014a} with a $LR$ value of $\sim 383$. It has been detected in the April 26 epoch with a flux density of $1500\mu$Jy and appears resolved with a $S$ over beam ratio of 6.88. As can be seen in Fig. \ref{fig:26_4}, the object appears elongated in the southwest to northeast direction with a small amount of structure extending also to the north of the source. Interestingly, despite its relatively large measured flux density in the April 26 observations, CLASC \#61 is not detected at all within the April 11 observations. From this epoch, a 3$\sigma$ upper limit of $231\mu$Jy is placed upon its flux density. The source is located towards the very eastern edge of the full-field images, explaining the large local RMS values of 77 and 45 $\mu$Jy in the April 11 and April 26 observations respectively. These COBRaS observations show that the 21\,cm emission from CLASC \#61 is highly variable over a 15-day period. The fact that this variation is so large over such a short period of time  suggests two possible scenarios for this source. It is either a short-period massive star binary whose colliding wind region (CWR) gives rise to a large amount of non-thermal (synchrotron) emission which is subsequently reduced (i.e. due to a large amount of free-free absorption along the line of sight) a short time later at a different orbital phase, or a flaring event due to a pre-main sequence type object. Whilst the presence of a strong variable IR detection is not a pre-requisite of a colliding wind system \citep[e.g. WR146]{dougherty_1996}, the apparent lack of any IR or optical counterpart for this source would however make a CW unlikely. It is clear that further observations are required in order to decipher the physical properties of this object.

\subsection{Unidentified objects in CLASC}
\label{unidentified_clasc}

Of the 61 sources detected within the COBRaS L-band observations, 34 were found without any previous identification. With measured flux densities at a single frequency, no spectral index information can be obtained for these objects. The April 11 observations provide 13 of the unknown sources (14 including separate components) with a second flux density measurement. For the remainder, one must rely on a 3$\sigma$ upper limit on the flux density to make a comparison between the two epochs. Due to the difference in the $u,v$ coverage between the two COBRaS observations (courtesy of differing integration times, missing telescopes, and data excision), the resulting images will be sensitive to different spatial scales on the sky. Therefore, for those sources that are resolved, that is, those with a source size to beam (in pixels) ratio of  $ \gtrsim 1.9$, this must be taken into consideration when comparing flux densities \footnote{This definition of an `extended' or resolved source was also used in the analysis of the Westerbork radio survey by \citet{setiagunawan_etal_2003}.}. However, the image-plane mosaicing performed should aid in this regard, providing more combined coverage than a single pointing.

Table \ref{tab:unknownsources} shows the flux densities (or 3$\sigma$ upper limit on the flux densities) and their $S$ to beam size ratio of the unknown sources across both observational epochs. The only possible classification of these sources at this juncture must come from a qualitative assessment of their morphologies. Each source has been categorised according to the shape of the detected 21\,cm emission across both epochs where possible. Objects with a spherical or slightly elongated structure have been denoted `S/E' whilst those representing a `bow-shape'-like structure have been denoted `B'. Sources in between either morphology have been given an `E/B' morphological type and finally those sources with extended structure are given an `Ex' type. A bow-shape-like structure is perhaps indicative of non-thermal (synchrotron) emission as a result of the collision between two massive star winds. This was found to be the case for Cyg OB2 \#9 when imaged with the VLBA. This latter object is now known to harbour a CWR between its two massive star components (see \citealt{dougherty_pittard_2006}). Extended structure on the other hand is more likely to indicate a possible extragalactic origin for a given source since large-scale extended radio emission is readily observed in AGN and radio galaxies. 

Of particular note amongst the unidentified sources is \#41 which, particularly in the April 26 images (see Fig. \ref{fig:both_7}), shows a striking potential jet morphology. Whilst little can be currently said about its true nature, this latter information, combined with a lack of identification with any known IR or optical source, suggests that \#41 is likely a radio galaxy.

\def\arraystretch{1.2}
\vspace{0.3cm}
\begin{table*}[htp]
\begin{footnotesize}
\caption[CLASC sources without identification]{CLASC sources without identification.}
\label{tab:unknownsources}
\begin{center}
\begin{tabular}{cccccccl}
\hline \hline
\vspace{0.1cm}
CLASC & $F_{21cm}^{26^{th}}$ & $S_{26^{th}}/\theta_{beam}$ & $F_{21cm}^{11^{th}}$ & $S_{11^{th}}/\theta_{beam}$ & Morphology  \\ 
ID & ($\mu$Jy) & &  ($\mu$Jy) & \\ 
\hline
1 & 147$\pm$35 & 1.48 & <192 & -- & E/B \\
2 & 1747$\pm$187 & 5.66 & 858$\pm$148 & 2.62 & S/E  \\ 
4 & 335$\pm$57 & 2.70 & <177 & -- & E/B  \\ 
5 & 187$\pm$38 & 1.65 & <204 & -- & S/E  \\ 
8 & <78 & -- & 237$\pm$64 & 1.03 & E/B \\ 
14 & 130$\pm$32 & 1.31 & <123 & -- & S/E  \\ 
15 & 117$\pm$32 & 1.09 & <138 & -- & S/E  \\ 
16 & 119$\pm$34 & 2.66 & <138 & -- & Ex  \\ 
17 & 193$\pm$44 & 1.83 & <135 & -- & S/E  \\
18 & 239$\pm$48 & 2.66 & <141 & -- & Ex \\
19 & 195$\pm$36 & 2.53 & <117 & -- & Ex \\
20 & 103$\pm$31 & 1.04 & <201 & -- & S/E  \\ 
21 & 556$\pm$77 & 6.40 & <141 & -- & Ex \\
23 & <81 & -- & 210$\pm$60 & 1.03 & E/B \\
24 & 137$\pm$32 & 1.13 & <153 & -- & S/E \\
26 & 120$\pm$29 & 1.39 & <117 & -- & S/E \\
27 & 104$\pm$29 & 1.31 & <99 & -- & S/E \\
29 & 103$\pm$31 & 4.48 & 214$\pm$58 & 1.79 & E/B  \\ 
30 & 81$\pm$22 & 1.04 & <123 & -- & S/E \\
32 & 301$\pm$64 & 2.31 & <219 & -- & E/B \\
35 & 314$\pm$52 & 1.96 & 259$\pm$63 & 1.14 & S/E  \\  
37 & 332$\pm$47 & 2.53 & 172$\pm$44 & 0.95 & S/E  \\ 
38 & 2957$\pm$302 & 5.44 & 2778$\pm$297 & 4.89 & S/E  \\ 
39 & 68$\pm$19 & 1.04 & <102 & -- & E/B \\
41a & 511$\pm$86 & 3.00 & 362$\pm$94 & 1.41 & Ex  \\ 
41b & 1239$\pm$161 & 6.79 & 891$\pm$160 & 3.46 & Ex  \\ 
42 & 883$\pm$109 & 6.23 &  583$\pm$101 & 3.12 & Ex  \\ 
43 & 246$\pm$45 & 2.09 & <150 & -- & B \\ 
44 & 105$\pm$30 & 1.04 & <117 & -- & E/B \\
45 & 3123$\pm$347 & 20.55 & 1500$\pm$212 & 5.46 & Ex \\ 
46 & <72 & -- & 319$\pm$77 & 1.45 & E/B \\
49 & 329$\pm$58 & 3.48 & <123 & -- & B \\
54 & 1230$\pm$144 & 5.18 & 871$\pm$129 & 2.34 & S/E  \\ 
55 & 2026$\pm$245 & 7.66 & 1539$\pm$225 & 3.20 & E/B  \\
58 & 99$\pm$27 & 1.04 & <123 & -- & E/B \\
\hline

\end{tabular}

\vspace{0.1cm}
\footnotesize{Morphologies: (S/E) spherical/elongated; (B) bow shaped; (E) elongated; (Ex) extended.}
\end{center}
\end{footnotesize}
\end{table*}%

\subsection{Transient objects}
\label{transients}

In regards to the radio variability of objects within CLASC, there exist two objects which stand out from the rest with flux densities that vary between epochs by $\gtrsim 5$. These are object IDs \#60 and \#61. Additionally, these objects are both heavily resolved in the April 26 images and show large amounts of extended emission at 21\,cm, similar to the case of CLASC \#22 (i.e. SBHW~90).

Short-period PACWB systems could potentially give rise to radio variability of these timescales, however they are generally seen to be compact objects, as evidenced by the four known PACWB systems detected in these e-MERLIN observations. Moreover, if these objects were indeed PACWB systems, such rapid variability (as observed) would suggest either a highly elliptical orbit or that the orbital separation is within the radio photosphere (and is therefore subject to a vast amount of free-free absorption). Young stellar objects are also known to be transient radio sources with  \citet{forbrich_etal_2008} for example showing variations of $\geq$ 40\,mJy in flaring YSOs over timescales of approximately 4\,hr. The similarity in the extended morphology and rapid 21\,cm flux density variations of these transient objects is suggestive that they are all of a similar nature. Follow-up observations are vital in order to decipher the puzzling physical attributes observed in these systems. 


The majority of the sources in the April 26 observations are detected at higher flux densities than in the April 11 data. 
In some cases (24 sources) the April 26 detections have a flux density $\leq$5$\sigma$ in the April 11 data and would likely not be detected with SEAC. There are also six sources that have April 11 flux densities roughly equal to or greater than the April 26 values. Unfortunately, there exists no detected source whose radio flux density can be reliably deemed to be constant between the two observational epochs. As a result, it is difficult to test the reliability concerning the observed variation of a number of sources within CLASC. In order to check the absolute amplitude calibration we applied it in two ways. We followed the process for each epoch separately, that is, using the amplitude calibrator to derive the flux densities for each independently. We also used the calculated calibrator flux densities from the April 26 data to set the values in the April 11 dataset during the amplitude calibration. The latter is the final solution we opted for on the basis of consistency between epochs and the better data quality of the April 26 observations. However, both of these routes provided very close resulting flux densities (within errors). It is therefore unlikely that the general trend of lower flux densities in the April 11 epoch is a result of the amplitude calibration.

One possibility that could account for some observed flux density variations between the two epochs is the difference between the $u,v$ coverage and resulting sensitivity of the data sets. 
For a number of pointings there is a significant difference in the regions in the $u,v$ plane sampled by either epoch. As discussed in section \ref{datareduction} this is due to differing observing times as well as missing antennas as a result of telescope problems and RFI. For compact, unresolved sources, this is not generally considered to be a problem. However, the majority of sources detected in CLASC are resolved and contain extended structure. In these cases, an under-sampled $u,v$ coverage will lead to images that are sensitive to flux on fewer spatial scales resulting in a net lowering of the integrated source flux density. The sensitivity of the images comes into play when considering the pixel-by-pixel flux-extraction method implemented by SEAC. For low-S/N sources, the true extent of the source can often be hidden within the noise of the image, and as a result their flux is underestimated. This however, is only the case for low-S/N sources (i.e. S/N $\le$ 7; \citealt{peck_thesis_2014}). This is also unlikely to be the cause of the variability for the two transient sources discussed above which lie predominantly within pointing C. The $u,v$ coverages for this pointing in both epochs are very similar both in spatial extent and coverage meaning that this is unlikely to account for their variation in flux density. Given the similarities between the observed properties of these transient objects, this may also suggest that differences in coverage is not the cause of variability for the majority of these sources.

\section{Summary and conclusions}
\label{conclusions}

The COBRaS 21\,cm observations were acquired during 2014 over two epochs separated by $\sim 15$ days. The observations consist of seven overlapping pointings covering an area of the sky of $\sim 0.48^{\circ} \times0.48^{\circ}$ centred on the core of the Cyg OB2 association. Following wide-field imaging, primary-beam-corrected, full-field mosaics were produced at each epoch. Source extraction was performed using a purpose-built source-extraction algorithm called SEAC, which uses a floodfill algorithm to detect and measure the positions and integrated flux densities of any group of associated pixels, or `islands', above a given threshold level. In total, 61 individual sources were detected within the 21\,cm COBRaS observations. 30 of which were detected within both epochs, and a total of 27 were matched to previously identified sources from various surveys, catalogues, and observations from within the literature. 

Five of the identified sources include the massive star (and LBV candidate) Cyg OB2 \#12, the observations of which are discussed at length in \citet{morford_etal_2016}. Discussion of the known massive star binary systems Cyg OB2 \#5,  Cyg OB2 \#8A,  Cyg OB2 \#9, and A11 is deferred to a future paper. 
No known single massive stars (with the possible exception of Cyg OB2 \#12) were detected within these COBRaS 21\,cm observations \citep{morford_etal_2016}. Clearly, our observations are sensitive to the non-thermal emission from colliding wind binary systems. Each of the 23 remaining identified objects (excluding the five massive star objects) have been discussed individually and suggestions as to their origin (i.e. either Galactic or extra-Galactic) have been made (where possible) by considering their flux densities between the two epochs, the extent to which they are resolved, their morphology, and their previous identification(s). For the 61 detected objects, we summarise the following findings:

\begin{enumerate}[label=(\roman*)]
\item{Five (CLASC \#3, \#17, \#50, \#51 and \#61) are of an unknown origin based on these COBRaS observations alone.} 
\item{CLASC \#7 however has been previously suggested as a hot spot candidate (HSC) associated with the relativistic jets of the micro-quasar Cyg X-3.}
\item{Six (CLASC \#10, \#12, \#22, \#25, \#40, and \#47) are suggested to be of an extragalactic origin, perhaps associated with an AGN. This hypothesis was based on their extended morphology, a lack of any significant variability between the two COBRaS 21\,cm observation epochs, and any potential X-ray spectrum as determined by previous observations.}
\item{Nine (CLASC \#7, \#11, \#28, \#34, \#36, \#52, \#53, \#57 and \#59) are likely situated within our Galaxy since they have either been matched (to a confident likelihood ratio) to at least one previous source also thought to be Galactic, or they have been found to vary significantly in their 21\,cm flux densities between the two COBRaS observation epochs. Furthermore, of these nine objects:
\begin{enumerate}
\item{Two sources are identified with YSOs. CLASC \#56 is a confirmed class \textsc{i} YSO and CLASC \#60 a class \textsc{ii} from the work of \citet{guarcello_etal_2013}.}
\item{CLASC \#61 shows extreme variation (an increase by a factor $> 6.5$) in its 21\,cm flux over the 15-day period between epochs, which is reminiscent of a possible flaring event in a pre-main sequence type object or a potential small-period massive star binary system imaged at significantly different orbital phases.}
\item{Three objects (CLASC \#11, \#50, and \#52) are consistent with the non-thermal 21\,cm emission observed from massive star binaries or YSOs.}
\item{Only one (CLASC \#57) was matched to a source from the 2MASS survey, however its position on the J-H vs. H-K diagram does not lie within the region covered by the OB candidate stars.}
\end{enumerate}}
\item{Two objects,  IDs \#57 and \#61,  share the same properties and have been labelled as transient objects since they appear as bright ($> 1$ mJy), resolved objects in the April 26 epoch yet are undetected 15 days earlier in the observations taken on April 11.}
\item{Thirty-three sources have been detected for the first time. A qualitative assessment of their morphologies has been made and is presented in Table \ref{tab:unknownsources}. However, without further information, a limited number of conclusions can be drawn for these objects.}
\end{enumerate}

The COBRaS C-band observations are anticipated to commence early 2020 and will provide 6\,cm flux densities down to 1$\sigma$ sensitivities of $\sim$ 3$\mu$Jy. These data will provide spectral index information for the 41 sources detected within the L-band observations, which will lead to a more quantitative assessment of the nature of each object. 

For those sources detected with an X-ray counterpart from the Chandra observations of \citet{wright_etal_2014a}, a comparison can be made between their X-ray and radio luminosities. Some of these objects have been branded as potential YSOs, extragalactic objects, or potential massive star binary systems. A comparison between their 21\,cm flux densities and X-ray luminosities may enable differentiation between these possibilities by comparing the two values to that predicted from the Gudel-Benz relation \citep{benz_gudel_1994} for active stars. It is clear that both the unknown and previously identified sources will benefit from follow-up observations in order to better monitor their radio variability, which will provide further evidence regarding their nature. 

\begin{acknowledgements}
        e-MERLIN is a national facility operated by The University of Manchester on behalf of the Science and Technology Facilities Council (STFC). {\textsc{parseltongue}} was developed in the context of the ALBUS project, which has benefited from research funding from the European Community's sixth Framework Programme under RadioNet R113CT 2003 5058187. J. Morford and D. Fenech wish to acknowledge funding from an STFC studentship and STFC consolidated grant (ST/M001334/1) respectively. We would like to thank the referee for their patience and a number of suggestions that have improved this paper. 
\end{acknowledgements}

%
\bibliographystyle{aa} 
\bibliography{aa_morford_ref} 

\begin{thebibliography}{97}
\expandafter\ifx\csname natexlab\endcsname\relax\def\natexlab#1{#1}\fi

\bibitem[{Argo(2014)}]{argo_2014}
Argo, M. 2014, Astrophys. Source Code Libr., 1, 07017

\bibitem[{{Beerer} {et~al.}(2010){Beerer}, {Koenig}, {Hora}, {Gutermuth},
  {Bontemps}, {Megeath}, {Schneider}, {Motte}, {Carey}, {Simon}, {Keto},
  {Smith}, {Allen}, {Fazio}, {Kraemer}, {Price}, {Mizuno}, {Adams},
  {Hern{\'a}ndez}, \& {Lucas}}]{beerer_etal_2010}
{Beerer}, I.~M., {Koenig}, X.~P., {Hora}, J.~L., {et~al.} 2010, ApJ, 720, 679

\bibitem[{Benz \& G{\"u}del(1994)}]{benz_gudel_1994}
Benz, A. \& G{\"u}del, M. 1994, A\&A, 285

\bibitem[{{Berlanas} {et~al.}(2018){Berlanas}, {Herrero}, {Comer{\'o}n},
  {Pasquali}, {Bertelli Motta}, \& {Sota}}]{berlanas_etal_2018}
{Berlanas}, S.~R., {Herrero}, A., {Comer{\'o}n}, F., {et~al.} 2018, \aap, 612,
  A50

\bibitem[{Blomme {et~al.}(2010)Blomme, De~Becker, Volpi, \&
  Rauw}]{blomme_etal_2010}
Blomme, R., De~Becker, M., Volpi, D., \& Rauw, G. 2010, A\&A, 519, A111

\bibitem[{Blomme {et~al.}(2013)Blomme, Naz{\'e}, Volpi, De~Becker, Prinja,
  Pittard, Parkin, \& Absil}]{blomme_etal_2013}
Blomme, R., Naz{\'e}, Y., Volpi, D., {et~al.} 2013, A\&A, 550, A90

\bibitem[{Caballero-Nieves {et~al.}(2014)Caballero-Nieves, Nelan, Gies,
  Wallace, DeGioia-Eastwood, Herrero, Jao, Mason, Massey, Moffat,
  {et~al.}}]{caballeronieves_etal_2014}
Caballero-Nieves, S.~M., Nelan, E.~P., Gies, D.~R., {et~al.} 2014, AJ, 147, 40

\bibitem[{{Cash} {et~al.}(1980){Cash}, {Charles}, {Bowyer}, {Walter},
  {Garmire}, \& {Riegler}}]{cash_etal_1980}
{Cash}, W., {Charles}, P., {Bowyer}, S., {et~al.} 1980, ApJL, 238, L71

\bibitem[{Chi {et~al.}(2013)Chi, Barthel, \& Garrett}]{chi_etal_2013}
Chi, S., Barthel, P., \& Garrett, M. 2013, A\&A, 550, A68

\bibitem[{Colombo {et~al.}(2007{\natexlab{a}})Colombo, Caramazza, Flaccomio,
  Micela, \& Sciortino}]{colombo_etal_2007b}
Colombo, J.~A., Caramazza, M., Flaccomio, E., Micela, G., \& Sciortino, S.
  2007{\natexlab{a}}, A\&A, 474, 495

\bibitem[{Colombo {et~al.}(2007{\natexlab{b}})Colombo, Flaccomio, Micela,
  Sciortino, \& Damiani}]{colombo_etal_2007a}
Colombo, J.~A., Flaccomio, E., Micela, G., Sciortino, S., \& Damiani, F.
  2007{\natexlab{b}}, A\&A, 464, 211

\bibitem[{{Comer{\'o}n} \& {Pasquali}(2012)}]{comeron_pasquali_2012}
{Comer{\'o}n}, F. \& {Pasquali}, A. 2012, A\&A, 543, A101

\bibitem[{Comer{\'o}n {et~al.}(2002)Comer{\'o}n, Pasquali, Rodighiero,
  Stanishev, De~Filippis, Mart{\'\i}, Ortiz, Stankov, \&
  Gredel}]{comeron_etal_2002}
Comer{\'o}n, F., Pasquali, A., Rodighiero, G., {et~al.} 2002, A\&A, 389, 874

\bibitem[{{Comer{\'o}n} \& {Torra}(2001)}]{comeron_toora_2001}
{Comer{\'o}n}, F. \& {Torra}, J. 2001, A\&A, 375, 539

\bibitem[{Condon \& Mitchell(1984)}]{condon_mitchell_1984}
Condon, J. \& Mitchell, K. 1984, AJ, 89, 610

\bibitem[{{Condon} {et~al.}(1998){Condon}, {Cotton}, {Greisen}, {Yin},
  {Perley}, {Taylor}, \& {Broderick}}]{condon_etal_1998}
{Condon}, J.~J., {Cotton}, W.~D., {Greisen}, E.~W., {et~al.} 1998, AJ, 115,
  1693

\bibitem[{{Contreras} {et~al.}(1997){Contreras}, {Rodr{\'{\i}}guez}, {Tapia},
  {Cardini}, {Emanuele}, {Badiali}, \& {Persi}}]{contreras_etal_1997}
{Contreras}, M.~E., {Rodr{\'{\i}}guez}, L.~F., {Tapia}, M., {et~al.} 1997,
  ApJL, 488, L153

\bibitem[{{Cutri} {et~al.}(2003){Cutri}, {Skrutskie}, {van Dyk}, {Beichman},
  {Carpenter}, {Chester}, {Cambresy}, {Evans}, {Fowler}, {Gizis}, {Howard},
  {Huchra}, {Jarrett}, {Kopan}, {Kirkpatrick}, {Light}, {Marsh}, {McCallon},
  {Schneider}, {Stiening}, {Sykes}, {Weinberg}, {Wheaton}, {Wheelock}, \&
  {Zacarias}}]{cutri_etal_2003}
{Cutri}, R.~M., {Skrutskie}, M.~F., {van Dyk}, S., {et~al.} 2003, VizieR Online
  Data Catalog, 2246

\bibitem[{De~Becker {et~al.}(2006)De~Becker, Rauw, Sana, Pollock, Pittard,
  Blomme, Stevens, \& Van~Loo}]{debecker_etal_2006}
De~Becker, M., Rauw, G., Sana, H., {et~al.} 2006, MNRAS, 371, 1280

\bibitem[{De~Ruiter {et~al.}(1977)De~Ruiter, Willis, \&
  Arp}]{deruiter_willis_arp_1977}
De~Ruiter, H., Willis, A., \& Arp, H. 1977, A\&ASS, 28, 211

\bibitem[{Dixon(1970)}]{dixon_1970}
Dixon, R.~S. 1970, ApJSS, 20, 1

\bibitem[{Dougherty \& Pittard(2006)}]{dougherty_pittard_2006}
Dougherty, S.~M. \& Pittard, J.~M. 2006, arXiv preprint astro-ph/0611088

\bibitem[{{Dougherty} \& {Williams}(2000)}]{dougherty_williams_2000}
{Dougherty}, S.~M. \& {Williams}, P.~M. 2000, MNRAS, 319, 1005

\bibitem[{{Dougherty} {et~al.}(1996){Dougherty}, {Williams}, {van der Hucht},
  {Bode}, \& {Davis}}]{dougherty_1996}
{Dougherty}, S.~M., {Williams}, P.~M., {van der Hucht}, K.~A., {Bode}, M.~F.,
  \& {Davis}, R.~J. 1996, \mnras, 280, 963

\bibitem[{{Drew} {et~al.}(2008){Drew}, {Greimel}, {Irwin}, \&
  {Sale}}]{drew_etal_2008}
{Drew}, J.~E., {Greimel}, R., {Irwin}, M.~J., \& {Sale}, S.~E. 2008, MNRAS,
  386, 1761

\bibitem[{Forbrich {et~al.}(2008)Forbrich, Menten, \&
  Reid}]{forbrich_etal_2008}
Forbrich, J., Menten, K.~M., \& Reid, M.~J. 2008, A\&A, 477, 267

\bibitem[{Forbrich {et~al.}(2016)Forbrich, Rivilla, Menten, Reid, Chandler,
  Rau, Bhatnagar, Wolk, \& Meingast}]{forbrich_etal_2016}
Forbrich, J., Rivilla, V.~M., Menten, K.~M., {et~al.} 2016, ApJ, 822, 93

\bibitem[{Fullerton {et~al.}(2006)Fullerton, Massa, \&
  Prinja}]{fullerton_etal_2006}
Fullerton, A.~W., Massa, D.~L., \& Prinja, R.~K. 2006, ApJ, 637, 1025

\bibitem[{Guarcello {et~al.}(2013)Guarcello, Drake, Wright, Drew, Gutermuth,
  Hora, Naylor, Aldcroft, Fruscione, Garc{\'\i}a-Alvarez,
  {et~al.}}]{guarcello_etal_2013}
Guarcello, M., Drake, J., Wright, N., {et~al.} 2013, AJ, 773, 135

\bibitem[{{Guarcello} {et~al.}(2015){Guarcello}, {Drake}, {Wright}, {Naylor},
  {Flaccomio}, {Kashyap}, \& {Garcia-Alvarez}}]{guarcello_etal_2015}
{Guarcello}, M.~G., {Drake}, J.~J., {Wright}, N.~J., {et~al.} 2015, ArXiv
  e-prints [\eprint[arXiv]{1501.03761}]

\bibitem[{{Guarcello} {et~al.}(2012){Guarcello}, {Wright}, {Drake},
  {Garc{\'{\i}}a-Alvarez}, {Drew}, {Aldcroft}, \&
  {Kashyap}}]{guarcello_etal_2012}
{Guarcello}, M.~G., {Wright}, N.~J., {Drake}, J.~J., {et~al.} 2012, ApJS, 202,
  19

\bibitem[{Hales {et~al.}(2012)Hales, Murphy, Curran, Middelberg, Gaensler, \&
  Norris}]{hales_etal_2012}
Hales, C.~A., Murphy, T., Curran, J.~R., {et~al.} 2012, MNRAS, 425, 979

\bibitem[{Hancock {et~al.}(2012)Hancock, Murphy, Gaensler, Hopkins, \&
  Curran}]{hancock_etal_2012}
Hancock, P., Murphy, T., Gaensler, B., Hopkins, A., \& Curran, J. 2012,
  Astrophysics Source Code Library, 1, 12009

\bibitem[{{Hanson}(2003)}]{hanson_2003}
{Hanson}, M.~M. 2003, ApJ, 597, 957

\bibitem[{Hopkins {et~al.}(1998)Hopkins, Mobasher, Cram, \&
  Rowan-Robinson}]{hopkins_etal_1998}
Hopkins, A.~M., Mobasher, B., Cram, L., \& Rowan-Robinson, M. 1998, MNRAS, 296,
  839

\bibitem[{{Huynh} {et~al.}(2005){Huynh}, {Jackson}, {Norris}, \&
  {Prandoni}}]{huynh_etal_2005}
{Huynh}, M.~T., {Jackson}, C.~A., {Norris}, R.~P., \& {Prandoni}, I. 2005, \aj,
  130, 1373

\bibitem[{Intema {et~al.}(2017)Intema, Jagannathan, Mooley, \&
  Frail}]{intema_etal_2017}
Intema, H., Jagannathan, P., Mooley, K., \& Frail, D. 2017, A\&A, 598, A78

\bibitem[{Kennedy {et~al.}(2010)Kennedy, Dougherty, Fink, \&
  Williams}]{kennedy_etal_2010}
Kennedy, M., Dougherty, S., Fink, A., \& Williams, P. 2010, ApJ, 709, 632

\bibitem[{Kiminki \& Kobulnicky(2012)}]{kiminki_kobulnicky_2012}
Kiminki, D.~C. \& Kobulnicky, H.~A. 2012, ApJ, 751, 4

\bibitem[{Kiminki {et~al.}(2015)Kiminki, Kobulnicky, Alexander, Lundquist,
  {et~al.}}]{kiminki_etal_2015}
Kiminki, D.~C., Kobulnicky, H.~A., Alexander, M.~J., Lundquist, M.~J., {et~al.}
  2015, ApJ, 811, 85

\bibitem[{Kiminki {et~al.}(2012)Kiminki, Kobulnicky, Ewing, Kiminki, Lundquist,
  Alexander, Vargas-Alvarez, Choi, \& Henderson}]{kiminki_etal_2012a}
Kiminki, D.~C., Kobulnicky, H.~A., Ewing, I., {et~al.} 2012, ApJ, 747, 41

\bibitem[{Kiminki {et~al.}(2009)Kiminki, Kobulnicky, Gilbert, Bird, \&
  Chunev}]{kiminki_etal_2009}
Kiminki, D.~C., Kobulnicky, H.~A., Gilbert, I., Bird, S., \& Chunev, G. 2009,
  AJ, 137, 4608

\bibitem[{Kiminki {et~al.}(2007)Kiminki, Kobulnicky, Kinemuchi, Irwin, Fryer,
  Berrington, Uzpen, Monson, Pierce, \& Woosley}]{kiminki_etal_2007}
Kiminki, D.~C., Kobulnicky, H.~A., Kinemuchi, K., {et~al.} 2007, ApJ, 664, 1102

\bibitem[{Kiminki {et~al.}(2008)Kiminki, McSwain, \&
  Kobulnicky}]{kiminki_etal_2008}
Kiminki, D.~C., McSwain, M.~V., \& Kobulnicky, H.~A. 2008, ApJ, 679, 1478

\bibitem[{Kn{\"o}dlseder(2000)}]{knodlseder_2000}
Kn{\"o}dlseder, J. 2000, A\&A, 360, 539

\bibitem[{Kobulnicky {et~al.}(2010)Kobulnicky, Gilbert, \&
  Kiminki}]{kobulnicky_etal_2010}
Kobulnicky, H.~A., Gilbert, I.~J., \& Kiminki, D.~C. 2010, ApJ, 710, 549

\bibitem[{{Kobulnicky} {et~al.}(2014){Kobulnicky}, {Kiminki}, {Lundquist},
  {Burke}, {Chapman}, {Keller}, {Lester}, {Rolen}, {Topel}, {Bhattacharjee},
  {Smullen}, {Vargas {\'A}lvarez}, {Runnoe}, {Dale}, \&
  {Brotherton}}]{kobulnicky_etal_2014}
{Kobulnicky}, H.~A., {Kiminki}, D.~C., {Lundquist}, M.~J., {et~al.} 2014, ApJS,
  213, 34

\bibitem[{Kobulnicky {et~al.}(2012)Kobulnicky, Smullen, Kiminki, Runnoe, Wood,
  Long, Alexander, Lundquist, \& Vargas-Alvarez}]{kobulnicky_etal_2012}
Kobulnicky, H.~A., Smullen, R.~A., Kiminki, D.~C., {et~al.} 2012, ApJ, 756, 50

\bibitem[{K{\"u}hr {et~al.}(1979)K{\"u}hr, Nauber, Pauliny-Toth, \&
  Witzel}]{kuehr_etal_1979}
K{\"u}hr, H., Nauber, U., Pauliny-Toth, I., \& Witzel, A. 1979, Preprint

\bibitem[{Linder {et~al.}(2009)Linder, Rauw, Manfroid, Damerdji, De~Becker,
  Eenens, Royer, \& Vreux}]{linder_etal_2009}
Linder, N., Rauw, G., Manfroid, J., {et~al.} 2009, A\&A, 495, 231

\bibitem[{Ling {et~al.}(2009)Ling, Zhang, \& Tang}]{ling_nan-zhang_tang_2009}
Ling, Z., Zhang, S.~N., \& Tang, S. 2009, ApJ, 695, 1111

\bibitem[{{Mahy} {et~al.}(2013){Mahy}, {Rauw}, {De Becker}, {Eenens}, \&
  {Flores}}]{mahy_etal_2013}
{Mahy}, L., {Rauw}, G., {De Becker}, M., {Eenens}, P., \& {Flores}, C.~A. 2013,
  A\&A, 550, A27

\bibitem[{Ma{\'\i}z-Apell{\'a}niz {et~al.}(2004)Ma{\'\i}z-Apell{\'a}niz,
  Walborn, Galu{\'e}, \& Wei}]{maiz-appelaniz_etal_2004}
Ma{\'\i}z-Apell{\'a}niz, J., Walborn, N.~R., Galu{\'e}, H.~{\'A}., \& Wei,
  L.~H. 2004, ApJSS, 151, 103

\bibitem[{Mart{\'\i} {et~al.}(2001)Mart{\'\i}, Paredes, \&
  Peracaula}]{marti_etal_2001}
Mart{\'\i}, J., Paredes, J., \& Peracaula, M. 2001, A\&A, 375, 476

\bibitem[{{Mart{\'{\i}}} {et~al.}(2007){Mart{\'{\i}}}, {Paredes}, {Ishwara
  Chandra}, \& {Bosch-Ramon}}]{marti_etal_2007}
{Mart{\'{\i}}}, J., {Paredes}, J.~M., {Ishwara Chandra}, C.~H., \&
  {Bosch-Ramon}, V. 2007, A\&A, 472, 557

\bibitem[{Mart{\'\i} {et~al.}(2005)Mart{\'\i}, P{\'e}rez-Ram{\'\i}rez, Garrido,
  Luque-Escamilla, \& Paredes}]{marti_etal_2005}
Mart{\'\i}, J., P{\'e}rez-Ram{\'\i}rez, D., Garrido, J., Luque-Escamilla, P.,
  \& Paredes, J. 2005, A\&A, 439, 279

\bibitem[{{Maryeva} {et~al.}(2016){Maryeva}, {Chentsov}, {Goranskij},
  {Dyachenko}, {Karpov}, {Malogolovets}, \& {Rastegaev}}]{maryeva_etal_2016}
{Maryeva}, O.~V., {Chentsov}, E.~L., {Goranskij}, V.~P., {et~al.} 2016, MNRAS,
  458, 491

\bibitem[{Massey \& Thompson(1991)}]{massey_thompson_1991}
Massey, P. \& Thompson, A. 1991, AJ, 101, 1408

\bibitem[{{McMullin} {et~al.}(2007){McMullin}, {Waters}, {Schiebel}, {Young},
  \& {Golap}}]{mcmullin_2007}
{McMullin}, J.~P., {Waters}, B., {Schiebel}, D., {Young}, W., \& {Golap}, K.
  2007, in Astronomical Society of the Pacific Conference Series, Vol. 376,
  Astronomical Data Analysis Software and Systems XVI, ed. R.~A. {Shaw},
  F.~{Hill}, \& D.~J. {Bell}, 127

\bibitem[{{Menten} {et~al.}(2007){Menten}, {Reid}, {Forbrich}, \&
  {Brunthaler}}]{menten_2007}
{Menten}, K.~M., {Reid}, M.~J., {Forbrich}, J., \& {Brunthaler}, A. 2007,
  Astronomy and Astrophysics, 474, 515

\bibitem[{Miller {et~al.}(2013)Miller, Bonzini, Fomalont, Kellermann, Mainieri,
  Padovani, Rosati, Tozzi, \& Vattakunnel}]{miller_etal_2013}
Miller, N.~A., Bonzini, M., Fomalont, E.~B., {et~al.} 2013, ApJSS, 205, 13

\bibitem[{Morford {et~al.}(2016)Morford, Fenech, Prinja, Blomme, \&
  Yates}]{morford_etal_2016}
Morford, J., Fenech, D., Prinja, R., Blomme, R., \& Yates, J. 2016, MNRAS, 463,
  763

\bibitem[{{Moss} {et~al.}(2007){Moss}, {Seymour}, {McHardy}, {Dwelly}, {Page},
  \& {Loaring}}]{moss_etal_2007}
{Moss}, D., {Seymour}, N., {McHardy}, I.~M., {et~al.} 2007, \mnras, 378, 995

\bibitem[{{M{\"u}nch} \& {Morgan}(1953)}]{munch_morgan_1953}
{M{\"u}nch}, L. \& {Morgan}, W.~W. 1953, ApJ, 118, 161

\bibitem[{Muxlow {et~al.}(2005)Muxlow, Richards, Garrington, Wilkinson,
  Anderson, Richards, Axon, Fomalont, Kellermann, Partridge,
  {et~al.}}]{muxlow_etal_2005}
Muxlow, T., Richards, A., Garrington, S., {et~al.} 2005, MNRAS, 358, 1159

\bibitem[{Offringa {et~al.}(2014)Offringa, McKinley, Hurley-Walker,
  {et~al.}}]{offringa_2014}
Offringa, A.~R., McKinley, B., Hurley-Walker, {et~al.} 2014, MNRAS, 444, 606

\bibitem[{Panagia \& Felli(1975)}]{panagia_feli_1975}
Panagia, N. \& Felli, M. 1975, A\&A, 39, 1

\bibitem[{Peck(2014)}]{peck_thesis_2014}
Peck, L.~W. 2014, PhD thesis, University College London

\bibitem[{Peck \& Fenech(2013)}]{peck_fenech_2013}
Peck, L.~W. \& Fenech, D.~M. 2013, Astron. \& Comput., 2, 54

\bibitem[{Perley \& Butler(2013)}]{perley_butler_2013}
Perley, R.~A. \& Butler, B.~J. 2013, ApJSS, 204, 19

\bibitem[{{Pigulski} \& {Ko{\l}aczkowski}(1998)}]{pigulski_kolaczkowski_1998}
{Pigulski}, A. \& {Ko{\l}aczkowski}, Z. 1998, MNRAS, 298, 753

\bibitem[{Poglitsch {et~al.}(2010)Poglitsch, Waelkens, Geis, Feuchtgruber,
  Vandenbussche, Rodriguez, Krause, Renotte, Van~Hoof, Saraceno,
  {et~al.}}]{poglitsch_etal_2010}
Poglitsch, A., Waelkens, C., Geis, N., {et~al.} 2010, A\&A, 518, L2

\bibitem[{Puls {et~al.}(2006)Puls, Markova, Scuderi, Stanghellini, Taranova,
  Burnley, \& Howarth}]{puls_etal_2006}
Puls, J., Markova, N., Scuderi, S., {et~al.} 2006, A\&A, 454, 625

\bibitem[{{Rauw}(2011)}]{rauw_2011}
{Rauw}, G. 2011, A\&A, 536, A31

\bibitem[{Rauw {et~al.}(1999)Rauw, Vreux, \& Bohannan}]{rauw_etal_1999}
Rauw, G., Vreux, J.-M., \& Bohannan, B. 1999, ApJ, 517, 416

\bibitem[{Rygl {et~al.}(2012)Rygl, Brunthaler, Sanna, Menten, Reid,
  Van~Langevelde, Honma, Torstensson, \& Fujisawa}]{rygl_etal_2012}
Rygl, K., Brunthaler, A., Sanna, A., {et~al.} 2012, A\&A, 539, A79

\bibitem[{{Schulte}(1958)}]{schulte_1958}
{Schulte}, D.~H. 1958, ApJ, 128, 41

\bibitem[{Setia~Gunawan {et~al.}(2003)Setia~Gunawan, de~Bruyn, van~der Hucht,
  \& Williams}]{setiagunawan_etal_2003}
Setia~Gunawan, D.~Y., de~Bruyn, A.~G., van~der Hucht, K.~A., \& Williams, P.~M.
  2003, ApJS, 149, 123

\bibitem[{Skiff(2009)}]{skiff_2010}
Skiff, B. 2009, VizieR Online Data Catalog, 1, 02023

\bibitem[{Sundqvist {et~al.}(2010)Sundqvist, Puls, \&
  Feldmeier}]{sundqvist_etal_2010}
Sundqvist, J., Puls, J., \& Feldmeier, A. 2010, A\&A, 510, A11

\bibitem[{Taylor {et~al.}(2003)Taylor, Gibson, Peracaula, Martin, Landecker,
  Brunt, Dewdney, Dougherty, Gray, Higgs, {et~al.}}]{taylor_etal_2003}
Taylor, A., Gibson, S., Peracaula, M., {et~al.} 2003, AJ, 125, 3145

\bibitem[{Taylor {et~al.}(1996)Taylor, Goss, Coleman, Van~Leeuwen, \&
  Wallace}]{taylor_etal_1996}
Taylor, A., Goss, W., Coleman, P., Van~Leeuwen, J., \& Wallace, B. 1996, ApJSS,
  107, 239

\bibitem[{{Uyan{\i}ker} {et~al.}(2001){Uyan{\i}ker}, {F{\"u}rst}, {Reich},
  {Aschenbach}, \& {Wielebinski}}]{uyaniker_etal_2001}
{Uyan{\i}ker}, B., {F{\"u}rst}, E., {Reich}, W., {Aschenbach}, B., \&
  {Wielebinski}, R. 2001, A\&A, 371, 675

\bibitem[{{Vink} {et~al.}(2008){Vink}, {Drew}, {Steeghs}, {Wright}, {Martin},
  {G{\"a}nsicke}, {Greimel}, \& {Drake}}]{vink_etal_2008}
{Vink}, J.~S., {Drew}, J.~E., {Steeghs}, D., {et~al.} 2008, MNRAS, 387, 308

\bibitem[{Wendker(1984)}]{wendker_1984}
Wendker, H. 1984, A\&ASS, 58, 291

\bibitem[{Wendker {et~al.}(1991)Wendker, Higgs, \&
  Landecker}]{wendker_etal_1991}
Wendker, H., Higgs, L., \& Landecker, T. 1991, A\&A, 241, 551

\bibitem[{Wolff {et~al.}(2007)Wolff, Strom, Dror, \& Venn}]{wolff_etal_2007}
Wolff, S., Strom, S., Dror, D., \& Venn, K. 2007, AJ, 133, 1092

\bibitem[{Wright \& Barlow(1975)}]{wright_barlow_1975}
Wright, A.~E. \& Barlow, M.~J. 1975, MNRAS, 170, 41

\bibitem[{Wright {et~al.}(2010{\natexlab{a}})Wright, Eisenhardt, Mainzer,
  Ressler, Cutri, Jarrett, Kirkpatrick, Padgett, McMillan, Skrutskie,
  {et~al.}}]{wright_etal_2010}
Wright, E.~L., Eisenhardt, P.~R., Mainzer, A.~K., {et~al.} 2010{\natexlab{a}},
  AJ, 140, 1868

\bibitem[{Wright {et~al.}(2016)Wright, Bouy, Drew, Sarro, Bertin, Cuillandre,
  \& Barrado}]{wright_etal_2016}
Wright, N.~J., Bouy, H., Drew, J.~E., {et~al.} 2016, MNRAS, stw1148

\bibitem[{{Wright} \& {Drake}(2009)}]{wright_drake_2009}
{Wright}, N.~J. \& {Drake}, J.~J. 2009, ApJS, 184, 84

\bibitem[{Wright {et~al.}(2010{\natexlab{b}})Wright, Drake, Drew, \&
  Vink}]{wright_etal_2010a}
Wright, N.~J., Drake, J.~J., Drew, J.~E., \& Vink, J.~S. 2010{\natexlab{b}},
  ApJ, 713, 871

\bibitem[{{Wright} {et~al.}(2014){Wright}, {Drake}, {Guarcello}, {Aldcroft},
  {Kashyap}, {Damiani}, {DePasquale}, \& {Fruscione}}]{wright_etal_2014}
{Wright}, N.~J., {Drake}, J.~J., {Guarcello}, M.~G., {et~al.} 2014, ArXiv
  e-prints [\eprint[arXiv]{1408.6579}]

\bibitem[{Wright {et~al.}(2014a)Wright, Drake, Guarcello, Aldcroft, Kashyap,
  Damiani, DePasquale, \& Fruscione}]{wright_etal_2014a}
Wright, N.~J., Drake, J.~J., Guarcello, M.~G., {et~al.} 2014a, arXiv preprint
  arXiv:1408.6579

\bibitem[{Wright {et~al.}(2015)Wright, Drew, \& Mohr-Smith}]{wright_etal_2015}
Wright, N.~J., Drew, J.~E., \& Mohr-Smith, M. 2015, MNRAS, 449, 741

\bibitem[{Wright {et~al.}(2014b)Wright, Parker, Goodwin, \&
  Drake}]{wright_etal_2014b}
Wright, N.~J., Parker, R.~J., Goodwin, S.~P., \& Drake, J.~J. 2014b, MNRAS,
  438, 639

\bibitem[{Zoonematkermani {et~al.}(1990)Zoonematkermani, Helfand, Becker,
  White, \& Perley}]{zoonematkermani_etal_1990}
Zoonematkermani, S., Helfand, D., Becker, R., White, R., \& Perley, R. 1990,
  ApJSS, 74, 181

\end{thebibliography}
%

\onecolumn

\appendix
\counterwithin{table}{section}
\clearpage
\section{Tables}


\footnotesize{
\begin{longtable}{ccccccccccc}
\caption{The COBRaS L-band All Source Catalogue.}\\
\label{tab:clasc_main}\\

\hline \hline
CLASC & Other & RA & DEC & F$_{11^{th}}$ & $\sigma_{11^{th}}$ & S/N & F$_{26^{th}}$ & $\sigma_{26^{th}}$ & S/N  \\ 
ID & ID & (J2000) & (J2000) & ($\mu$Jy) & ($\mu$Jy/beam) & ${11^{th}}$ &  ($\mu$Jy) & ($\mu$Jy/beam) & ${26^{th}}$ \\
(1) & (2) & (3) & (4) & (5) & (6) & (7) & (8) & (9) & (10) \\
\hline
\noalign{\vskip 0.05in}
\endfirsthead
\caption{continued.}\\
\hline\hline
CLASC & Other & RA & DEC & F$_{11^{th}}$ & $\sigma_{11^{th}}$ & S/N & F$_{26^{th}}$ & $\sigma_{26^{th}}$ & S/N  \\ 
ID & ID & (J2000) & (J2000) & ($\mu$Jy) & ($\mu$Jy/beam) & ${11^{th}}$ &  ($\mu$Jy) & ($\mu$Jy/beam) & ${26^{th}}$ \\
(1) & (2) & (3) & (4) & (5) & (6) & (7) & (8) & (9) & (10) \\
\hline
\noalign{\vskip 0.05in}
\endhead
\hline
\endfoot
\endlastfoot
1 & -- & 20 32 03.05 & 41 10 30.10 & -- & 64 & -- & 147$\pm$35 & 26 & 5.3 \\
2 & -- & 20 32 06.13 & 41 14 07.54 & 858$\pm$148 & 74 & 7.5 & 1747$\pm$187 & 28 & 33.4 \\
3 & -- & 20 32 08.29 & 41 13 11.49 & -- & 62 & -- & 255$\pm$43 & 24 & 8.1 \\
4 & -- & 20 32 14.23 & 41 16 38.26 & -- & 59 & -- & 335$\pm$57 & 28 & 6.3 \\
5 & -- & 20 32 15.47 & 41 13 31.34 & -- & 68 & -- & 187$\pm$38 & 25 & 7.5 \\
6a & Cyg 5 (SW) & 20 32 22.42 & 41 18 18.89 & -- & 65 & -- & 862$\pm$111 & 27 & 12.3 \\
6b & Cyg 5 (NE) & 20 32 22.48 & 41 18 19.38 & 448$\pm$96 & 61 & 5.3 & 1438$\pm$167 & 27 & 13.2 \\
7a & HSC N & 20 32 26.87 & 41 04 32.97 & 3643$\pm$398 & 73 & 33.1 & 3807$\pm$393 & 38 & 55.8 \\
7b & -- & 20 32 26.82 & 41 04 32.79 & 337$\pm$87 & 72 & 5.2 & -- & 33 & -- \\
8 & -- & 20 32 29.53 & 41 04 23.81 & 237$\pm$64 & 59 & 5.5 & -- & 26 & -- \\
9 & A11 & 20 32 31.53 & 41 14 08.14 & -- & 53 & -- & 281$\pm$42 & 23 & 13.1 \\
10 & W14 2638 & 20 32 34.21 & 41 24 14.17 & 222$\pm$57 & 51 & 5.3 & 597$\pm$83 & 32 & 12.5 \\
11a & SBHW 81 & 20 32 36.65 & 41 14 47.36 & 1895$\pm$236 & 63 & 12.3 & 4193$\pm$436 & 23 & 48.9 \\
11b & -- & 20 32 36.69 & 41 14 47.86 & -- & 68 & -- & 160$\pm$38 & 24 & 4.5 \\
12a & SBHW 83 & 20 32 38.17 & 41 23 37.18 & 1537$\pm$217 & 61 & 7.1 & 3676$\pm$404 & 36 & 12.4 \\
12b & SBHW 83 & 20 32 38.18 & 41 23 38.66 & -- & 48 & -- & 1118$\pm$156 & 39 & 6.3 \\
12c & SBHW 83 & 20 32 38.20 & 41 23 39.43 & 246$\pm$72 & 64 & 4.6 & 1519$\pm$202 & 39 & 5.3 \\
13 & Cyg 12 & 20 32 40.95 & 41 14 28.99 & 802$\pm$134 & 60 & 7.3 & 2564$\pm$277 & 23 & 23.5 \\
14 & -- & 20 32 41.30 & 41 23 39.54 & -- & 41 & -- & 130$\pm$32 & 25 & 5.8 \\
15 & -- & 20 32 42.09 & 41 24 00.25 & -- & 46 & -- & 117$\pm$32$^\dag$ & 28 & 4.8 \\
16 & -- & 20 32 43.44 & 41 07 37.79 & -- & 46 & -- & 119$\pm$34 & 26 & 4.4 \\
17 & -- & 20 32 43.98 & 41 19 10.78 & -- & 45 & -- & 193$\pm$44 & 29 & 5.3 \\
18 & -- & 20 32 44.90 & 41 23 23.52 & -- & 47 & -- & 239$\pm$48 & 26 & 4.9 \\
19 & -- & 20 32 46.58 & 41 15 29.58 & -- & 39 & -- & 195$\pm$36 & 19 & 6.0 \\
20 & -- & 20 32 48.35 & 41 03 03.99 & -- & 67 & -- & 103$\pm$31$^\ddag$ & 29 & 4.8 \\
21 & -- & 20 32 49.23 & 41 14 49.48 & -- & 47 & -- & 556$\pm$77 & 21 & 7.2 \\
22a & SBHW 90 & 20 32 56.79 & 41 08 53.35 & 13522$\pm$1372 & 58 & 74.3 & 15547$\pm$1569 & 33 & 115.5 \\
22b & SBHW 90 & 20 32 56.80 & 41 08 52.35 & -- & 71 & -- & 305$\pm$65 & 34 & 4.9 \\
23 & -- & 20 32 57.69 & 41 03 55.35 & 210$\pm$60 & 56 & 4.9 & -- & 27 & -- \\
24 & -- & 20 32 58.49 & 41 03 07.50 & -- & 51 & -- & 137$\pm$32 & 28 & 6.5 \\
25 & W14 3683 & 20 32 59.07 & 41 04 58.66 & 566$\pm$96 & 58 & 10.3 & 898$\pm$106 & 27 & 21.6 \\
26 & -- & 20 33 00.17 & 41 06 40.64 & -- & 39 & -- & 120$\pm$29 & 22 & 5.1 \\
27 & -- & 20 33 05.31 & 41 21 20.45 & -- & 33 & -- & 104$\pm$29 & 23 & 4.7 \\
28 & -- & 20 33 07.66 & 41 08 54.58 & 422$\pm$72 & 40 & 9.3 & 478$\pm$60 & 19 & 16.2 \\
29 & -- & 20 33 09.59 & 41 05 05.73 & 430$\pm$77 & 47 & 8.7 & 776$\pm$92 & 24 & 18.2 \\
30 & -- & 20 33 10.56 & 41 14 46.46 & -- & 41 & -- & 81$\pm$22 & 20 & 5.0 \\
31 & Cyg 9 & 20 33 10.73 & 41 15 08.14 & 1037$\pm$132 & 47 & 21.6 & 1320$\pm$142 & 28 & 45.7 \\
32 & -- & 20 33 14.65 & 41 00 12.86 & -- & 73 & -- & 301$\pm$64 & 38 & 5.2 \\
33b & Cyg 8A & 20 33 15.06 & 41 18 49.63 & -- & 38 & -- & 102$\pm$26 & 22 & 5.5 \\
33a & Cyg 8A & 20 33 15.07 & 41 18 50.43 & 2577$\pm$274 & 44 & 40.7 & 1151$\pm$123 & 22 & 49.4 \\
34 & W14 4790 & 20 33 18.28 & 41 17 39.38 & -- & 38 & -- & 96$\pm$21 & 18 & 7.7 \\
35 & -- & 20 33 18.29 & 41 02 11.50 & 259$\pm$63 & 54 & 5.9 & 314$\pm$52 & 30 & 9.4 \\
36 & -- & 20 33 23.58 & 41 27 25.59 & 443$\pm$110 & 96 & 5.9 & 1135$\pm$155 & 51 & 10.2 \\
37 & -- & 20 33 23.59 & 41 09 17.45 & 172$\pm$44 & 42 & 5.9 & 332$\pm$47 & 21 & 13.0 \\
38 & -- & 20 33 26.99 & 41 08 53.26 & 2778$\pm$297 & 47 & 42.1 & 2957$\pm$302 & 26 & 79.2 \\
39 & -- & 20 33 31.98 & 41 15 47.68 & -- & 34 & -- & 68$\pm$19 & 18 & 4.9 \\
40 & W14 5440 & 20 33 32.09 & 41 05 57.65 & 234$\pm$52 & 45 & 7.4 & 354$\pm$51 & 23 & 11.7 \\
41a & -- & 20 33 33.79 & 40 59 42.63 & 362$\pm$94 & 73 & 4.7 & 511$\pm$86 & 40 & 7.6 \\
41b & -- & 20 33 33.86 & 40 59 43.26 & 891$\pm$160 & 71 & 5.0 & 1239$\pm$161 & 39 & 10.1 \\
42 & -- & 20 33 44.69 & 41 05 46.10 & 583$\pm$101 & 47 & 6.6 & 883$\pm$109 & 26 & 12.3 \\
43 & -- & 20 33 47.36 & 41 04 56.78 & -- & 50 & -- & 246$\pm$45 & 26 & 7.2 \\
44 & -- & 20 33 50.19 & 41 03 51.33 & -- & 39 & -- & 105$\pm$30 & 28 & 4.8 \\
45 & -- & 20 33 51.97 & 41 21 51.54 & 1500$\pm$212 & 64 & 7.1 & 3123$\pm$347 & 33 & 9.1 \\
46 & -- & 20 33 52.01 & 41 21 43.26 & 319$\pm$77 & 58 & 5.0 & -- & 24 & -- \\
47a & SBHW 109 & 20 33 52.30 & 41 15 46.94 & -- & 45 & -- & 2304$\pm$268 & 27 & 4.8 \\
47b & SBHW 109 & 20 33 52.43 & 41 15 42.12 & 320$\pm$77$^\ddag$ & 45 & 4.0 & 1382$\pm$168 & 25 & 6.1 \\
48 & -- & 20 33 54.69 & 41 08 21.32 & 142$\pm$43$^\ddag$ & 41 & 4.5 & 230$\pm$41 & 22 & 7.6 \\
49 & -- & 20 33 55.22 & 41 06 39.34 & -- & 41 & -- & 329$\pm$58 & 26 & 4.7 \\
50 & SBHW 110 & 20 33 55.48 & 41 02 53.45 & 3856$\pm$415 & 73 & 39.4 & 3832$\pm$394 & 40 & 70.0 \\
51 & -- & 20 33 57.99 & 41 18 02.53 & -- & 38 & -- & 88$\pm$24 & 21 & 5.1 \\
52 & SBHW 112 & 20 33 58.33 & 41 09 14.57 & 2433$\pm$276 & 50 & 23.5 & 3181$\pm$328 & 27 & 61.6 \\
53a & -- & 20 34 01.12 & 41 18 16.74 & -- & 43 & -- & 204$\pm$43 & 26 & 5.3 \\
53b & -- & 20 34 01.19 & 41 18 16.72 & -- & 43 & -- & 199$\pm$42 & 25 & 5.2 \\
54 & -- & 20 34 01.47 & 41 03 41.96 & 871$\pm$129 & 62 & 11.4 & 1230$\pm$144 & 33 & 16.8 \\
55 & -- & 20 34 02.43 & 41 24 57.33 & 1539$\pm$225 & 92 & 9.7 & 2026$\pm$245 & 50 & 12.7 \\
56 & G13 318006 & 20 34 06.76 & 41 16 00.56 & 139$\pm$44$^{\ddag}$ & 48 & 5.6 & 649$\pm$85 & 26 & 10.9 \\
57 & W14 6616 & 20 34 07.30 & 41 07 24.60 & -- & 51 & -- & 419$\pm$81$^\dag$ & 31 & 4.0 \\
58 & -- & 20 34 14.00 & 41 11 42.86 & -- & 41 & -- & 99$\pm$27 & 25 & 5.0 \\
59 & -- & 20 34 18.13 & 41 15 44.75 & 352$\pm$90$^\dag$ & 59 & 4.1 & 2496$\pm$277 & 32 & 12.2 \\
60 & G13 319565 & 20 34 22.21 & 41 17 00.65 & -- & 55 & -- & 147$\pm$38 & 33 & 5.6 \\
61 & W14 7183 & 20 34 32.71 & 41 11 16.90 & -- & 77 & -- & 1500$\pm$191 & 45 & 8.4 \\
\bottomrule
\caption*{\footnotesize Notes: other identifiers - {\textbf Cyg}: \citet{schulte_1958}; {\textbf HSC N}: \citet{marti_etal_2005}; {\textbf A11}: \citet{comeron_etal_2002}; {\textbf SBHW}: \citet{setiagunawan_etal_2003}; {\textbf W14}: \citet{wright_etal_2014a}; {\textbf G13}: \citet{guarcello_etal_2013}.\\  ${\dagger}$ these sources are faint detections and their flux density has been measured with SEAC using threshold values, $\sigma_s = 4.5 $ and $\sigma_f = 3$.\\
${\ddagger}$ these sources are faint detections and their flux density has been measured with SEAC using threshold values, $\sigma_s = 4.0 $ and $\sigma_f = 2.6$.\\}\\
\end{longtable}
}
\normalsize

\vspace{0.5in}
\noindent Table \ref{tab:clasc_main} is accompanied by the following information.

\begin{itemize}
\item{\textit{Column(1)}: the identification number given to the sources from this work, sequentially increasing in order of right ascension (RA).}
\item{\textit{Column(2)}: any other identifier the source may have from a previous identification.}
\item{\textit{Column((3)-(4))}: the J2000 positional information of each source as measured from the April 26 observations. Each position was calculated from SEAC by taking a weighted mean of the detected pixel area (see Section \ref{sourceposition}). The positions are all accurate to $< 0.032$ arcseconds in both RA and DEC.}
\item{\textit{Columns(5)-(7)}: the integrated flux density, $F_{11^{th}}$ and its associated error in $\mu$Jy, the 1$\sigma$ RMS noise level, $\sigma_{11^{th}}$, within an 3" radius surrounding the edge of the source and the S/N ($F_{p}/\sigma_{11^{th}}$, where $F_p$ is the peak flux density) of each of the sources detected at 21 cm from the April 11 observations.}
\item{\textit{Columns(8)-(10)}: the same as for columns $(5)-(7)$ but in relation to the objects as detected within the 21\,cm observations taken on April 26.}
\end{itemize}


\clearpage
\setlength\tabcolsep{4pt}
\def\arraystretch{1.3}

\footnotesize{
\begin{longtable}{cccccccl}
\caption[Source identification information]{Source identification information.}\\
\label{tab:identifications}\\
\hline \hline
CLASC & Common & Offset & $\lambda$ & $r$ & $LR$ & Catalogue & \multicolumn{1}{c}{Flux Density/Mag/X-ray Flux} \\
ID & ID & (") & & & & Ref &  \\ 
(1) & (2) & (3) & (4) & (5) & (6) & (7) & \multicolumn{1}{c}{(8)} \\
\hline
\noalign{\vskip 0.05in}
\endfirsthead
\caption{continued.}\\
\hline\hline
CLASC & Common & Offset & $\lambda$ & $r$ & $LR$ & Catalogue & \multicolumn{1}{c}{Flux Density/Mag/X-ray Flux} \\
ID & ID & (") & & & & Ref &  \\ 
(1) & (2) & (3) & (4) & (5) & (6) & (7) & \multicolumn{1}{c}{(8)} \\
\hline
\noalign{\vskip 0.05in}
\endhead
\hline
\endfoot
\endlastfoot
3 & --  & 1.193 & 4.72E-05  & 2.5 & 493.6 & M07 &  $F_{49\,cm}$=0.96 mJy \\
6 & Cyg 5 & 0.132 & 7.88E-05  & 0.1 & >10$^3$ & S09 &   $V = 9.3$  \\ 
  &   & 2.452 & 9.75E-05  & 3.9 & 2.6 & R11 &    \\ 
  &   & 2.222 & 3.30E-05  & 1.2 & >10$^3$ & S03 &   $F_{21cm}$ = 3.6 mJy  \\ 
  &   & 0.376 & 1.11E-03  & 0.3 & 429.6 & W14 &   XR$_{f} = 2.46\times10^{-12}$ erg cm$^{-2}$ s$^{-1}$  \\ 
  &   & 0.724 & 9.15E-04  & 0.6 & 455.5 & G13 &   $3.6_{\mu m}$ = 14.4, 4.5$_{\mu m}$ = 13.3, 5.8$_{\mu m}$ = 12.4, 8.0$_{\mu m}$ = 11.8, 24$_{\mu m}$ = 7.1\,mag  \\ 
  &   & 0.198 & 6.49E-05  & 2.8 & 145.8 & 2MASS &   $J = 5.2, H = 4.7, K = 4.3$  \\
  &   & 0.231 & 1.35E-05  & 4.5 & 1.7 & WISE  &   $3.4_{\mu m}$ = 4.1, 4.6$_{\mu m}$ = 3.5, 12$_{\mu m}$ = 3.5, 22$_{\mu m}$ = 2.8\,mag  \\ 
7 & HSC N & 0.249 & 4.63E-07  & 0.9 & >10$^3$ & M05 &   $F_{6cm}$ = 1.9 mJy, $F_{21cm}$ = 3.3 mJy  \\ 
9 & A11 & 0.108 & 1.11E-03  & 0.2 & 443.8 & W14 &   XR$_{f} = 5.08\times10^{-13}$ erg cm$^{-2}$ s$^{-1}$  \\ 
  &   & 0.205 & 1.59E-03  & 0.2 & 306.7 & W09 &  Log L$_{0.5-8 kev} =  32.4 \times 10^{-7}$\, W \\
  &   & 0.596 & 5.60E-05  & 3.7 & 11.0  & 2MASS &   $J = 7.8, H = 7.1, K = 6.7$  \\ 
  &   & 0.373 & 1.33E-03  & 1.7 & 88.0  & SPITZER &   $3.6_{\mu m}$ = 6.4, 4.5$_{\mu m}$ = 6.3, 5.8$_{\mu m}$ = 6.3, 8.0$_{\mu m}$ = 6.2, 24$_{\mu m}$ = 5.4\,mag  \\ 
  &   & 0.186 & 9.52E-06  & 0.7 & >10$^3$ & WISE  &   $3.4_{\mu m}$ = 6.4, 4.6$_{\mu m}$ = 6.1, 12$_{\mu m}$ = 6.2, 22$_{\mu m}$ = 5.6\,mag \\ 
  &   & 1.973 &   &   & 0.0 & M07 &   $F_{49cm}$ = 1.45 mJy  \\ 
11  & SBHW 81 & 0.872 & 2.98E-05  & 0.6 & >10$^3$ & S03 &   $F_{21cm}$ = 3.6 mJy  \\ 
12  & SBHW 83 & 0.762 & 6.00E-06  & 1.2 & >10$^3$ & S03 &   $F_{21cm}$ = 8.4 mJy, $F_{86cm}$ = 62 mJy  \\ 
13  & Cyg 12  & 1.413 & 2.41E-05  & 4.1 & 5.3 & M07 &   $F_{49cm}$ = 0.93 mJy  \\ 
  &   & 0.333 & 7.88E-05  & 0.3 & >10$^3$ & S09 &   $V = 11.5$  \\ 
  &   & 0.554 & 9.75E-05  & 3.9 & 2.3 & R11 &    \\ 
  &   & 0.364 & 1.11E-03  & 0.6 & 375.5 & W14 &   XR$_{f} = 2.37\times10^{-12}$ erg cm$^{-2}$ s$^{-1}$  \\ 
  &   & 0.247 & 1.59E-03  & 0.3 & 303.2 & W09 &  Log L$_{0.5-8 kev} =  33.4 \times 10^{-7}$\, W \\
  &   & 1.180 & 1.10E-03  & 0.8 & 330.9 & 2MASS &   $J = 4.7, H = 3.5, K = 2.7$  \\ 
  &   & 0.796 & 3.95E-07  & 4.7 & 16.6  & SPITZER &   $3.6_{\mu m}$ = N/A, 4.5$_{\mu m}$ = NA, 5.8$_{\mu m}$ = 2.2, 8.0$_{\mu m}$ = 2.3, 24$_{\mu m}$ = 1.7\,mag \\ 
  &   & 0.949 & 1.33E-03  & 2.0 & 52.2  & HERSCHEL  &   $100_{\mu m}$ = 349.2 mJy \\ 
  &   & 0.109 & 7.11E-06  & 2.9 & 937.5 & WISE  &   $3.4_{\mu m}$ = 0.9, 4.6$_{\mu m}$ = 0.7, 12$_{\mu m}$ = 2.2, 22$_{\mu m}$ = 1.8\,mag  \\ 
17  & --  & 2.400 & 5.92E-05  & 3.4 & 25.1  & M07 &   $F_{49cm}$ = 0.67 mJy  \\ 
22  & SBHW 90 & 1.208 & 5.27E-06  & 1.6 & >10$^3$ & S03 &   $F_{21cm}$ = 14.7 mJy, $F_{86cm}$ = 54 mJy  \\ 
25  & W14 3683  & 0.489 & 1.11E-03  & 0.7 & 353.1 & W14 &   XR$_{f} = 4.36\times10^{-15}$ erg cm$^{-2}$ s$^{-1}$  \\
  &   & 0.195 & 1.33E-03  & 1.0 & 233.8 & SPITZER &   $3.6_{\mu m}$ = N/A, 4.5$_{\mu m}$ = 14.5, 5.8$_{\mu m}$ = 13.6, 8.0$_{\mu m}$ = 12.7, 24$_{\mu m}$ = 9.2\,mag  \\
28  & --  & 0.068 & 1.33E-03  & 0.3 & 353.7 & SPITZER &   $3.6_{\mu m}$ = N/A, 4.5$_{\mu m}$ = 14.6, 5.8$_{\mu m}$ = 13.8, 8.0$_{\mu m}$ = 13.2, 24$_{\mu m}$ = 8.3\,mag \\ 
31  & Cyg 9 & 0.292 & 7.88E-05  & 0.2 & >10$^3$ & S09 &   $V = 11.0$  \\ 
  &   & 0.194 & 2.80E-04  & 0.3 & >10$^3$ & R11 &    \\ 
  &   & 0.120 & 1.11E-03  & 0.2 & 442.3 & W14 &   XR$_{f} = 1.39\times10^{-12}$ erg cm$^{-2}$ s$^{-1}$  \\
  &   & 0.837 & 1.59E-03  & 0.2 & 309.3 & W09 &  Log L$_{0.5-8 kev} =  32.7 \times 10^{-7}$\, W \\
  &   & 0.428 & 5.54E-05  & 4.2 & 1.2 & 2MASS &   $J = 6.5, H = 5.9, K = 5.6$  \\ 
  &   & 0.810 & 1.33E-03  & 3.3 & 1.5 & SPITZER &   $3.6_{\mu m}$ = 5.6, 4.5$_{\mu m}$ = 5.2, 5.8$_{\mu m}$ = 5.1, 8.0$_{\mu m}$ = 5.0, 24$_{\mu m}$ = 4.8\,mag \\ 
  &   & 0.132 & 1.38E-05  & 2.5 & >10$^3$ & WISE  &   $3.4_{\mu m}$ = 5.3, 4.6$_{\mu m}$ = 5.0, 12$_{\mu m}$ = 5.0, 22$_{\mu m}$ = 4.7\,mag  \\ 
33  & Cyg 8A  & 0.292 & 7.88E-05  & 0.1 & >10$^3$ & S09 &   $V = 9.1$  \\ 
  &   & 0.749 & 2.80E-04  & 1.2 & 881.2 & R11 &    \\ 
  &   & 0.150 & 1.11E-03  & 0.3 & 437.2 & W14 &   XR$_{f} = 3.84\times10^{-12}$ erg cm$^{-2}$ s$^{-1}$  \\
  &   & 0.410 & 1.59E-03  & 0.1 & 311.5 & W09 &  Log L$_{0.5-8 kev} =  33.3 \times 10^{-7}$\, W \\
  &   & 0.347 & 5.54E-05  & 2.8 & 156.3 & 2MASS &   $J = 6.1, H = 5.7, K = 5.5$  \\ 
  &   & 0.280 & 1.33E-03  & 1.4 & 140.8 & SPITZER &   $3.6_{\mu m}$ = 5.5, 4.5$_{\mu m}$ = 5.3, 5.8$_{\mu m}$ = 5.3, 8.0$_{\mu m}$ = 5.2, 24$_{\mu m}$ = 4.2\,mag \\ 
34  & W14 4790  & 0.148 & 1.11E-03  & 0.2 & 437.5 & W14 &   XR$_{f} = 1.31\times10^{-14}$ erg cm$^{-2}$ s$^{-1}$  \\
  &   & 0.704 & 2.80E-04  & 1.1 & 956.6 & R11 &    \\ 
  &   & 0.731 & 1.59E-03  & 0.2 & 305.6 & W09 &  Log L$_{0.5-8 kev} =  30.9 \times 10^{-7}$\, W \\
  &   & 0.587 & 5.76E-05  & 2.4 & 543.9 & 2MASS &   $J = 11.5, H = 11.0, K = 10.7$  \\ 
  &   & 0.918 & 1.34E-03  & 2.6 & 14.5  & SPITZER &   $3.6_{\mu m}$ = 10.4, 4.5$_{\mu m}$ = 10.3, 5.8$_{\mu m}$ = 10.0, 8.0$_{\mu m}$ = 9.4\,mag  \\ 
36  & --  & 0.422 & 1.33E-03  & 2.1 & 40.8  & SPITZER &   $3.6_{\mu m}$ = 14.5, 4.5$_{\mu m}$ = 14.2  \\ 
40  & W14 5440  & 0.465 & 1.11E-03  & 0.7 & 358.9 & W14 &   XR$_{f} = 1.26\times10^{-14}$ erg cm$^{-2}$ s$^{-1}$  \\ 
47  & SBHW 109  & 1.918 & 3.23E-05  & 1.1 & >10$^3$ & S03 &   $F_{21cm}$ = 3.0 mJy  \\ 
48  & --  & 0.208 & 1.34E-03  & 0.3 & 357.8 & SPITZER &   $3.6_{\mu m}$ = 16.1, 4.5$_{\mu m}$ = 15.8   \\ 
50  & SBHW 110  & 2.978 & 1.35E-05  & 2.6 & >10$^3$ & S03 &   $F_{21cm}$ = 2.2 mJy  \\ 
  &   & 0.148 & 1.11E-03  & 0.2 & 437.6 & W14 &   XR$_{f} = 1.01\times10^{-14}$ erg cm$^{-2}$ s$^{-1}$  \\
  &   & 0.076 & 1.33E-03  & 0.4 & 348.9 & SPITZER &   $3.6_{\mu m}$ = N/A, 4.5$_{\mu m}$ = 15.1, 5.8$_{\mu m}$ = N/A, 8.0$_{\mu m}$ = 13.1, 24$_{\mu m}$ = N/A\,mag\\ 
51  & --  & 2.401 & 1.59E-03  & 2.1 & 38.2  & W09 &  Log L$_{0.5-8 kev} =  31.4 \times 10^{-7}$\, W \\
52  & SBHW 112  & 1.276 & 3.10E-05  & 0.8 & >10$^3$ & S03 &   $F_{21cm}$ = 2.2 mJy  \\ 
53  & --  & 0.820 & 1.34E-03  & 2.7 & 9.0 & SPITZER &   $4.5_{\mu m}$ = 14.7, 8.0$_{\mu m}$ = 12.8   \\ 
56  & G13 318006  & 0.655 & 9.15E-04  & 0.2 & 532.6 & G13 &   $3.6_{\mu m}$ = 14.4, 4.5$_{\mu m}$ = 13.3, 5.8$_{\mu m}$ = 12.4, 8.0$_{\mu m}$ = 11.6, 24$_{\mu m}$ = 7.1\,mag  \\ 
  &   & 0.877 & 8.89E-05  & 2.9 & 78.3  & WISE  &   $3.4_{\mu m}$ = 14.7, 4.6$_{\mu m}$ = 13.5, 12$_{\mu m}$ = 10.4, 22$_{\mu m}$ = 6.8\,mag  \\ 
57  & W14 6616  & 0.755 & 1.11E-03  & 1.3 & 204.6 & W14 &   XR$_{f} = 4.78\times10^{-15}$ erg cm$^{-2}$ s$^{-1}$  \\ 
59  & --  & 0.492 & 1.33E-03  & 2.5 & 18.4  & SPITZER &   $3.6_{\mu m}$ = 15.5, 4.5$_{\mu m}$ = 15.5\,mag   \\ 
60  & G13 319565  & 1.323 & 9.15E-04  & 0.5 & 469.8 & G13 &   $5.8_{\mu m}$ = 8.7, 8.0$_{\mu m}$ = 7.1, 24$_{\mu m}$ = 1.3\,mag   \\ 
 &   & 2.796 & 9.76E-05  & 3.9 & 2.6 & R11 &  \\
61  & W14 7183  & 0.344 & 1.11E-03  & 0.6 & 382.8 & W14 &   XR$_{f} = 1.37\times10^{-14}$ erg cm$^{-2}$ s$^{-1}$  \\ 
\bottomrule
\caption*{\footnotesize \textbf{M07:} \citet{marti_etal_2007}; \textbf{S09:} \citet{skiff_2010}; \textbf{R11:} \citet{rauw_2011}; \textbf{S03:} \citet{setiagunawan_etal_2003}; \textbf{2MASS:} \citet{cutri_etal_2003}; \textbf{M05:} \citet{marti_etal_2005}; \textbf{W14:} \citet{wright_etal_2014a}; \textbf{W09:} \citet{wright_drake_2009}; \textbf{G12:} \citet{guarcello_etal_2012}; \textbf{G13:} \citet{guarcello_etal_2013}.}\\
\end{longtable}
}
\normalsize

\vspace{0.5in}
\noindent Table \ref{tab:identifications} is accompanied by the following information:

\begin{itemize}
\item{\textit{Column(1)}: the ID number of each source from CLASC as given in Table \ref{tab:clasc_main}.}
\item{\textit{Column(2)}: any existing common identifier, the references for which are as follows: `Cyg \#': \citet{schulte_1958}; `A11': \citet{comeron_etal_2002}; `SBHW \#': \citet{setiagunawan_etal_2003}.}
\item{\textit{Column(3)}: the offset between the position of the CLASC source and that of the identification in arcseconds.}
\item{\textit{Columns(4)-(6)}: the calculated values of $\lambda$, $r$ and $LR$ (see Equation \ref{eqn:lr_xc}).}
\item{\textit{Column(7)}: the reference to the observations or survey from which each identification is taken (see footnote of Table \ref{tab:identifications}).}
\item{ \textit{Column(8)}: any additional information regarding the flux density, magnitude, or X-ray flux of each identification.}
\end{itemize}


\clearpage
\setlength\tabcolsep{2pt}
\def\arraystretch{1.3}

\footnotesize{
\begin{longtable}{cccc@{\hskip -0.01in}ccccc@{\hskip -0.01in}cc}
\caption[Resolution Analysis]{Information regarding the spatial extent of each of the 61 CLASC sources on the sky as measured with the e-MERLIN interferometer.}\\
\label{tab:resolution_analysis}\\
\hline \hline
CLASC & ${\rm{Area}}/\theta_{beam}$ & $S_{int}/S_{peak}$ & $\theta_{maj}\theta_{min}/b_{maj}b_{min}$ & \multicolumn{2}{c}{Angular Size (\arcsec; $11^{th}$)} & ${\rm{Area}}/\theta_{beam}$ & $S_{int}/S_{peak}$ & $\theta_{maj}\theta_{min}/b_{maj}b_{min}$ & \multicolumn{2}{c}{Angular Size (\arcsec; $26^{th}$)} \\ 
ID & $11^{th}$ & $11^{th}$  &  $11^{th}$ & Major Axis & Minor Axis & $26^{th}$ & $26^{th}$ & $26^{th}$ & Major Axis & Minor Axis \\
(1) & (2) & (3) & (4) & (5) & (6) & (7) & (8) & (9) & (10) & (11) \\
\hline
\noalign{\vskip 0.05in}
\endfirsthead
\caption{continued.}\\
\hline\hline
CLASC & ${\rm{Area}}/\theta_{beam}$ & $S_{int}/S_{peak}$ & $\theta_{maj}\theta_{min}/b_{maj}b_{min}$ & \multicolumn{2}{c}{Angular Size (\arcsec; $11^{th}$)} & ${\rm{Area}}/\theta_{beam}$ & $S_{int}/S_{peak}$ & $\theta_{maj}\theta_{min}/b_{maj}b_{min}$ & \multicolumn{2}{c}{Angular Size (\arcsec; $26^{th}$)} \\ 
ID & $11^{th}$ & $11^{th}$  &  $11^{th}$ & Major Axis & Minor Axis & $26^{th}$ & $26^{th}$ & $26^{th}$ & Major Axis & Minor Axis \\
(1) & (2) & (3) & (4) & (5) & (6) & (7) & (8) & (9) & (10) & (11) \\
\hline
\noalign{\vskip 0.05in}
\endhead
\hline
\endfoot
\endlastfoot
1 & -- -- & -- -- & -- -- & -- & -- & 1.48 (N) & 1.04 (N) & 1.13 (N) & 0.24 & LAS \\
2 & 2.62 (Y) & 1.54 (N) & 4.85 (Y) & 0.72 & LAS & 5.66 (Y) & 1.84 (N) & 2.12 (Y) & 0.20$\pm$0.01 & 0.16$\pm$0.01 \\
3 & -- -- & -- -- & -- -- & -- & -- & 2.18 (Y) & 1.33 (N) & 2.58 (Y) & 0.32$\pm$0.05 & 0.12$\pm$0.06 \\
4 & -- -- & -- -- & -- -- & -- & -- & 2.70 (Y) & 1.88 (N) & 3.23 (Y) & 0.37 & LAS \\
5 & -- -- & -- -- & -- -- & -- & -- & 1.65 (N) & 0.98 (N) & 2.22 (Y) & 0.28$\pm$0.05 & 0.12$\pm$0.05 \\
6a & -- -- & -- -- & -- -- & -- & -- & 6.57 (Y) & 2.58 (Y) & 123.23 (Y) & 4.73$\pm$0.00 & 0.79$\pm$0.00 \\
6b & 1.91 (Y) & 1.38 (N) & 2.54 (Y) & 0.52 & LAS & 9.62 (Y) & 3.99 (Y) & 13.84 (Y) & 0.77 & LAS \\
7a & 4.84 (Y) & 0.90 (N) & 28.11 (Y) & 0.22$\pm$0.02 & 0.16$\pm$0.03 & 6.49 (Y) & 1.80 (N) & 1.97 (Y) & 0.18$\pm$0.01 & 0.16$\pm$0.01 \\
7b & 1.24 (N) & 1.51 (N) & 32.02 (Y) & 0.25$\pm$0.18 & 0.19$\pm$0.16 & -- -- & -- -- & -- -- & -- & -- \\
8 & 1.03 (N) & 0.73 (N) & 1.58 (N) & 0.37 & LAS & -- -- & -- -- & -- -- & -- & -- \\
9 & -- -- & -- -- & -- -- & -- & -- & 1.92 (Y) & 0.94 (N) & 1.22 (N) & 0.12$\pm$0.03 & 0.01$\pm$0.03 \\
10 & 1.05 (N) & 0.81 (N) & 1.31 (N) & 0.36 & LAS & 3.31 (Y) & 1.51 (N) & 2.85 (Y) & 0.31 & LAS \\
11a & 4.98 (Y) & 2.43 (Y) & 6.34 (Y) & 0.70 & LAS & 25.60 (Y) & 3.66 (Y) & 36.69 (Y) & 1.36 & LAS \\
11b & -- -- & -- -- & -- -- & -- & -- & 2.00 (Y) & 1.48 (N) & 2.11 (Y) & 0.26 & LAS \\
12a & 6.20 (Y) & 3.57 (Y) & 19.16 (Y) & 1.87 & LAS & 22.33 (Y) & 8.37 (Y) & 21.20 (Y) & 1.12 & LAS \\
12b & -- -- & -- -- & -- -- & -- & -- & 7.75 (Y) & 4.54 (Y) & 6.64 (Y) & 0.49 & LAS \\
12c & 1.12 (N) & 0.84 (N) & 1.17 (N) & 0.34 & LAS & 11.54 (Y) & 7.34 (Y) & 4.84 (Y) & 0.43 & LAS \\
13 & 3.15 (Y) & 1.84 (N) & 13.03 (Y) & 1.14 & LAS & 21.11 (Y) & 4.78 (Y) & 22.40 (Y) & 1.11 & LAS \\
14 & -- -- & -- -- & -- -- & -- & -- & 1.31 (N) & 0.88 (N) & 1.77 (N) & 0.21$\pm$0.07 & 0.08$\pm$0.08 \\
15 & -- -- & -- -- & -- -- & -- & -- & 1.09 (N) & 0.86 (N) & 1.50 (N) & 0.27 & LAS \\
16 & -- -- & -- -- & -- -- & -- & -- & 2.66 (Y) & 1.06 (N) & 28.69 (Y) & 1.52 & LAS \\
17 & -- -- & -- -- & -- -- & -- & -- & 1.83 (N) & 1.25 (N) & 1.63 (N) & 0.21 & LAS \\
18 & -- -- & -- -- & -- -- & -- & -- & 2.66 (Y) & 1.88 (N) & 6.21 (Y) & 0.70 & LAS \\
19 & -- -- & -- -- & -- -- & -- & -- & 2.53 (Y) & 1.72 (N) & 2.44 (Y) & 0.29 & LAS \\
20 & -- -- & -- -- & -- -- & -- & -- & 1.04 (N) & 0.74 (N) & 1.59 (N) & 0.20$\pm$0.08 & 0.04$\pm$0.07 \\
21 & -- -- & -- -- & -- -- & -- & -- & 6.40 (Y) & 3.71 (Y) & 5.19 (Y) & 0.48 & LAS \\
22a & 16.22 (Y) & 3.15 (Y) & 22.39 (Y) & 1.57 & LAS & 38.88 (Y) & 4.03 (Y) & 63.99 (Y) & 2.10 & LAS \\
22b & -- -- & -- -- & -- -- & -- & -- & 2.87 (Y) & 1.82 (N) & 1.86 (N) & 0.25 & LAS \\
23 & 1.03 (N) & 0.76 (N) & 1.51 (N) & 0.41 & LAS & -- -- & -- -- & -- -- & -- & -- \\
24 & -- -- & -- -- & -- -- & -- & -- & 1.13 (N) & 0.76 (N) & 1.78 (N) & 0.29$\pm$0.06 & 0.00$\pm$0.03 \\
25 & 1.76 (N) & 0.94 (N) & 2.31 (Y) & 0.46 & LAS & 4.40 (Y) & 1.53 (N) & 1.94 (Y) & 0.19$\pm$0.02 & 0.15$\pm$0.02 \\
26 & -- -- & -- -- & -- -- & -- & -- & 1.39 (N) & 1.05 (N) & 1.75 (N) & 0.29 & LAS \\
27 & -- -- & -- -- & -- -- & -- & -- & 1.31 (N) & 0.95 (N) & 2.23 (Y) & 0.28$\pm$0.08 & 0.11$\pm$0.09 \\
28 & 2.12 (Y) & 1.14 (N) & 31.00 (Y) & 0.27$\pm$0.07 & 0.14$\pm$0.12 & 3.40 (Y) & 1.53 (N) & 1.86 (N) & 0.24$\pm$0.02 & 0.04$\pm$0.04 \\
29 & 1.79 (N) & 1.04 (N) & 29.46 (Y) & 0.29$\pm$0.07 & 0.09$\pm$0.09 & 4.48 (Y) & 1.80 (N) & 2.27 (Y) & 0.21$\pm$0.03 & 0.18$\pm$0.03 \\
30 & -- -- & -- -- & -- -- & -- & -- & 1.04 (N) & 0.79 (N) & 1.33 (N) & 0.21 & LAS \\
31 & 2.96 (Y) & 1.02 (N) & 3.90 (Y) & 0.59 & LAS & 3.35 (Y) & 1.01 (N) & 1.06 (N) & 0.06$\pm$0.01 & 0.00$\pm$0.00 \\
32 & -- -- & -- -- & -- -- & -- & -- & 2.31 (Y) & 1.54 (N) & 1.38 (N) & 0.27 & LAS \\
33a & 4.70 (Y) & 1.45 (N) & 9.51 (Y) & 1.04 & LAS & 3.79 (Y) & 1.05 (N) & 2.90 (Y) & 0.22 & LAS \\
33b & -- -- & 1.45 (N) & -- -- & -- & -- & 1.18 (N) & 0.85 (N) & 0.42 (N) & 0.04$\pm$0.06 & 0.00$\pm$0.00 \\
34 & -- -- & -- -- & -- -- & -- & -- & 1.04 (N) & 0.69 (N) & 1.30 (N) & 0.20 & LAS \\
35 & 1.14 (N) & 0.82 (N) & 24.91 (Y) & 0.22$\pm$0.16 & 0.00$\pm$0.12 & 1.96 (Y) & 1.12 (N) & 1.83 (N) & 0.17$\pm$0.06 & 0.14$\pm$0.05 \\
36 & 1.12 (N) & 0.79 (N) & 1.74 (N) & 0.38 & LAS & 4.35 (Y) & 2.19 (Y) & 4.35 (Y) & 0.38 & LAS \\
37 & 0.95 (N) & 0.69 (N) & 1.44 (N) & 0.33 & LAS & 2.53 (Y) & 1.24 (N) & 1.84 (N) & 0.22$\pm$0.03 & 0.07$\pm$0.06 \\
38 & 4.89 (Y) & 1.39 (N) & 5.45 (Y) & 0.64 & LAS & 5.44 (Y) & 1.46 (N) & 1.54 (N) & 0.15$\pm$0.00 & 0.10$\pm$0.01 \\
39 & -- -- & -- -- & -- -- & -- & -- & 1.04 (N) & 0.78 (N) & 1.06 (N) & 0.20 & LAS \\
40 & 0.95 (N) & 0.70 (N) & 20.15 (Y) & 0.15$\pm$0.10 & 0.00$\pm$0.12 & 2.53 (Y) & 1.30 (N) & 1.91 (Y) & 0.22$\pm$0.03 & 0.11$\pm$0.04 \\
41a & 1.41 (N) & 1.06 (N) & 1.63 (N) & 0.39 & LAS & 3.00 (Y) & 1.70 (N) & 2.30 (Y) & 0.22 & LAS \\
41b & 3.46 (Y) & 2.43 (Y) & 6.91 (Y) & 1.07 & LAS & 6.79 (Y) & 3.12 (Y) & 7.11 (Y) & 0.78 & LAS \\
42 & 3.12 (Y) & 1.90 (N) & 4.70 (Y) & 0.71 & LAS & 6.23 (Y) & 2.78 (Y) & 5.58 (Y) & 0.41 & LAS \\
43 & -- -- & -- -- & -- -- & -- & -- & 2.09 (Y) & 1.30 (N) & 2.58 (Y) & 0.27$\pm$0.06 & 0.17$\pm$0.05 \\
44 & -- -- & -- -- & -- -- & -- & -- & 1.04 (N) & 0.79 (N) & 1.04 (N) & 0.20 & LAS \\
45 & 5.46 (Y) & 3.28 (Y) & 7.02 (Y) & 0.75 & LAS & 20.55 (Y) & 10.32 (Y) & 21.28 (Y) & 0.90 & LAS \\
46 & 1.45 (N) & 1.10 (N) & 2.70 (Y) & 0.56 & LAS & -- -- & -- -- & -- -- & -- & -- \\
47a & -- -- & -- -- & -- -- & -- & -- & 24.73 (Y) & 17.62 (Y) & 41.61 (Y) & 1.33 & LAS \\
47b & 2.38 (Y) & 1.75 (N) & 3.10 (Y) & 0.68 & LAS & 14.80 (Y) & 9.14 (Y) & 18.81 (Y) & 0.85 & LAS \\
48 & 0.95 (N) & 0.76 (N) & 1.30 (N) & 0.32 & LAS & 2.35 (Y) & 1.36 (N) & 1.87 (N) & 0.20 & LAS \\
49 & -- -- & -- -- & -- -- & -- & -- & 3.48 (Y) & 2.71 (Y) & 3.88 (Y) & 0.47 & LAS \\
50 & 4.41 (Y) & 1.34 (N) & 6.67 (Y) & 0.67 & LAS & 5.09 (Y) & 1.36 (N) & 1.41 (N) & 0.12$\pm$0.00 & 0.09$\pm$0.01 \\
51 & -- -- & -- -- & -- -- & -- & -- & 1.09 (N) & 0.80 (N) & 1.23 (N) & 0.23 & LAS \\
52 & 6.80 (Y) & 2.07 (Y) & 13.15 (Y) & 1.34 & LAS & 9.53 (Y) & 1.94 (Y) & 1.95 (Y) & 0.18$\pm$0.01 & 0.16$\pm$0.01 \\
53a & -- -- & -- -- & -- -- & -- & -- & 2.22 (Y) & 1.48 (N) & 2.89 (Y) & 0.33$\pm$0.07 & 0.15$\pm$0.07 \\
53b & -- -- & -- -- & -- -- & -- & -- & 2.13 (Y) & 1.55 (N) & 2.57 (Y) & 0.40 & LAS \\
54 & 2.34 (Y) & 1.23 (N) & 29.86 (Y) & 0.26$\pm$0.05 & 0.10$\pm$0.10 & 5.18 (Y) & 2.20 (Y) & 2.95 (Y) & 0.27$\pm$0.03 & 0.22$\pm$0.03 \\
55 & 3.20 (Y) & 1.73 (N) & 4.36 (Y) & 0.70 & LAS & 7.66 (Y) & 3.20 (Y) & 5.59 (Y) & 0.40 & LAS \\
56 & 1.05 (N) & 0.52 (N) & 16.97 (Y) & 0.21$\pm$0.08 & 0.00$\pm$0.00 & 4.35 (Y) & 2.26 (Y) & 5.10 (Y) & 0.38 & LAS \\
57 & -- -- & -- -- & -- -- & -- & -- & 5.09 (Y) & 3.38 (Y) & 26.96 (Y) & 1.23 & LAS \\
58 & -- -- & -- -- & -- -- & -- & -- & 1.04 (N) & 0.78 (N) & 1.59 (N) & 0.26 & LAS \\
59 & 1.96 (Y) & 1.46 (N) & 1.40 (N) & 0.40 & LAS & 14.54 (Y) & 6.51 (Y) & 10.70 (Y) & 0.59 & LAS \\
60 & -- -- & -- -- & -- -- & -- & -- & 1.13 (N) & 0.79 (N) & 1.78 (N) & 0.19$\pm$0.08 & 0.12$\pm$0.08 \\
61 & -- -- & -- -- & -- -- & -- & -- & 6.88 (Y) & 3.97 (Y) & 7.55 (Y) & 0.57 & LAS \\
\bottomrule
\caption*{\footnotesize Notes: ${\rm{Size}}/\theta_{beam}$ has been measured within SEAC (i.e. down to a level of $3\sigma$) and gives a quantitative measure of the source size in pixels compared to that of the beam size ($\theta_{beam}$). Major and minor axis measurements when given are deconvolved. LAS is the convolved largest angular size.} 
\end{longtable}
}
\normalsize

\vspace{0.5in}
\noindent Table \ref{tab:resolution_analysis} and is accompanied by the following information:

\begin{itemize}
\item{\textit{Column(1)}: the ID number of each source from CLASC as given in Table \ref{tab:clasc_main}.}
\item{\textit{Columns(2) and (7)}: the source area ($S$: in pixels) divided by the beam area ($\theta_{beam}$: in pixels), as measured by SEAC for the sources detected in the April 11 and April 26 epochs respectively. An indication of whether each source is resolved (Y) or not (N) is given in brackets following each measurement.}
\item{\textit{Columns(3) and (8)}: the ratio of the integrated flux density ($S_{int}$) to the peak pixel flux density ($S_{peak}$) of each of the CLASC sources (as measured by SEAC), for the April 11 and April 26 epochs respectively. Similarly to \textit{columns(2) and (7)}, an indication of whether each source is resolved (Y) or not (N) is given in brackets following each measurement (please see to Figure \ref{fig:resolution_total} for further information).}
\item{\textit{Columns(4) and (9)}: the convolved angular area of each CLASC source as measured from JMFIT, i.e. $\theta_{maj}$ and $\theta_{min}$, divided by the beam area, i.e. $b_{maj}$ and $b_{min}$, for the April 11 and April 26 epochs respectively.}
\item{ \textit{Columns(5)-(6) and (10)-(11))}: the major and minor axes of the deconvolved angular sizes of each of the sources within the CLASC, across both the April 11 and April 26 epochs. }
\end{itemize}



\clearpage
\newpage

\counterwithin{figure}{section}
\clearpage
\section{Figures}

\begin{center}
\begin{figure*}[!htp]
\begin{center}
\begin{subfigure}[b]{\textwidth}
\hspace*{-0.4cm}
\includegraphics[height=22cm,width=\textwidth]{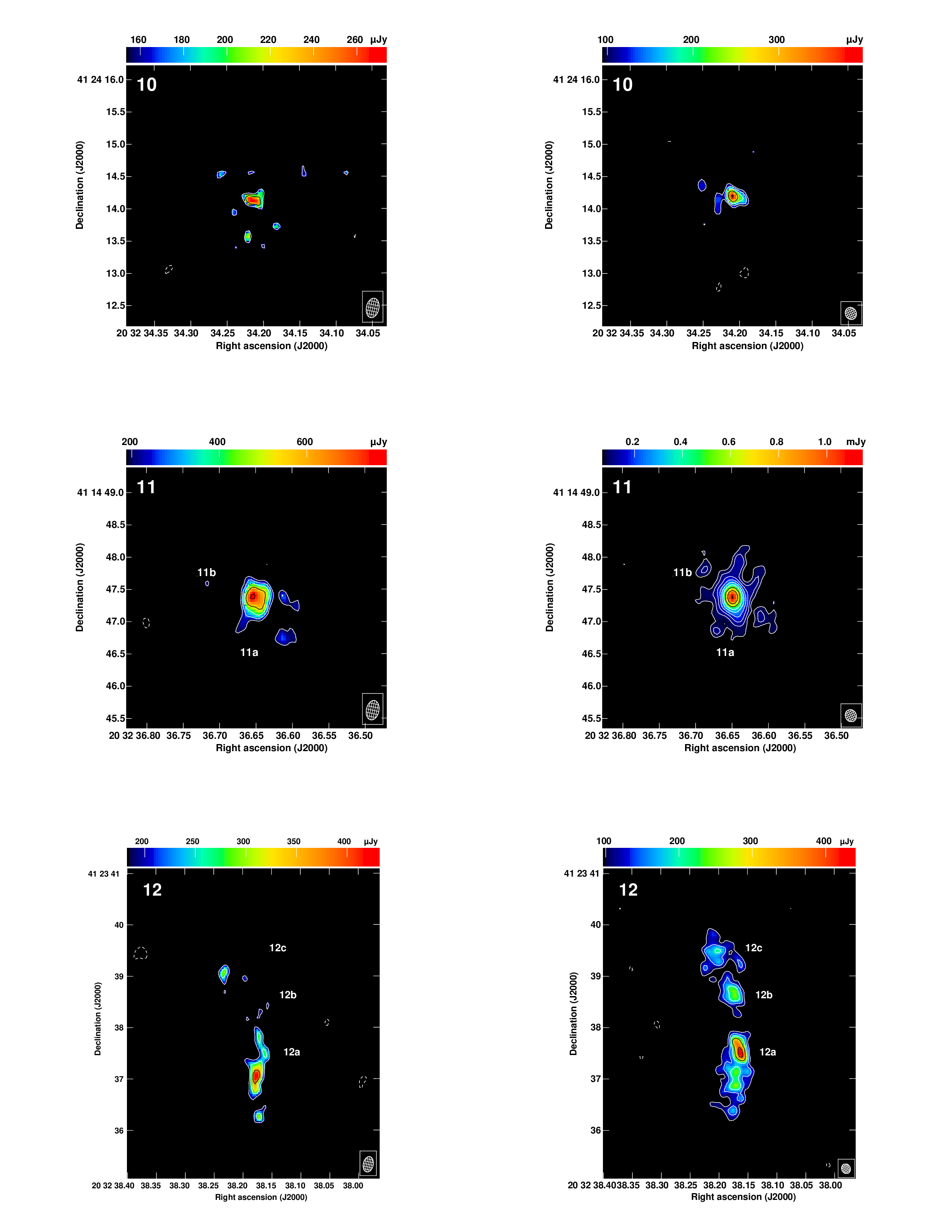}
\caption{}
\label{fig:both_2}
\end{subfigure}
\caption[CLASC sources detected in both epochs]{CLASC sources detected in both April 11 (left) and 26 (right) epochs continued from Fig. \ref{fig:both_1}. The colour scale ranges from the 3$\times \sigma_{RMS}$ to the peak flux density. The contours are plotted at -1, 1, 1.4, 2, 2.8, 4, 5.6, 8, 11.3, 16) $\times$ 3$\sigma_{RMS}$. }
\label{fig:both_app}
\end{center}
\end{figure*}
\end{center}

\begin{figure*}[!htp]\ContinuedFloat
\begin{center}
\begin{subfigure}[b]{\textwidth}
\hspace*{-0.4cm}
\includegraphics[scale=0.85]{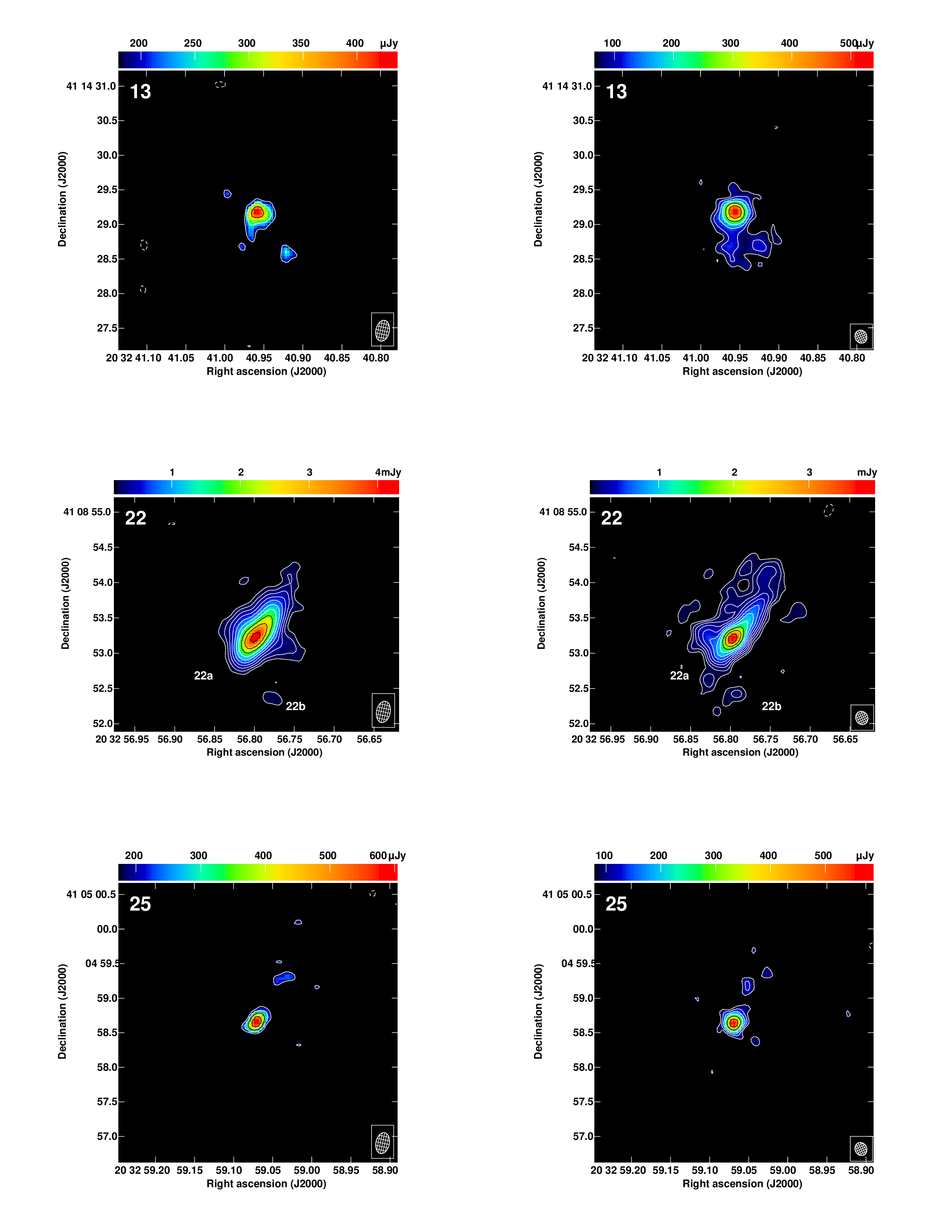}
\caption[CLASC sources detected in both epochs]{Continued from Fig. \ref{fig:both_1}. }
\label{fig:both_3}
\end{subfigure}
\end{center}
\end{figure*}

\begin{figure*}[!htp]\ContinuedFloat
\begin{center}
\begin{subfigure}[b]{\textwidth}
\hspace*{-0.4cm}
\includegraphics[scale=0.85]{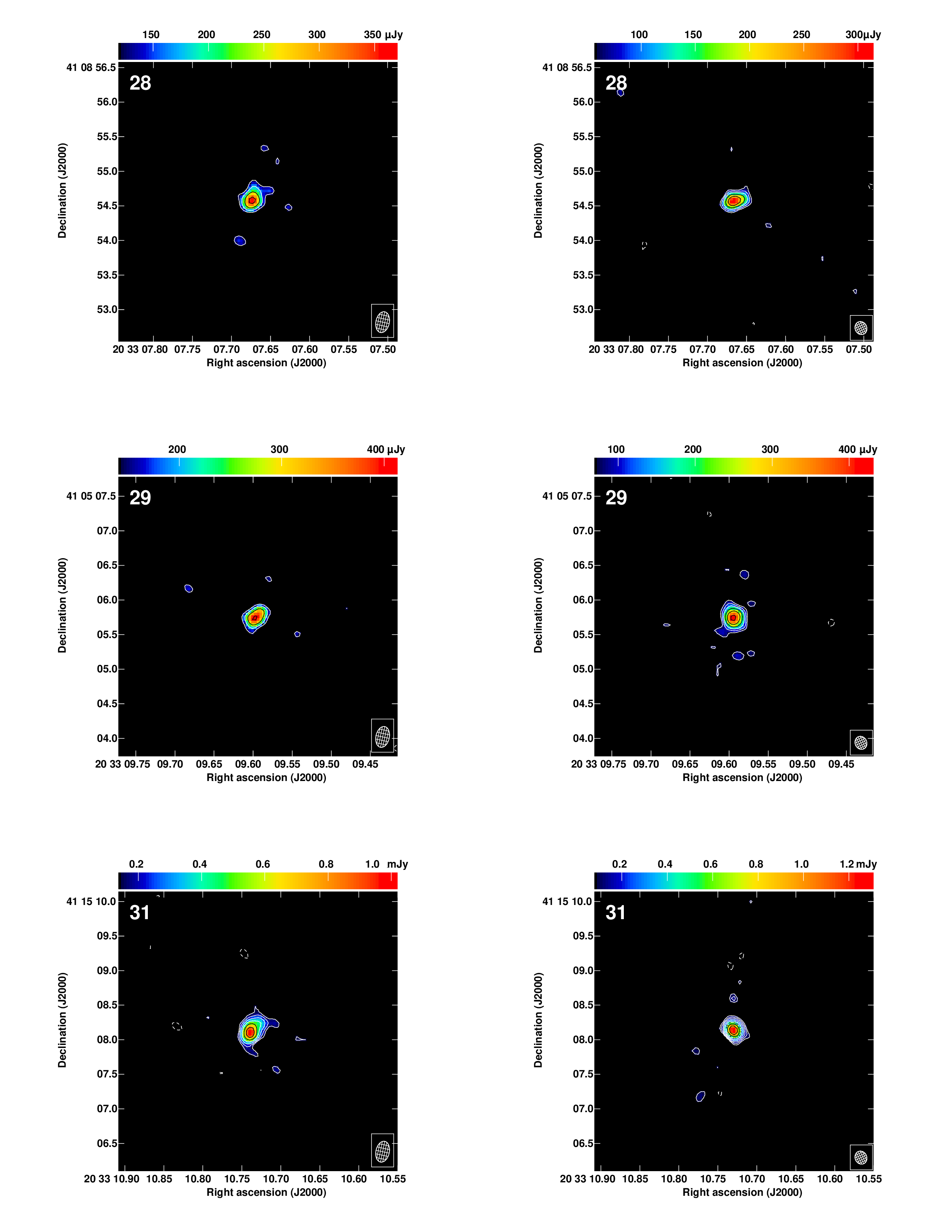}
\caption[CLASC sources detected in both epochs]{Continued from Fig. \ref{fig:both_1}. }
\label{fig:both_4}
\end{subfigure}
\end{center}
\end{figure*}

\begin{figure*}[!htp]\ContinuedFloat
\begin{center}
\begin{subfigure}[b]{\textwidth}
\hspace*{-0.4cm}
\includegraphics[scale=0.85]{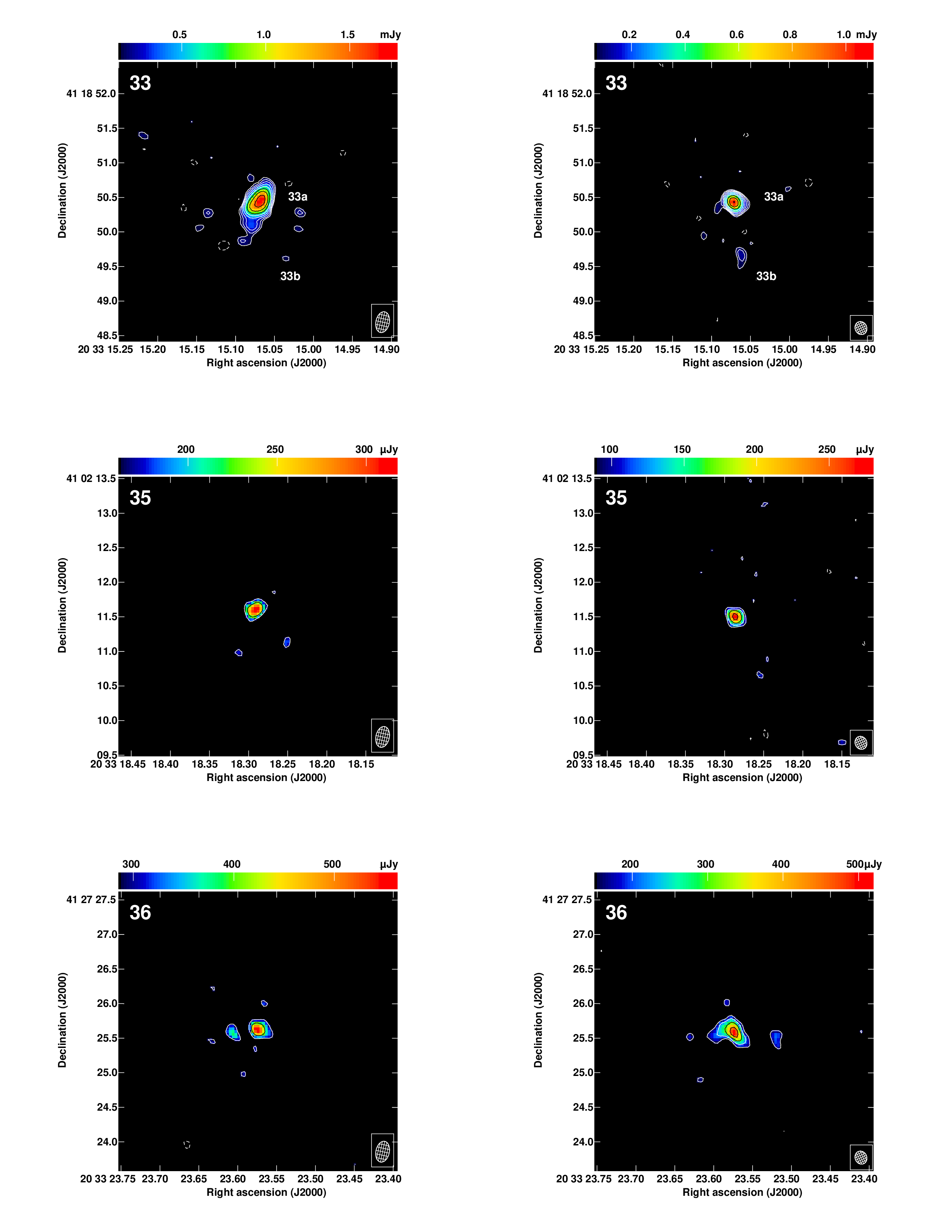}
\caption[CLASC sources detected in both epochs]{Continued from Fig. \ref{fig:both_1}. }
\label{fig:both_5}
\end{subfigure}
\end{center}
\end{figure*}

\begin{figure*}[!htp]\ContinuedFloat
\begin{center}
\begin{subfigure}[b]{\textwidth}
\hspace*{-0.4cm}
\includegraphics[scale=0.85]{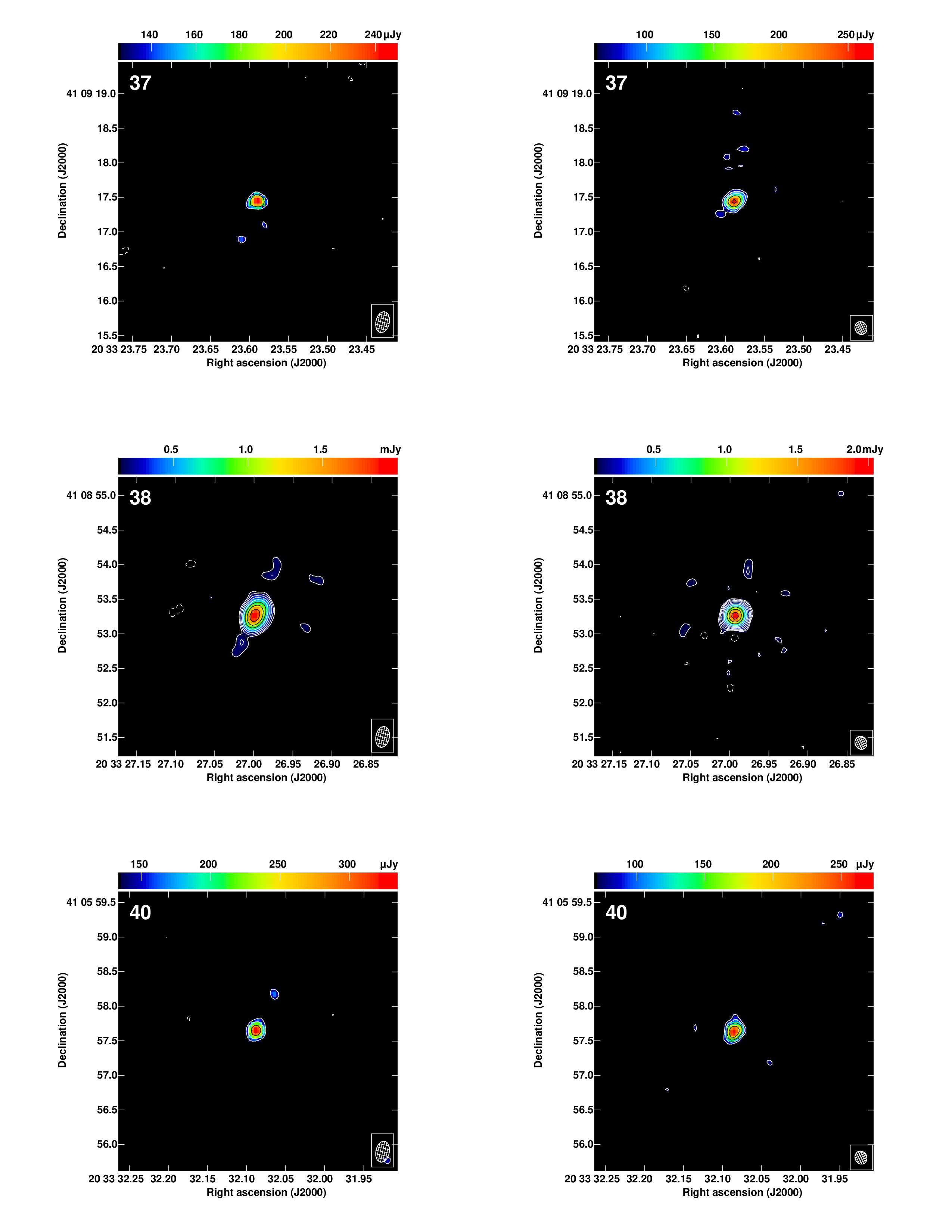}
\caption[CLASC sources detected in both epochs]{Continued from Fig. \ref{fig:both_1}. }
\label{fig:both_6}
\end{subfigure}
\end{center}
\end{figure*}

\begin{figure*}[!htp]\ContinuedFloat
\begin{center}
\begin{subfigure}[b]{\textwidth}
\hspace*{-0.4cm}
\includegraphics[scale=0.85]{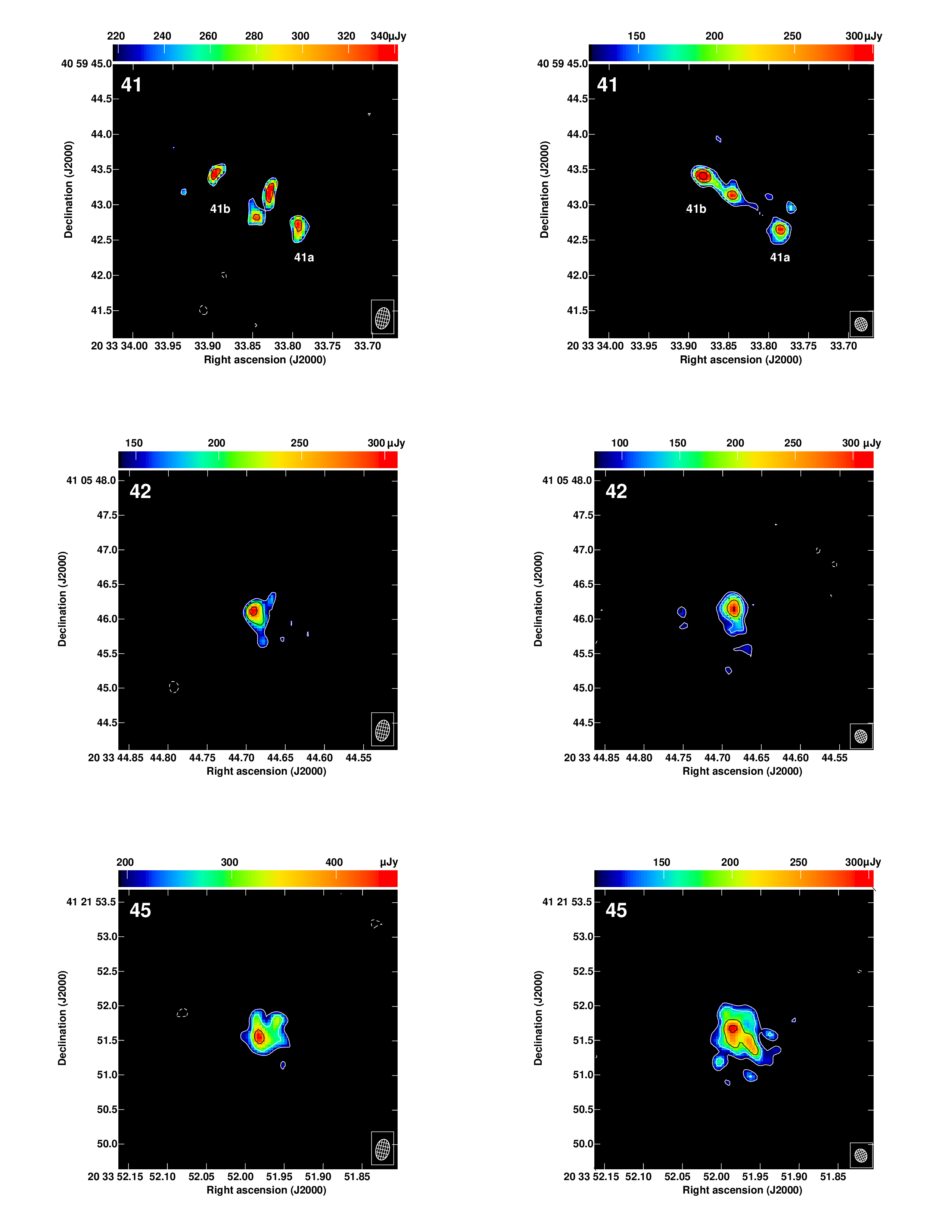}
\caption[CLASC sources detected in both epochs]{Continued from Fig. \ref{fig:both_1}. }
\label{fig:both_7}
\end{subfigure}
\end{center}
\end{figure*}

\begin{figure*}[!htp]\ContinuedFloat
\begin{center}
\begin{subfigure}[b]{\textwidth}
\hspace*{-0.4cm}
\includegraphics[scale=0.85]{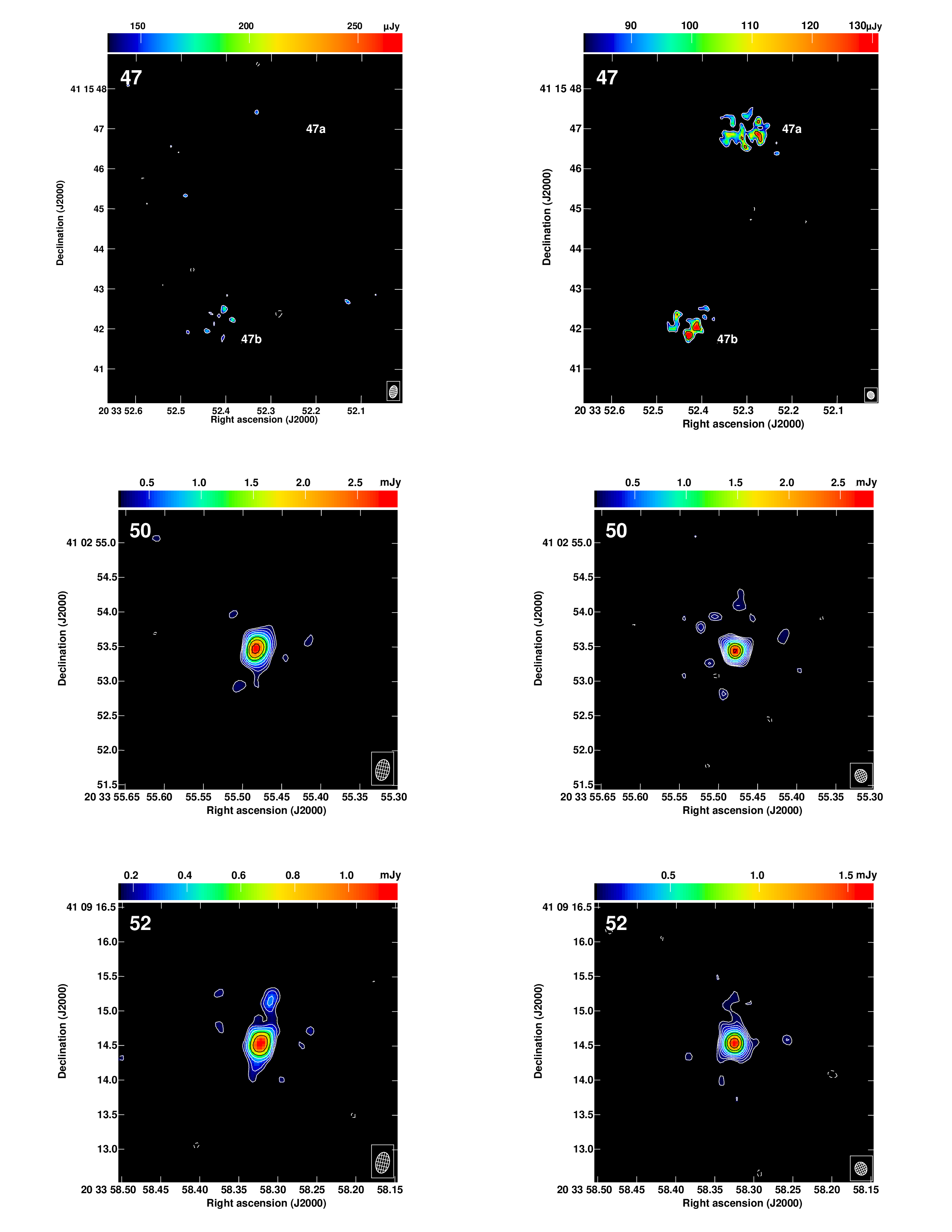}
\caption[CLASC sources detected in both epochs]{Continued from Fig. \ref{fig:both_1}. }
\label{fig:both_8}
\end{subfigure}
\end{center}
\end{figure*}

\begin{figure*}[!htp]\ContinuedFloat
\begin{center}
\begin{subfigure}[b]{\textwidth}
\hspace*{-0.4cm}
\includegraphics[scale=0.91]{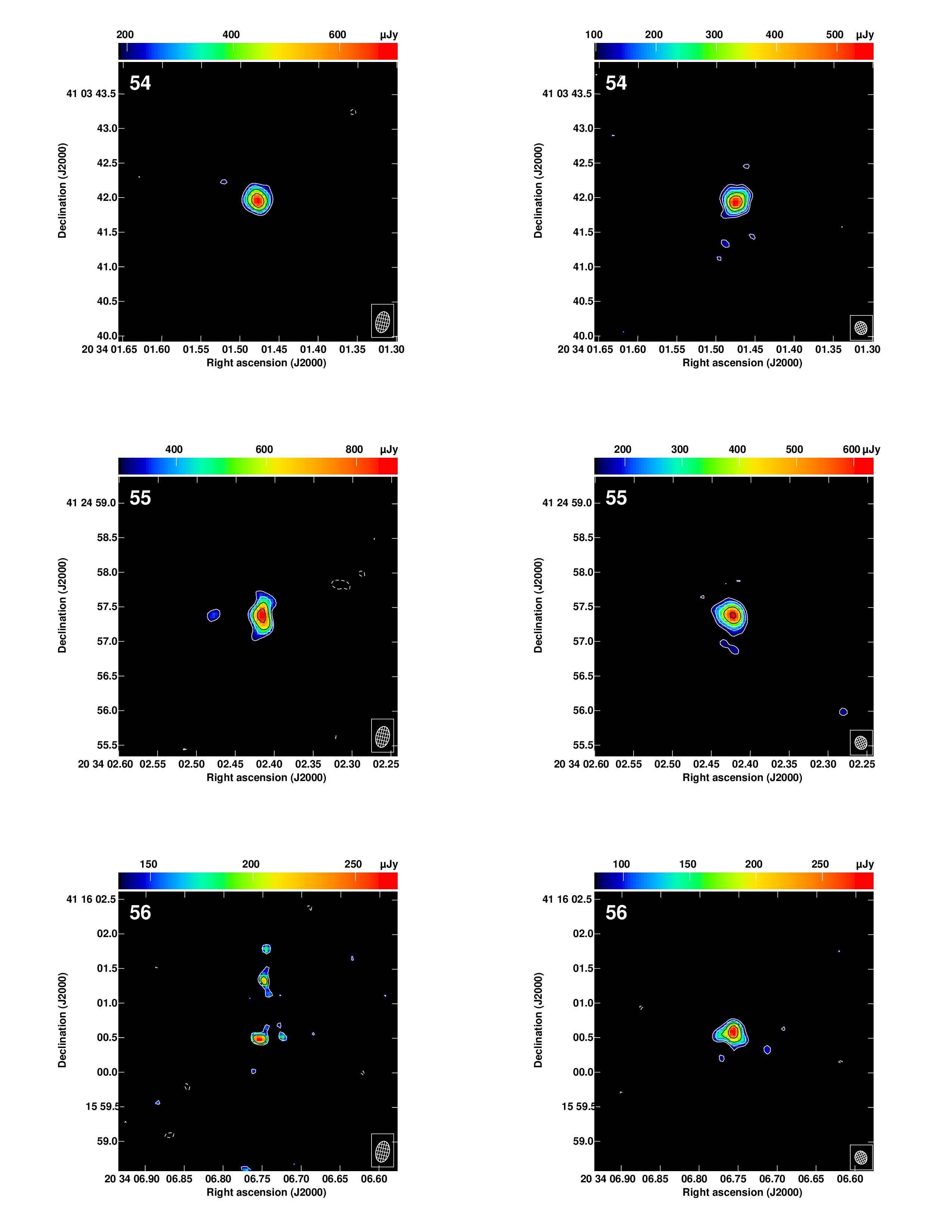}
\caption[CLASC sources detected in both epochs]{Continued from Fig. \ref{fig:both_1}. }
\label{fig:both_9}
\end{subfigure}
\end{center}
\end{figure*}


\begin{figure*}[!htp]\ContinuedFloat
\begin{center}
\begin{subfigure}[b]{\textwidth}
\hspace*{-0.4cm}
\includegraphics[scale=0.91]{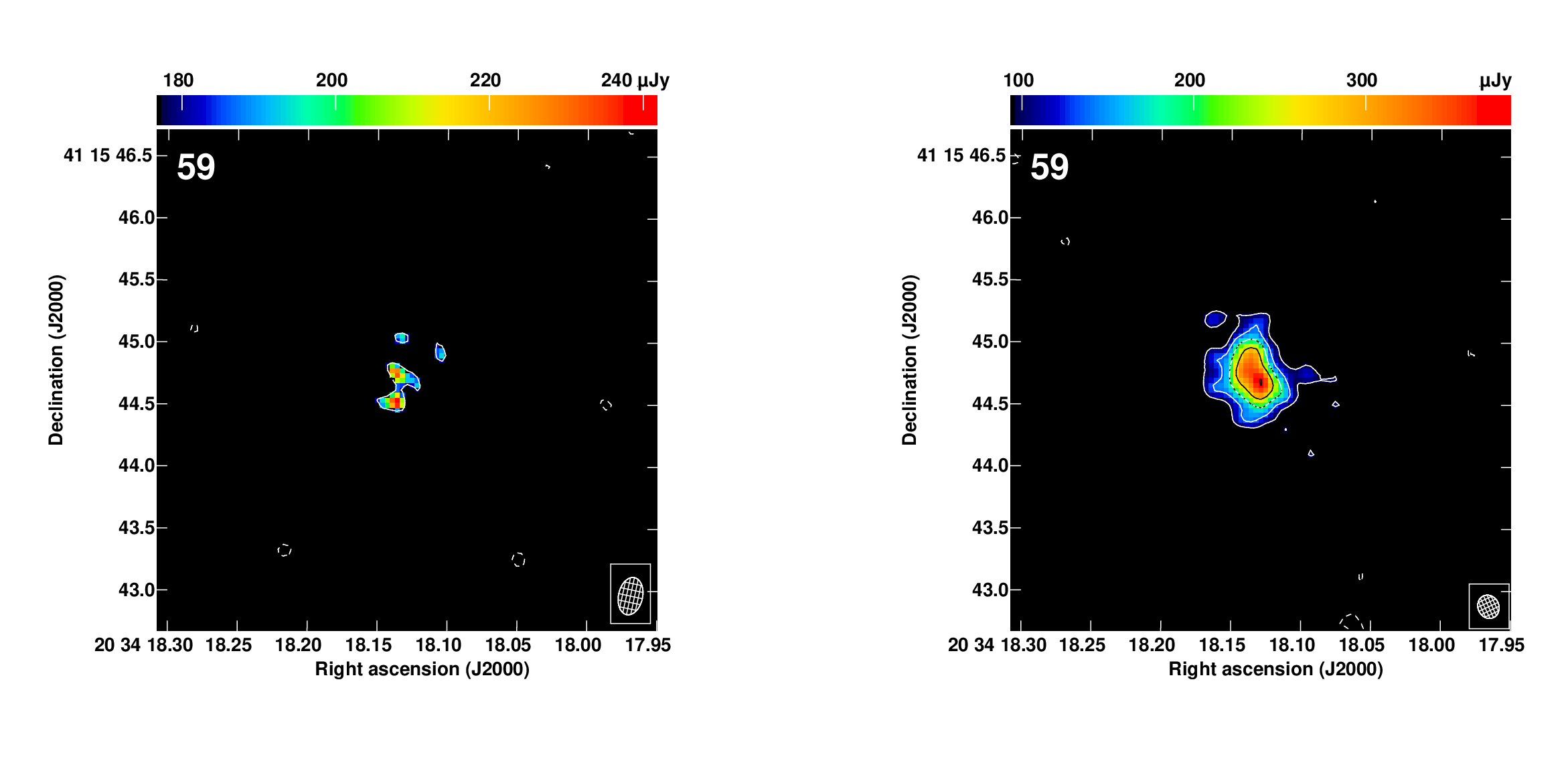}
\caption[CLASC sources detected in both epochs]{Fig. \ref{fig:both_1} continued.}
\label{fig:both_10}
\end{subfigure}
\end{center}
\end{figure*}

\newpage

\begin{figure*}[htp]
\begin{center}
\begin{subfigure}[b]{\textwidth}
\includegraphics[scale=0.82]{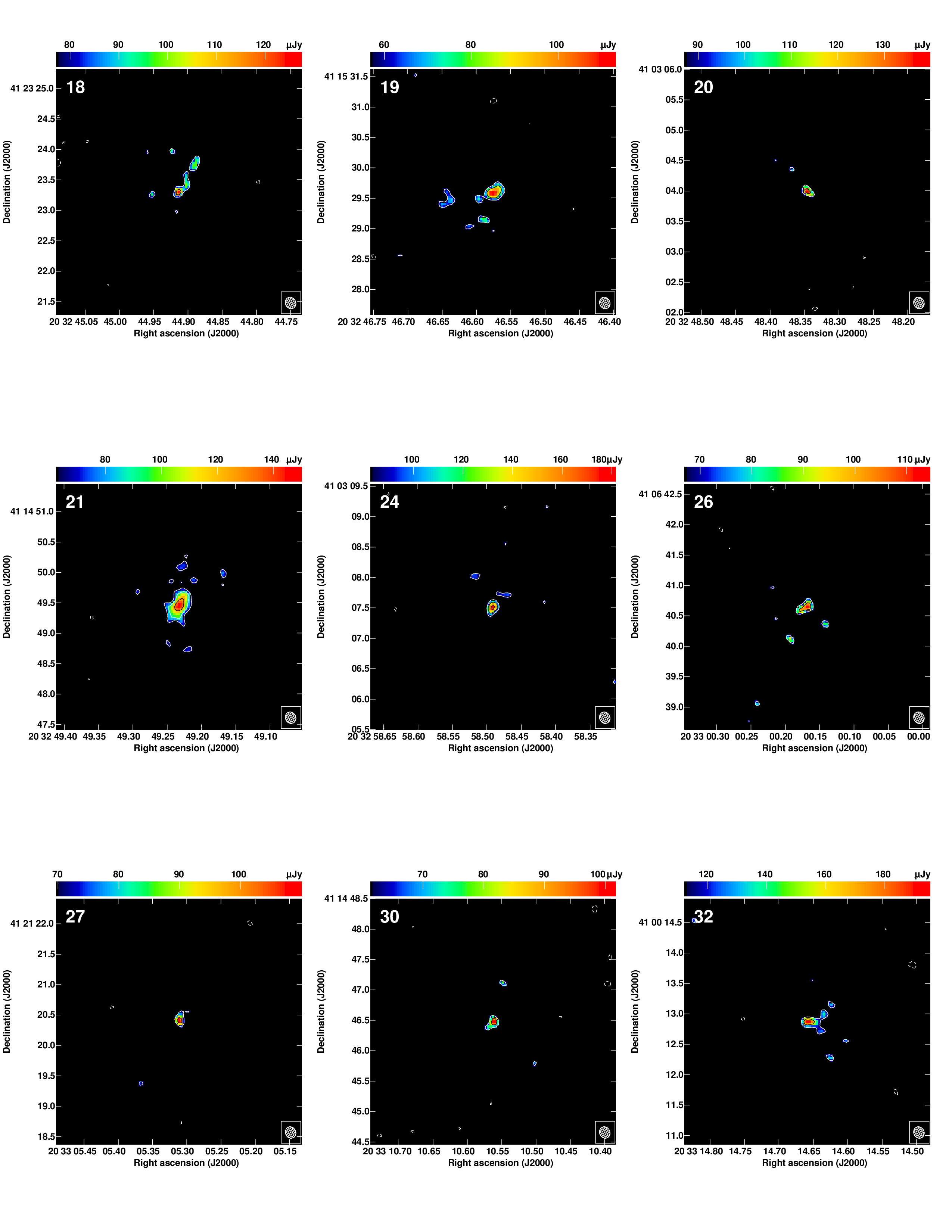}
\caption{}
\label{fig:26_2}
\end{subfigure}
\caption[CLASC sources detected in only the April 26 epoch]{CLASC sources detected in only the April 26 epoch continued from Fig. \ref{fig:26_1}. The colour-scale ranges from the 3$\times \sigma_{RMS}$ to the peak flux density. The contours are plotted at -1, 1, 1.4, 2, 2.8, 4, 5.6, 8, 11.3, 16) $\times$ 3$\times\sigma_{RMS}$.}
\label{fig:26_app}
\end{center}
\end{figure*}

\begin{figure*}[htp]\ContinuedFloat
\begin{center}
\begin{subfigure}[b]{\textwidth}
\includegraphics[scale=0.85]{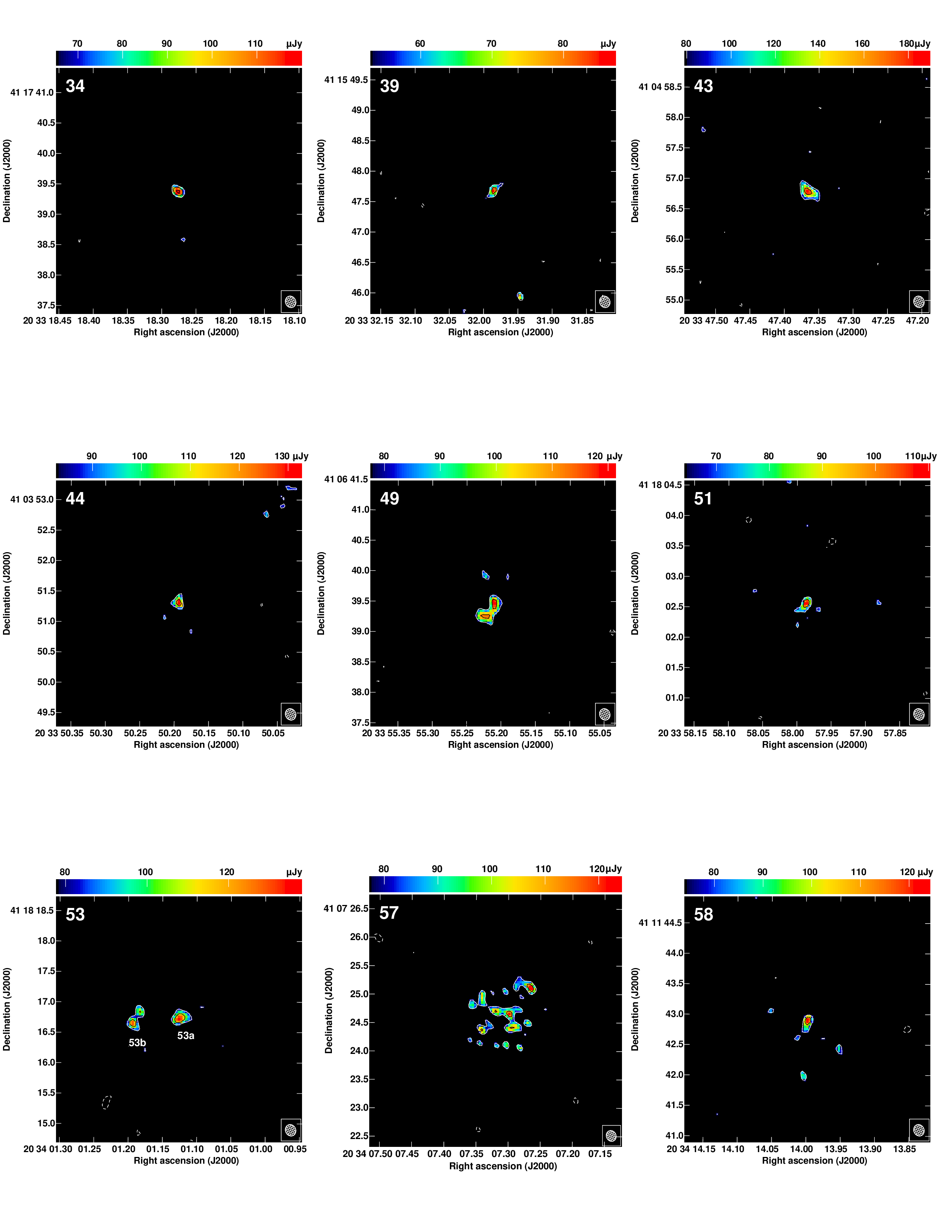}
\caption[CLASC sources detected in only the April 26 epochs]{Continued from Fig. \ref{fig:26_1}.}
\label{fig:26_3}
\end{subfigure}
\end{center}
\end{figure*}

\begin{figure*}[htp]\ContinuedFloat
\begin{center}
\begin{subfigure}[b]{\textwidth}
\centering
\includegraphics[scale=0.85]{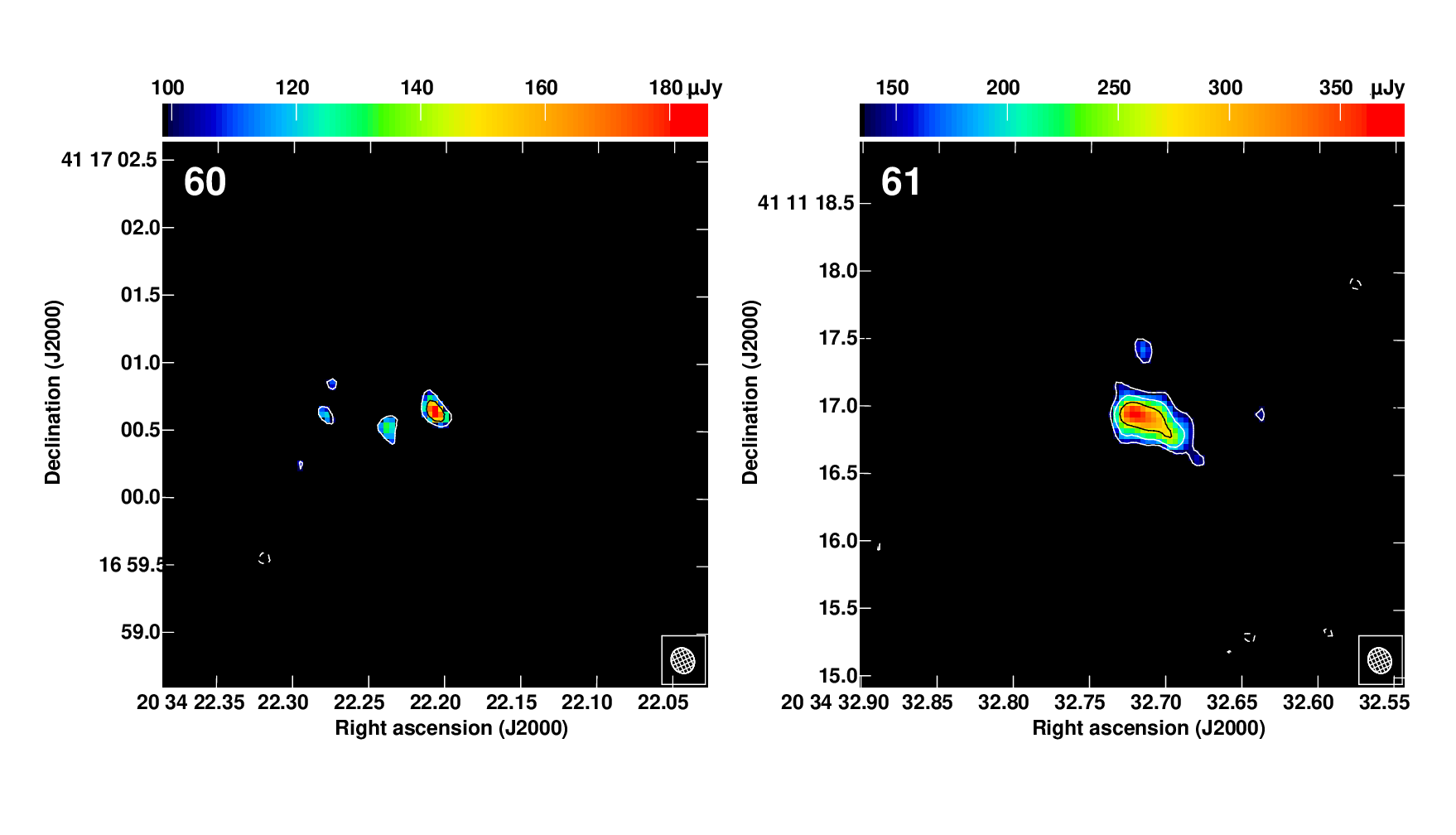}
\caption[CLASC sources detected in only the April 26 epochs]{Continued from Fig. \ref{fig:26_1}.}
\label{fig:26_4}
\end{subfigure}
\end{center}
\end{figure*}

\end{document}